\documentclass{aa} 
\usepackage{graphicx}
\usepackage{psfig}
\usepackage{epsfig}
\begin{document}
\title{The Geneva-Copenhagen Survey of the Solar neighbourhood II
\thanks{Based in part on observations made with the Danish 0.5-m and 1.5-m
telescopes at ESO, La Silla, Chile. The complete Table 1 is only 
available electronically from the CDS via anonymous ftp to 
cdsarc.u-strasbg.fr or via 
http://cdsweb.u-strasbg.fr/cgi-bin/qcat?J/A+A/???/???
}
}
\subtitle{New {\it uvby} calibrations and rediscussion of stellar ages, the 
G dwarf problem, age-metallicity diagram, and heating mechanisms of the disk}

\titlerunning{The Geneva-Copenhagen Survey of the Solar neighbourhood II}

\author{J. Holmberg \inst{1,2,3,4}
 \and
 B. Nordstr\"om \inst{3}
 \and
 J. Andersen \inst{3,4}
}
\offprints{Johan Holmberg (E-mail address: {\it johan@astro.ku.dk})} 
\institute{
   Max-Planck-Institut f{\"u}r Astronomie, K{\"o}nigstuhl 17, 
   DE-69117 Heidelberg, Germany
 \and
    Tuorla Observatory, V\"ais\"al\"antie 20, 
    FI-21500 Piikki\"o, Finland
  \and
    The Niels Bohr Institute, Astronomy Group, Juliane Maries Vej 30,
    DK-2100 Copenhagen, Denmark
  \and
    Nordic Optical Telescope Scientific Association, Apartado 474,
    ES-38700 Santa Cruz de La Palma, Spain.
}
\date{Received February 2, 2007; accepted June 19, 2007}

\abstract 
{Ages, metallicities, space velocities, and Galactic orbits of stars in the
Solar neighbourhood are fundamental observational constraints on models of 
galactic disk evolution. Understanding and minimising systematic errors and 
sample selection biases in the data is crucial for their interpretation.
}
{We aim to consolidate the calibrations of {\it uvby}$\beta$ photometry into 
$\rm T_{eff}$, [Fe/H], distance, and age for F and G stars and rediscuss the 
results of the Geneva-Copenhagen Survey (Nordstr{\"o}m et al. 2004; GCS) in 
terms of the evolution of the disk.
}
{We use recent V-K photometry, angular diameters, high-resolution spectroscopy, 
Hipparcos parallaxes, and extensive numerical simulations to re-examine and 
verify the temperature, metallicity, distance, and reddening calibrations for 
the {\it uvby}$\beta$ system. We also highlight the selection effects inherent 
in the apparent-magnitude limited GCS sample.
}
{We substantially improve the $\rm T_{eff}$ and [Fe/H] calibrations for early F 
stars, where spectroscopic temperatures have large systematic errors. 
A slight offset of the GCS photometry and the non-standard helium abundance 
of the Hyades invalidate its use for checking metallicity or age scales; 
however, the distances, reddenings, metallicities, and age scale for GCS field 
stars require minor corrections only. Our recomputed ages are in excellent 
agreement with the independent determinations by Takeda et al. (2007), 
indicating that isochrone ages can now be reliably determined.
}
{The revised G-dwarf metallicity distribution remains incompatible with
closed-box models, and the age-metallicity relation for the thin disk remains 
almost flat, with large and real scatter at all ages ($\sigma_{intrinsic}$= 
0.20 dex). Dynamical heating of the thin disk continues throughout its life; 
specific in-plane dynamical effects dominate the evolution of the $U$ and $V$ 
velocities, while the $W$ velocities remain random at all ages. When assigning 
thick and thin-disk membership for stars from kinematic criteria, parameters for 
the oldest stars should be used to characterise the thin disk.
}

\maketitle
%

\section{Introduction}
Models for the evolution of spiral galaxy disks describe their star formation 
history, nucleosynthesis, chemical enrichment, and dynamical evolution. 
Traditional parameterised models yield single-valued relations for the increase 
in the (total or individual) heavy-element abundances of stars at a given 
position in the disk, radial gradients in elemental abundances, and the 
kinematic heating of the local disk, all as functions of time. The underlying 
paradigm is the efficient mixing and recycling of interstellar gas, such that 
mean values of abundances and kinematics as functions of time and radial 
position in the disk describe the underlying physical processes adequately (see 
Casuso \& Beckman \cite{casuso04}, Naab \& Ostriker \cite{naab06}, or Cescutti 
at al. \cite{cescutti07} for recent examples). Any dispersion of the observed 
values around the mean relations is assumed to be due to observational error. 

Under realistic conditions, however, local variations in the rate of chemical 
enrichment must have occurred (see e.g. Brook et al. \cite{brook07} for such an 
approach). The question of interest is therefore how well these relations and 
their intrinsic scatter can be determined from the observations.

The Milky Way is the one galaxy in which these predictions can be tested in
detail. Thus, complete, accurate information on the stellar content of the
Solar  neighbourhood remains a fundamental observational constraint on any
set of models. Much of the discussion focuses on the age-metallicity relation
(AMR) for the Solar neighbourhood, and the key questions are two-fold
(Twarog \cite{twarog}, Carlberg et al. \cite{carlberg}, Meusinger,
Stecklum \& Reimann \cite{meusinger}, Edvardsson et al. \cite{edv93}, 
Rocha-Pinto et al \cite{rocha-pinto00}, Feltzing, Holmberg \& Hurley 
\cite{feltzing01}, and Nordstr{\"o}m et al. \cite{nordstrom04}): 
{\it (i)} Does the average relation show
 the expected rise in metallicity from the formation of the (thin) disk to the 
present time? and {\it (ii)} 
how large is the intrinsic dispersion in metallicity at any given age? 
Given the wide-ranging
ramifications of these questions, the observational data used for the test 
must be prepared, selected, and discussed with the utmost care.

The most comprehensive recent study of nearby stars in the solar 
neighbourhood is the Geneva-Copenhagen Survey (Nordstr{\"o}m et al. 
\cite{nordstrom04}; GCS in the following). 
The GCS provides metallicities, ages, kinematics, 
and Galactic orbits for a complete, magnitude-limited, all-sky sample 
of $\sim$14,000 F and G dwarfs brighter than $V\sim8.3$. The basic 
observational data are {\it uvby}$\beta$ photometry, Hipparcos/Tycho-2
parallaxes and proper motions, and some 63,000 new, accurate radial 
velocity observations, supplemented by earlier data. 
The best calibrations then available were used to derive $\rm T_{eff}$, 
[Fe/H], and distances from the photometry. The astrometry and radial 
velocities were then used to compute space motions and identify binaries in the 
sample. Finally, unbiased ages and error estimates were computed from a set 
of theoretical isochrones by a sophisticated Bayesian 
technique (J{\o}rgensen \& Lindegren \cite{jorgensen05}), and Galactic orbits 
were computed from the present positions and velocity vectors of the stars 
and a Galactic potential model. 

The GCS is large enough to yield adequate statistics for subsets of stars
defined by age, metallicity, or abundance, and is essentially free of the 
kinematic and/or metallicity biases affecting most earlier samples. However, 
systematic selection effects may still remain in the data or be introduced 
by the calibrations used to derive astrophysical parameters 
from the observations. Accordingly, the purpose of the present paper is to
critically re-examine the observational determination of $\rm T_{eff}$, 
[Fe/H], and age for F- and G-type dwarf stars in the light of the most 
recent developments in the field. We also compare our results with those
of the recent papers by Haywood (\cite{haywood06}; H06 in the following), 
Valenti \& Fischer (\cite{valenti}; VF05 in the following), and Takeda et 
al. (\cite{takeda}). 

We begin the paper by briefly describing the possible impact of systematic 
and random errors in these parameters on the overarching theme of spiral 
galaxy evolution in Sect. \ref{why}. The reader mostly interested in our 
new {\it uvby}$\beta$ calibrations in terms of effective temperature, 
metallicity, distance, reddening, and age computations will find these discussed 
in Sect. \ref{tempcal} -- \ref{stellarages}. The reader primarily interested 
in the evolution of the Galactic disk may skip directly to Sect. \ref{sample}, 
where we discuss the end-to-end simulations we have used to verify the 
robustness of the results reached with the new calibrations. We rediscuss 
the ``G dwarf problem'', age-metallicity, and age-velocity relations for 
the Solar neighbourhood in Sect. \ref{dGproblem} -- \ref{diskheating} and
compare the thick and thin disks in Sect. \ref{TDtd}. Finally, we summarise 
our findings and conclusions in Sect. \ref{endchap}.

\section{Astrophysical implications of calibration errors}\label{why}

The key parameters to be compared with models are the masses 
(i.e. main-sequence lifetimes), ages, heavy-element
abundances, spatial distributions, and space velocities or Galactic 
orbits for sufficiently large samples of stars that are representative of the
general population of disk stars. However, except for positions and 
velocities, these parameters cannot be determined by direct observation. 

Typical observational data are multiband colour indices and perhaps spectra,
from which $\rm T_{eff}$, [Fe/H], and absolute magnitude or log $g$ are
derived, using theoretical or empirical calibrations. From $\rm T_{eff}$, 
[Fe/H], and $\rm M_V$, in turn, the age and mass of each star can be 
derived by comparison with stellar models. 

This is straightforward in theory.
In practice, it is highly non-trivial because the problem is very
non-linear, and significant uncertainties in the calibrations remain. In 
particular, uncertainties in the theory of stellar atmospheres continue to
play a major role, both for the predicted effective temperatures of 
stellar models and for the transformations between $\rm T_{eff}$, [Fe/H], 
and log $g$ and the observed colour indices. Because of the strong 
correlations between these parameters, calibration errors in one may 
bias the determination of another and lead to spurious correlations 
between the derived quantities, and incorrect conclusions about the evolution
of the Galactic disk. 

As one example, assume that the adopted calibration yields too high 
$\rm T_{eff}$ for the hotter stars (as indeed we find below). [Fe/H] as 
determined from equivalent widths of 
spectral lines will then be overestimated. The hotter $\rm T_{eff}$ 
corresponds to a lower age, the higher [Fe/H] to cooler models and thus 
an even lower derived age for the observed star. A spurious age-metallicity 
relation results. 

As another example, assume that the $\rm T_{eff}$ scale of the models is
correct for the solar abundance, but too hot at lower [Fe/H], as was found 
in GCS. This will lead to overestimated ages for the metal-poor stars, 
i.e. again a spurious age-metallicity relation. A similar -- or additional 
-- error is introduced if the enhanced [$\alpha$/Fe] ratio of metal-poor 
stars (e.g. Edvardsson et al. \cite{edv93}) is ignored; if the 
heavy-element abundance parameter $Z$ of the models is assumed to scale 
simply as [Fe/H], too hot models are selected, and the resulting age is 
overestimated again. 

Finally, the importance of a clear understanding of the definition of the
stellar sample cannot be over-emphasised: Taking the AMR as the prototype 
again, the choice of stellar sample can determine the outcome even before 
a single observation is made. E.g., the F-dwarf sample studied by 
Edvardsson et al. (\cite{edv93}) excluded any old, metal-rich stars that 
might exist and {\it a priori} decided the shape of their AMR - a fact 
emphasised in the paper, but largely ignored in later references. More
complete and accurate observational data are always useful, but 
temperature calibrations, the choice and verification of stellar models, 
and the methods used to compute ages and their uncertainties are far more 
urgent issues at present. 

\section{Temperature calibration}\label{tempcal}

$\rm T_{eff}$ is the most critical parameter in the determination of isochrone
ages, but also affects the spectroscopic metallicity determinations. 
Two qualitatively different methods to determine $\rm T_{eff}$ exist, based 
on measurements of spectral lines or on colour indices. When using spectra, 
the determination of $\rm T_{eff}$ is usually based 
on the excitation balance of iron, but depth ratios of sets of 
spectral lines have also been used. When using photometry, the determination 
of $\rm T_{eff}$ is usually tied to the infrared flux of the stars as
estimated from models (the IRFM technique). 

For a fundamental test, we return to the basic definition of $\rm T_{eff}$:
$f_{\rm bol}=\frac{\phi^{2}}{4}\sigma T_{\rm eff}^{4}$, where $f_{\rm bol}$
is the bolometric flux and $\phi$ the angular diameter of the star. The 
practical problem in applying the formula is that solar-type main-sequence 
stars have very small angular diameters that are difficult to measure 
accurately. The situation is quickly improving, however, as new interferometers 
enter operation, notably the ESO VLTI. Kervella et al. (\cite{kervella04}) 
summarise the situation and give diameters for 20 A-M main-sequence stars and 
8 A-K0 sub-giants. The angular diameters are of excellent quality, especially 
those from the VLTI, which have errors down to 1\%. 

Ram\'{i}rez \& Mel\'endez (\cite{ramirez05a}) combined these diameters with 
bolometric flux measurements to derive $\rm T_{eff}$ directly 
for 10 dwarfs and 2 sub-giants. Propagating the errors in the
diameters and fluxes through to the temperatures, the mean error is 57K.
Ram\'{i}rez \& Mel\'endez (\cite{ramirez05a}) also give (IRFM) 
estimates for these stars, with a mean difference 
$<T_{\rm eff}^{\rm IRFM}-T_{\rm eff}^{\rm dir}>$= 10K, dispersion 
98K, and mean error 28K. 

As regards spectroscopic $\rm T_{eff}$ determinations, Santos et al. 
(\cite{santos05} and references therein) give results for stars both with and 
without detected planetary companions, including 7 stars with measured
angular diameters. The mean difference is 
$<T_{\rm eff}^{\rm Santos}-T_{\rm eff}^{\rm dir}>$= 92K, with a dispersion of 
91K and mean error 34K. This is in agreement with Santos et al. 
(\cite{santos05}) themselves, who find their $\rm T_{eff}$ scale to be 139K 
hotter than that by Alonso et al. (\cite{alonso96}), based on the IRFM and 
calibrated to B-V and [Fe/H]. 

Another valuable comparison is with the recent paper by VF05. They give
spectroscopic data for $\sim$1000 stars on the 
Keck/Lick/AAT planet search program, based on fits of synthetic spectra to their 
high-resolution data and correcting their zero-points from observations of 
Vesta as a proxy for the Sun. For the 8 stars with measured diameters, the 
mean difference is $<T_{\rm eff}^{\rm V\&F}-T_{\rm eff}^{\rm dir}>$= 64K, 
with a dispersion of 124K and mean error 44K.

The reasons for the failure of spectroscopic $\rm T_{eff}$
determinations based on the excitation equilibrium in 1D static LTE models 
are given in e.g. Asplund (\cite{asplund}). Basically, because real stars 
are spherical, hydrodynamical systems with lines formed in NLTE, the three 
fundamental assumptions underlying traditional model atmospheres are 
inadequate and lead to biased results. The correction procedure used by 
VF05 seems to have eliminated at least part 
of this bias.

In the following, we review the GCS and other recent temperature scales and
derive an improved calibration for the {\it uvby}$\beta$ system.

\subsection{$\rm T_{eff}$ in the GCS}

The marked offset of the Santos et al. (\cite{santos05}) $\rm T_{eff}$ values
is confirmed for the 160 stars common to Santos et al. and the GCS, which
employed temperatures based on the Alonso et al. (\cite{alonso96}) 
calibration 
of {\it b-y}, $m_{1}$ and $c_{1}$ to the IRFM scale. The mean difference 
is $<T_{\rm eff}^{\rm Santos}-T_{\rm eff}^{\rm GC}>$= 127K (dispersion 84K,
mean error 7K). For the spectral type range in question, this shows that 
the GCS $\rm T_{eff}$ scale is in excellent agreement {\it in the mean} 
with temperatures derived directly from the definition of $\rm T_{eff}$.

However, the Alonso et al. (\cite{alonso96}) calibration of {\it b-y}, 
$m_{1}$ and $c_{1}$ may be valid only in the ranges in $\rm T_{eff}$ 
within which enough calibration stars exist. Fig. 11a of Alonso et al. 
(\cite{alonso96}) shows the fitted temperatures as a function of {\it b-y} 
and makes clear that there is a marked dearth of calibration stars blueward 
of {\it b-y} $\approx$ 0.3 or $\rm T_{eff}$ $\approx$ 6500K; a similar lack 
is seen for the reddest stars. Alonso et al. (\cite{alonso96}) give a
dispersion for their relation of $\sigma(\theta_{\rm eff})=0.019$ (110K), 
while Ram\'irez \& Mel\'endez (\cite{ramirez05b}) give 87K for their new 
{\it b-y} relation, albeit at the price of not being valid over the whole 
parameter range of the GCS stars. 

The potential problem in the Alonso et al. (\cite{alonso96}) {\it uvby}
calibration for blue and red stars is clearly demonstrated by a comparison 
with other calibrations of Str\"omgren photometry. They all agree rather 
well between 0.3$<${\it b-y}$<$0.6, but diverge outside these limits, see 
e.g. Fig. 24 in Clem et al. (\cite{clem04}). 

The situation is unfortunately similar for stars with direct 
$\rm T_{eff}$ determinations, where Kervella et al. (\cite{kervella04}) 
have only one star between 6000 and 8500K (Procyon at 6500K), compared to 5
stars between 8500 and 10000K and 11 between 5000 and 6000K. 

\begin{figure}[htbp] 
\resizebox{\hsize}{!}{\includegraphics[angle=-90]{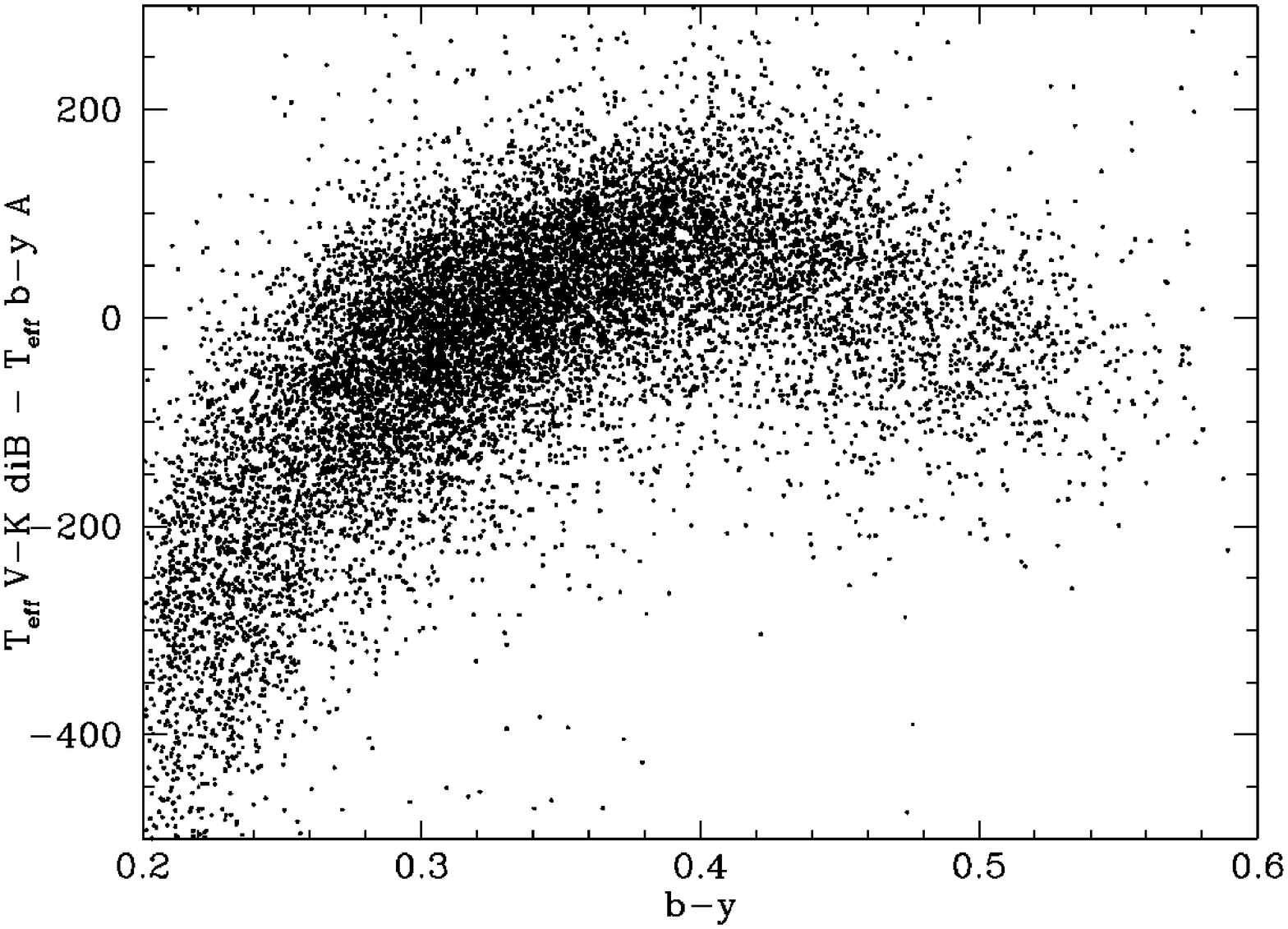}} 
\caption{Differences between $\rm T_{eff}$ from the Alonso et al. 
(\cite{alonso96}) {\it b-y} calibration, used in the GCS, and $\rm T_{eff}$ 
as derived from the V-K calibration by di Benedetto (\cite{diBenedetto}).} 
\label{teffdiff} 
\end{figure}

\subsection{Calibrations based on V-K}

Alternatively, temperatures based on V-K can be used, because 2MASS 
photometry is available and well suited to our colour and magnitude ranges: 
Of the 16682 GCS stars, 16139 have 2MASS K-magnitudes of the highest quality 
class ``A''. We combine these with the $V$ magnitude from the GCS {\it 
uvby}$\beta$ photometry. 

In contrast to calibrations based on only visible colours, $\rm T_{eff}$ 
estimates based on V-K are remarkably robust; the small differences 
are probably mostly related to the different existing definitions of the 
K-band. As an example, the V-K calibrations of Alonso (\cite{alonso96}), 
Ram\'irez \& Mel\'endez (\cite{ramirez05b}), and di Benedetto 
(\cite{diBenedetto}) agree pairwise to within $\sim$20K for the stars in the 
GCS. In contrast, the Alonso V-K and {\it b-y} calibrations differ by 180K 
rms!

Today, the K-band system of choice is that defined by the 2MASS catalogue, 
which also gives K magnitudes of high quality for most of the stars in the
GCS. We have used the calibration of di Benedetto (\cite{diBenedetto}), which 
is in excellent agreement with the IRFM temperature scale of Ram\'{i}rez 
\& Mel\'endez (\cite{ramirez05a}) and is valid for the whole colour range of 
the GCS. In order to transform the Johnson V-K colours used by di Benedetto 
to the Ks system used by 2MASS, we use the relation: 
$$(V-K)_{J}=1.007[(V-K_{s}-0.044)-0.01]$$

\begin{figure}[htbp] 
\resizebox{\hsize}{!}{\includegraphics[angle=-90]{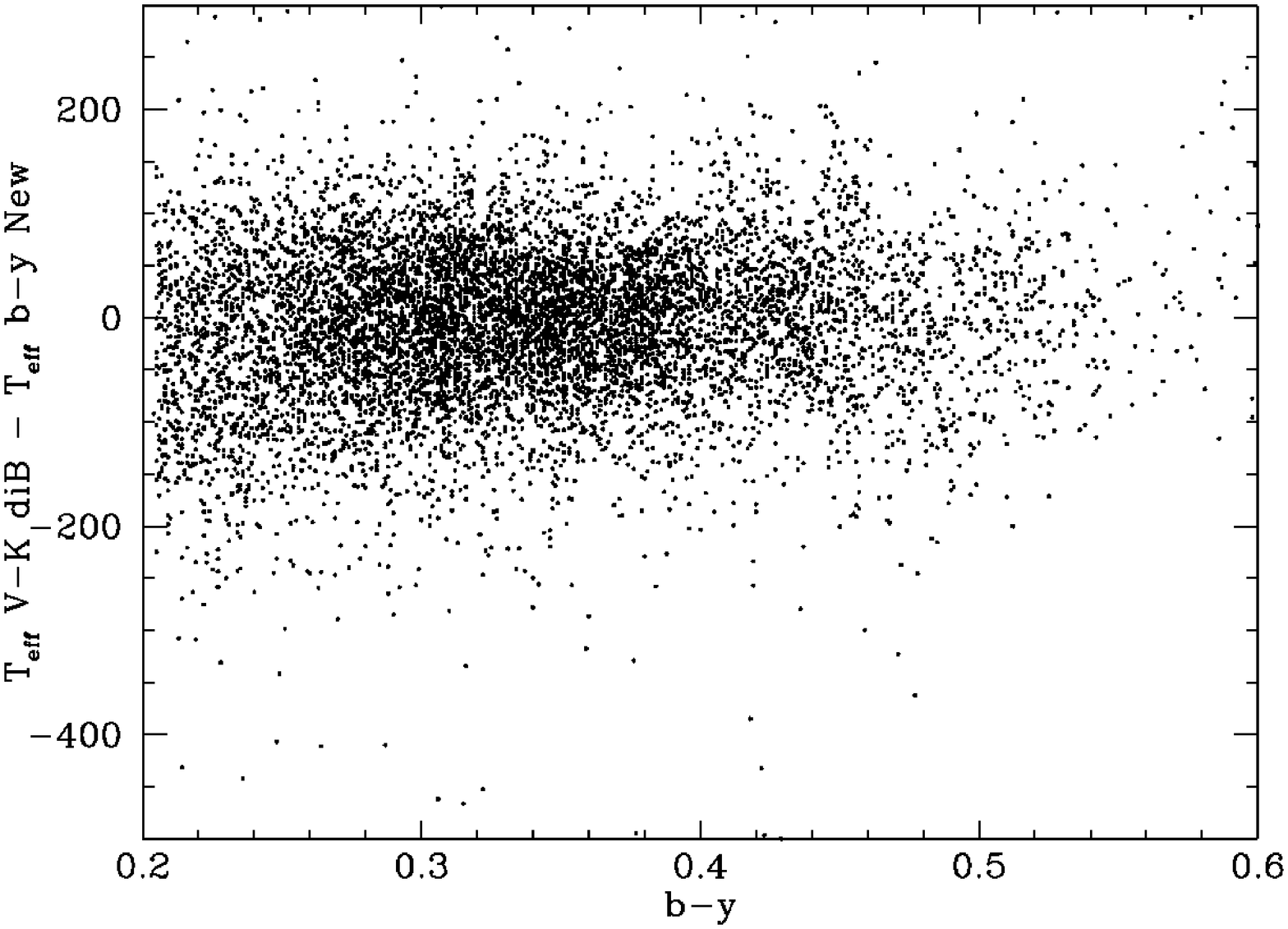}}
\resizebox{\hsize}{!}{\includegraphics[angle=-90]{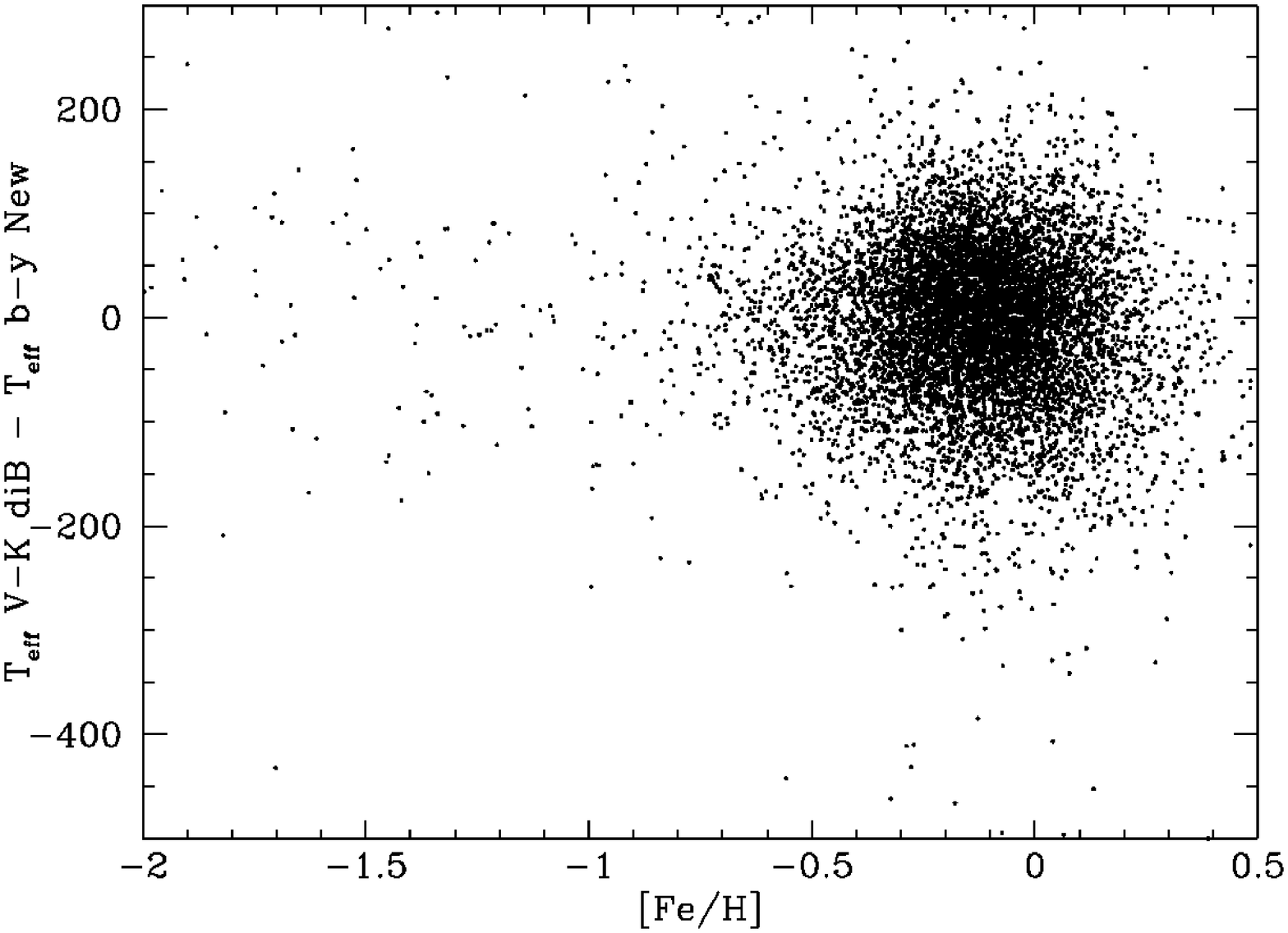}}
\caption{Differences between the $\rm T_{eff}$ of GCS stars as derived from 
our new {\it b-y} calibration and from the V-K calibration by di Benedetto 
(\cite{diBenedetto}).} 
\label{teffdiff2} 
\end{figure}

Fig. \ref{teffdiff} shows the difference between the $\rm T_{eff}$ values 
from the original GCS catalogue and those derived from the 2MASS photometry 
and the corresponding calibration by di Benedetto (\cite{diBenedetto}). The 
systematic differences are quite marked, especially for the hotter stars. 

The trend seen in Fig. \ref{teffdiff} is qualitatively similar to that found 
by H06 (Fig. 7), also by comparison with values from di Benedetto 
(\cite{diBenedetto}), but is better determined here, due to our much larger 
sample and because V-K is a much more reliable temperature indicator 
than B-V.

\subsection{A new {\it b-y} -- $\rm T_{eff}$ 
calibration}

Because of the larger number of suitable calibration stars available, the V-K 
calibration has less systematic error. However, the observational accuracy of 
the {\it b-y} index is substantially better than that of V-K. We have
therefore derived a new $\rm T_{eff}$ calibration for {\it b-y}, based on the 
V-K temperature scale set by di Benedetto (\cite{diBenedetto}). The best 
result is achieved when the sample is divided into three temperature 
ranges. In the blue range (0.20$<${\it b-y}$<$0.33) the relation is:\\ 

\noindent ${\rm \theta_{eff}}= 0.649+0.095(b-y)+0.034{\rm [Fe/H]}+
1.207(b-y)^{2}\\
~~~~~~~~~-0.005{\rm [Fe/H]}^{2}-0.181(b-y){\rm [Fe/H]}$\\

\noindent where $\rm \theta_{eff}$ = 5040K/$\rm T_{eff}$. The dispersion 
of the fit is $\rm \sigma(\theta_{eff})=0.007$ when 2.5-$\sigma$
outliers are removed, corresponding to 60K or 0.004 in $\rm \log T_{eff}$ for 
the mean $\rm \log T_{eff}$. 

In the middle range (0.33$<${\it b-y}$<$0.50), the relation is:\\ 

\noindent ${\rm \theta_{eff}}= 0.754-0.365(b-y)-0.001{\rm [Fe/H]}+
1.635(b-y)^{2}\\
~~~~~~~~~-0.011{\rm [Fe/H]}^{2}-0.091(b-y){\rm [Fe/H]}$\\

The dispersion of the fit is $\rm \sigma(\theta_{eff}) = 0.009$ when 
2.5-$\sigma$ outliers are removed, corresponding to 57K or again 0.004 in 
$\rm \log T_{eff}$ for the mean $\rm \log T_{eff}$.

\noindent In the red range (0.50$<${\it b-y}$<$0.60), the relation is:\\ 

\noindent ${\rm \theta_{eff}}= 0.255+1.656(b-y)+0.018{\rm [Fe/H]}-
0.397(b-y)^{2}\\
~~~~~~~~~-0.011{\rm [Fe/H]}^{2}-0.101(b-y){\rm [Fe/H]}$\\

The dispersion of the fit is $\rm \sigma(\theta_{eff}) = 0.012$ when 
2.5-$\sigma$ outliers are removed, corresponding to 55K or 0.005 in 
$\rm \log T_{eff}$ for the mean $\rm \log T_{eff}$

We note that this calibration yields 5777K for the Sun when using {\it 
(b-y)}$_{\odot}$= 0.403 from Holmberg, Flynn \& Portinari (\cite{holmberg06}), 
although this was not forced on the fit. The two main calibrations are also 
very well connected: At their common colour, {\it (b-y)}= 0.33, the difference 
in temperature is 2, 6, and 7K at [Fe/H]= 0, -0.50, and -1.00. Due to the 
small number of very red calibration stars, the difference in $\rm T_{eff}$ 
grows to -19, 18, and 52K at [Fe/H]= 0, -0.50, and -1.00 at {\it (b-y)}= 
0.50 -- still very small. 

Fig. \ref{teffdiff2} 
demonstrates the agreement of the resulting $\rm T_{eff}$ values with those 
derived from V-K and the di Benedetto (\cite{diBenedetto}) calibration. 

\begin{figure}[htbp] 
\resizebox{\hsize}{!}{\includegraphics[angle=-90]{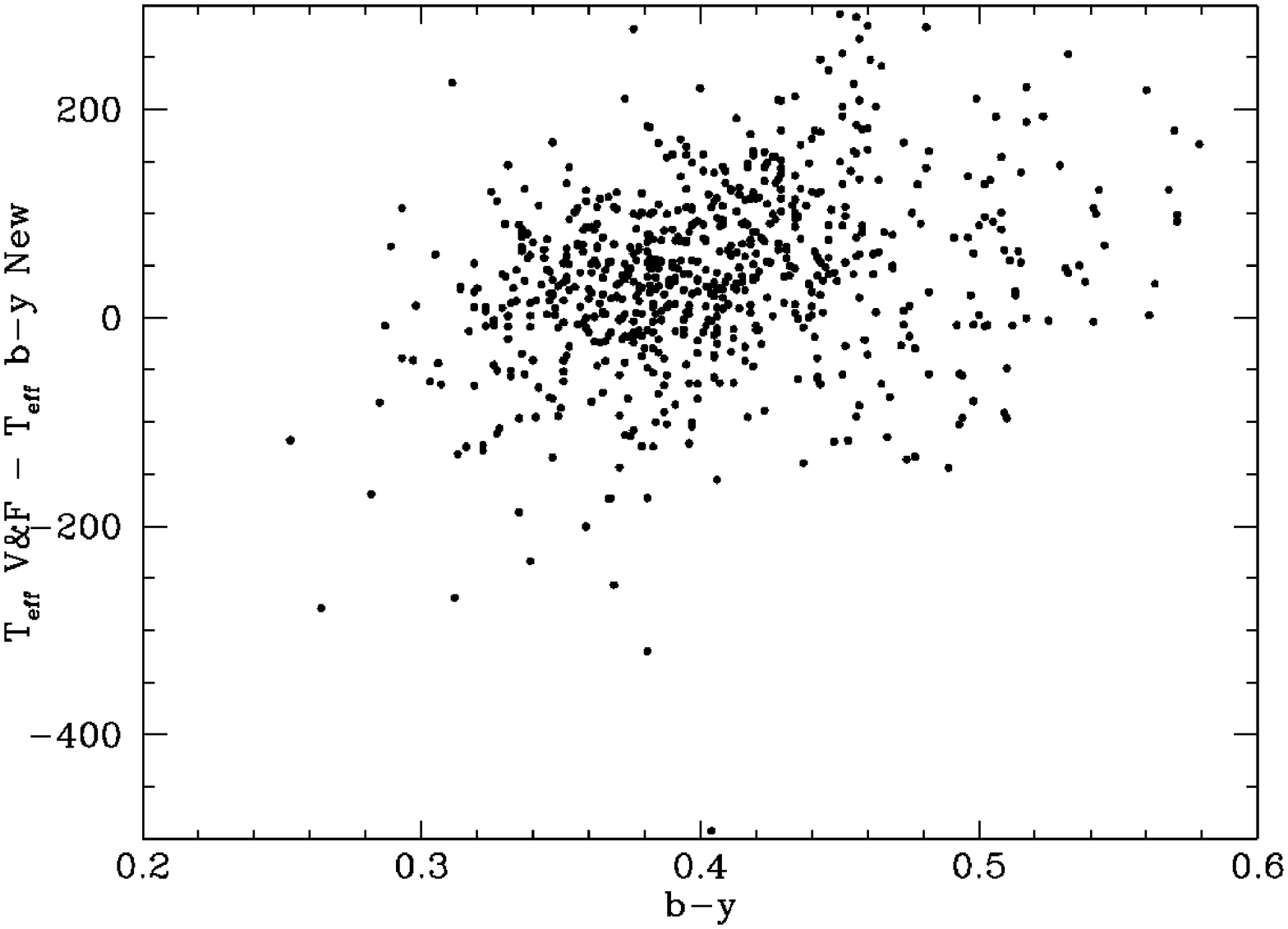}} 
\caption{Differences between $\rm T_{eff}$ from our new {\it b-y} 
calibration and $\rm T_{eff}$ as derived from spectrum fitting by VF05.} 
\label{teffvalenti} 
\end{figure}

Finally, we compare our new photometric temperatures to the spectroscopic 
results by VF05. Fig. \ref{teffvalenti} shows 
the temperature differences for the 697 single stars in common with the GCS. 
The mean difference is similar to that from the stars with angular diameters: 
$<T_{\rm eff}^{\rm V\&F}-T_{\rm eff}^{\rm b-y}>$= 43K, with a dispersion of 
91K.

\section{Metallicity calibration} 

\begin{figure}[htbp] 
\resizebox{\hsize}{!}{\includegraphics[angle=0]{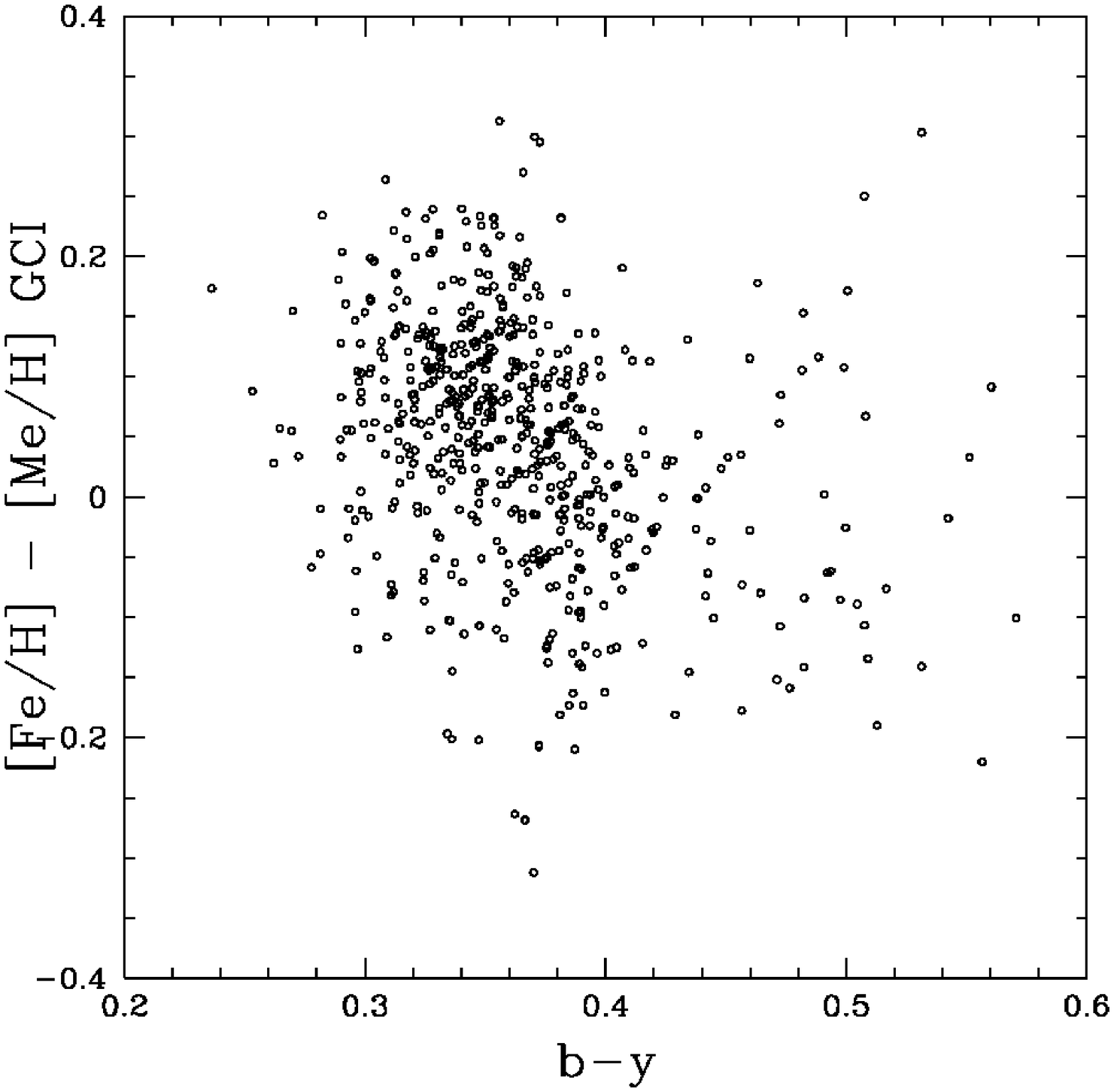}
                      \includegraphics[angle=0]{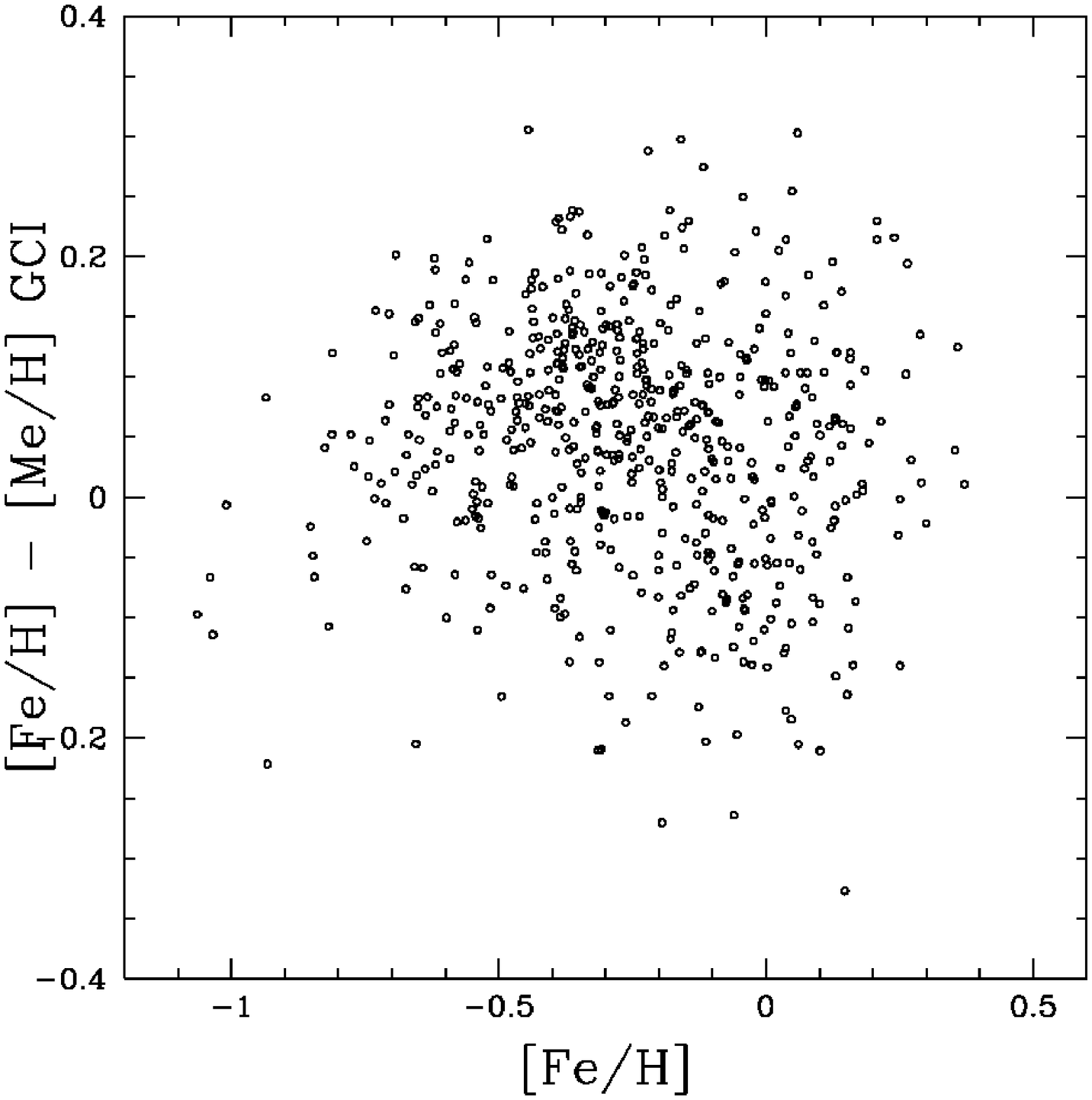}} 
\caption{ Modern spectroscopic [Fe/H] determinations vs. the photometric 
[Me/H] of the GCS catalogue. } 
\label{fehdiff} 
\end{figure}

[Fe/H] is important, both in itself as a diagnostic 
of the chemical evolution of the disk, and because it enters into the 
determination of stellar ages from theoretical isochrones. However,
a comparison of the GCS photometric metallicities with the several 
high-quality spectroscopic studies available today (Fig. \ref{fehdiff}) 
shows significant differences, even when considering only spectroscopic 
studies using a photometric temperature scale.

For the four major spectroscopic studies (in the order Edvardsson et al. 
\cite{edv93}, Chen et al. \cite{chenyq00}, Reddy et al. \cite{reddy03}, and 
Allende Prieto et al. \cite{allende04}), the mean differences 
$<[\rm Fe/H]-[Me/H]>$ are, for the F-calibration range: 0.064, 0.038, 0.114 
and 0.013; and for the G-calibration range: 0.019, -0.038, 0.022 and -0.044. 
Only Allende Prieto et al. (\cite{allende04}) have stars in the range of the 
``K'' calibration (as defined in GCS), with $<[\rm Fe/H]-[Me/H]>$= -0.005. 

As regards Santos et al. (\cite{santos05}), who use a spectroscopic 
$\rm T_{eff}$ scale, we find $<[\rm Fe/H]-[Me/H]>$= 0.162 for stars in F 
range, $<[\rm Fe/H]-[Me/H]>$= 0.082 for stars in the G range, and  
$<[\rm Fe/H]-[Me/H]>$= 0.022 for their stars in the K range. We conclude 
that metallicities derived using the spectroscopic $\rm T_{eff}$ scale must 
be reduced by about 0.10, 0.09 and 0.03 dex for the F, G and K temperature 
ranges. This is in fair agreement with the estimated effect on [Fe/H] of a 
temperature shift of about 100K (see e.g. Feltzing \& Gustafsson 
\cite{feltz98}).

We conclude that, when comparing metallicity scales from different sources, 
it is crucial to verify how these scales are established. It is especially 
important to check which $\rm T_{eff}$ scale is used, because an erroneous 
$\rm T_{eff}$ scale can introduce large biases in the abundances, especially 
for solar or somewhat hotter temperatures.

\subsection{The original GCS calibration}

The determination of accurate metallicities for F and G stars is one of the 
strengths of the Str{\"o}mgren {\it uvby}$\beta$ system. Among the then 
available {\it uvby}$\beta$ calibrations, the GCS used that by 
Schuster \& Nissen (\cite{schuni89}) for the majority of the stars. 

However, this calibration was found to give substantial systematic errors for 
the very reddest G and K dwarfs ($b-y >$ 0.46), where very few spectroscopic 
calibrators were available at that time. In the GCS we therefore derived a 
new relation, based on a sample of 72 dwarf stars in the colour range 0.44 
$\leq b-y \leq$ 0.59, using the same terms in the calibration equation as the 
Schuster \& Nissen (\cite{schuni89}) G-star calibration. The resulting 
equation was: \\

\noindent ${\rm [Fe/H]}= -2.06+24.56m_{1}-31.61m_{1}^{2}-53.64m_{1}(b-y) \\ 
~~~+73.50m_{1}^{2}(b-y)+[26.34m_{1}-0.46c_{1}-17.76m_{1}^{2}]c_{1}$ \\

\begin{figure}[hbtp] 
\resizebox{\hsize}{!}{\includegraphics[angle=0]{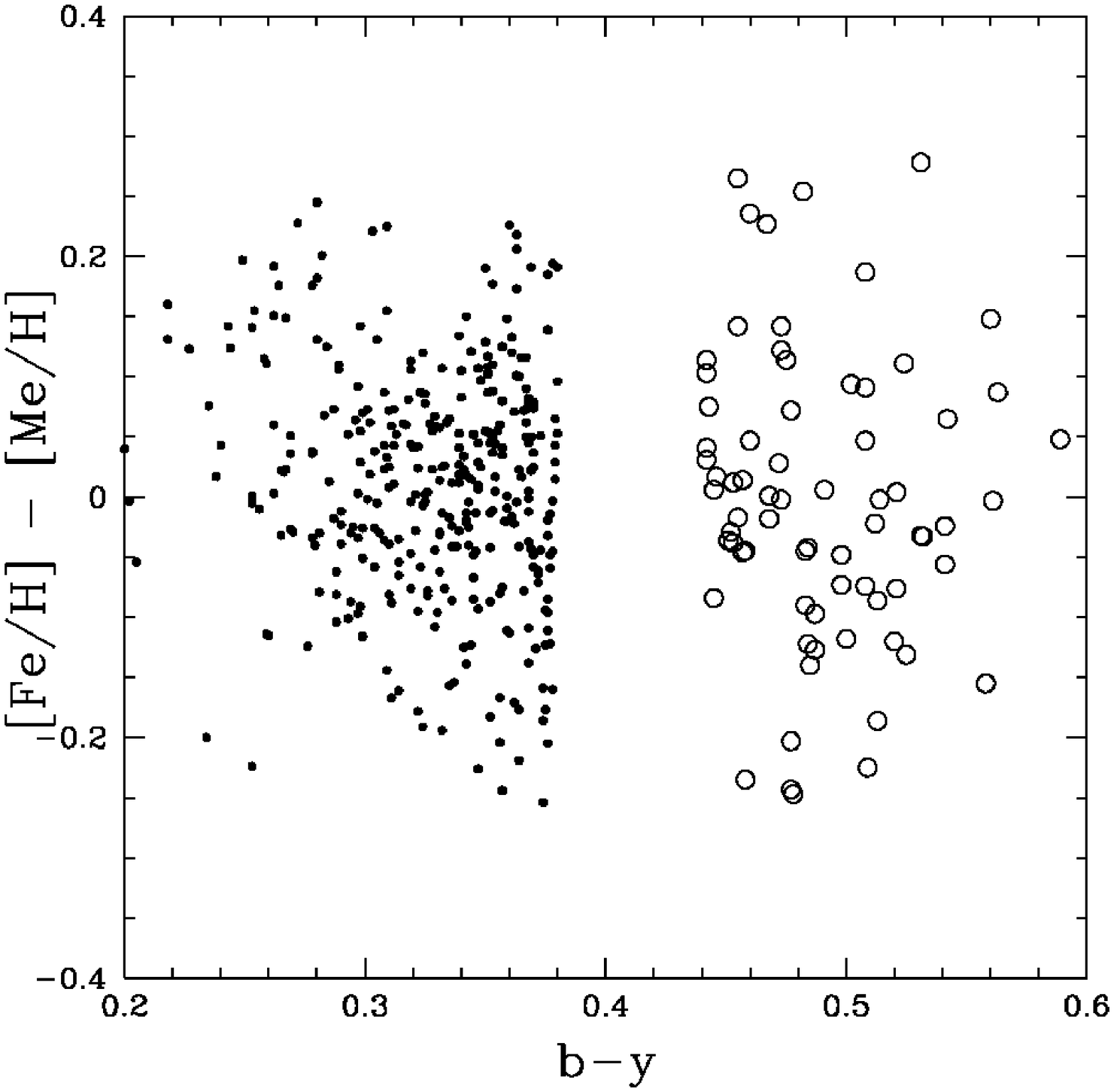}
                      \includegraphics[angle=0]{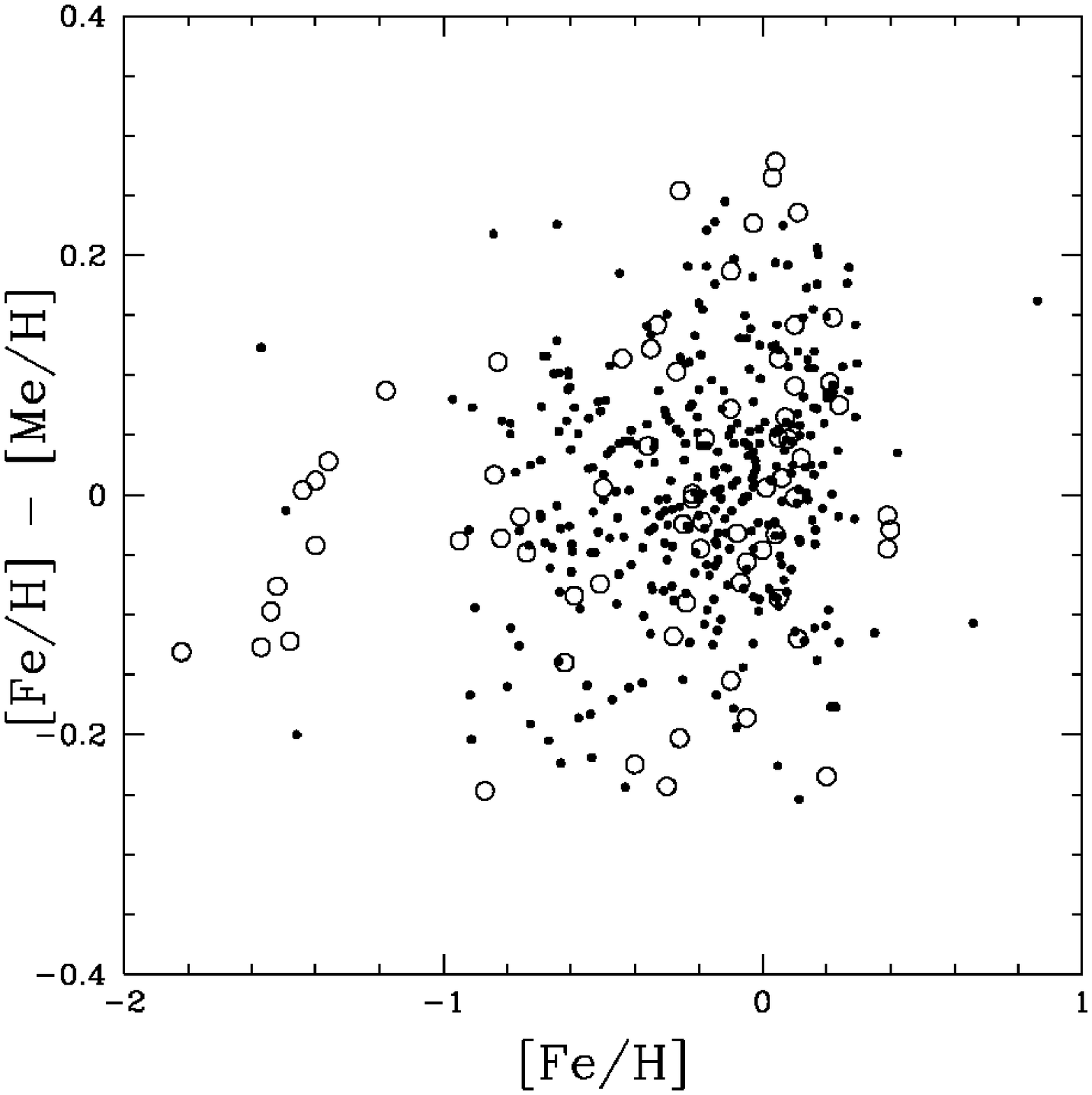}} 
\caption{ GCS photometric metallicities ([Me/H]) vs. 
the spectroscopic [Fe/H] values used to establish the calibrations. Open 
circles depict the cool stars, dots the hot stars (see text).}
\label{fehcomp} 
\end{figure}

The metallicities from this calibration are compared to 
the new spectroscopic reference values in Fig.~\ref{fehcomp}. The 
dispersion around the (zero) mean is 0.12 dex.

For the $\sim$600 stars in the interval 0.44$<${\it b-y}$<$ 0.46, the new
calibration agreed with that by the Schuster \& Nissen (\cite{schuni89}) to
within 0.00 dex in the mean (s.d. 0.12).

About 2400 GCS stars with high $\rm T_{eff}$ and low log $g$ were outside
the range covered by the Schuster \& Nissen (\cite{schuni89}) calibration.
For these stars, the calibration of $\beta$ and $m_1$ by Edvardsson et al. 
(\cite{edv93}) was used when valid. For the stars in common, the two 
calibrations agree very well (mean difference of 0.00 dex, dispersion only 
0.05). 

For stars outside the limits of both calibrations, we derived a new relation, 
using the same terms in the equation as the 
Schuster \& Nissen (\cite{schuni89}) 
calibration for F stars. From 342 stars in the ranges: 0.18 $\leq b-y
\leq$ 0.38, 0.07 $\leq m_{1} \leq$ 0.26, 0.21 $\leq c_{1} \leq$ 0.86, and -
1.5
$\leq [\rm Fe/H] \leq$ 0.8, we found the following calibration equation:\\

\noindent ${\rm [Fe/H]}= 9.60-61.16m_{1}+81.25m_{1}(b-y)\\
~~~~~~~~~+224.65m_{1}^{2}(b-y) -153.18m_{1}(b-y)^{2}\\
~~~~~~~~~+[12.23-90.23m_{1}+38.70(b-y)]\log(m_{1}-c_{3}),$ \\

\noindent where $c_3 = 0.45-3.98(b-y)+5.08(b-y)^{2}$. The dispersion of the 
fit is 0.10 dex.

\begin{figure}[htbp] 
\resizebox{\hsize}{!}{\includegraphics[angle=0]{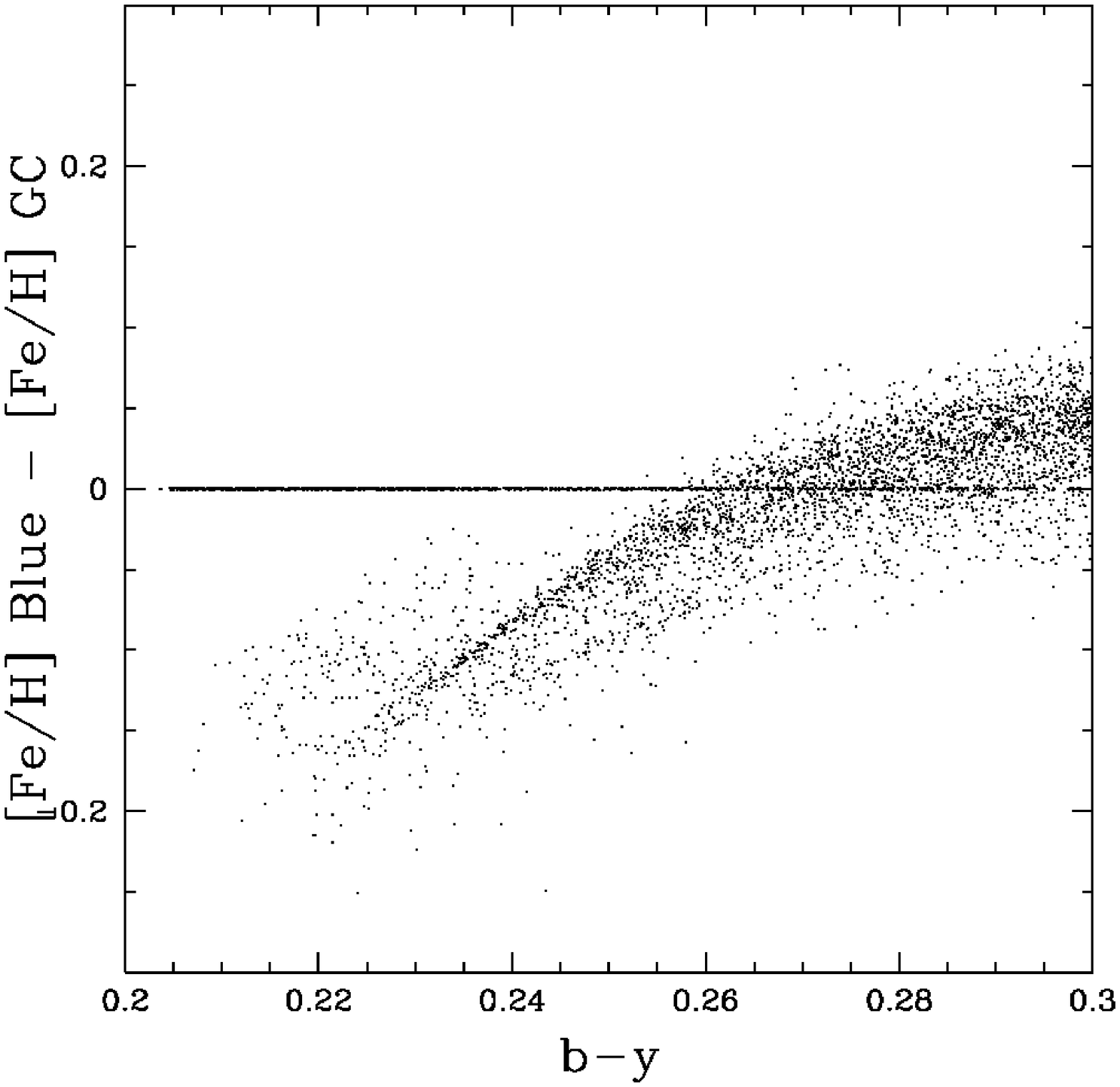}
                      \includegraphics[angle=0]{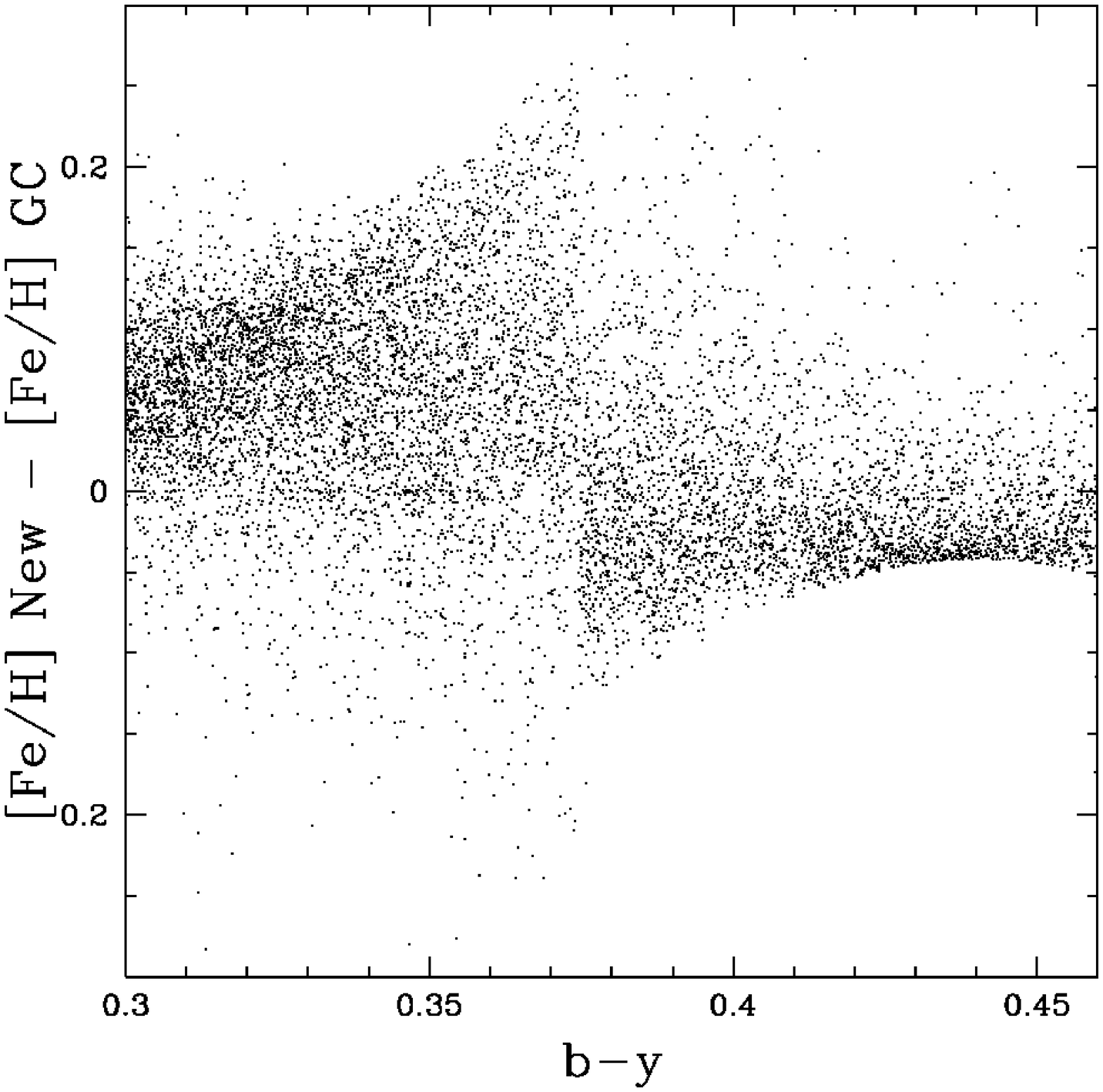}}
\caption{{\it Left:} Differences between the GCS metallicities for all stars
and those now derived from the GCS blue calibration (the sharp line at 
$\Delta$[Fe/H]= 0 corresponds to those GCS stars that already used this 
calibration). 
{\it Right:} The difference between metallicities from the new calibration 
derived in this paper and the original GCS metallicities, based on the 
Schuster \& Nissen (\cite{schuni89}) calibration.}
\label{mehdiff} 
\end{figure}
 
Fig.~\ref{fehcomp} compares the photometric [Fe/H] from this calibration 
with the spectroscopic values; for the stars in common, they again agree 
very well those from the Schuster \& Nissen (\cite{schuni89}) calibration 
(mean difference 0.02 dex, $\sigma$ only 0.04).

\subsection{An improved metallicity calibration}

In order to define an improved photometric metallicity calibration that 
is valid for {\it all} GCS stars, we select only spectroscopic 
investigations using a photometric temperature scale (unlike the compilation 
of literature values by Cayrel de Strobel et al. (\cite{cayrel01}) used 
by H06). This yields a total sample of 573 stars from Edvardsson et al. 
(\cite{edv93}), Chen et al. \cite{chenyq00}, Reddy et al. \cite{reddy03}, 
Allende Prieto et al. (\cite{allende04}), and Feltzing \& Gustafsson 
(\cite{feltz98}). They all 
have a very consistent metallicity scale: Comparing stars in common yields 
differences of of 0.02 dex or smaller in the mean, with a dispersion of 
$\sim$0.07 dex. The 573 stars span the range 0.24 $\leq b-y
\leq$ 0.63, 0.10 $\leq m_{1} \leq$ 0.70, 0.17 $\leq c_{1} \leq$ 0.53, 
and -1.0$\leq [\rm Fe/H] \leq$ 0.37.

A new fit of the {\it uvby} indices to the spectroscopic [Fe/H] values from 
this sample was performed, using a calibration equation containing all 
possible combinations of $b-y, m_{1}$, and $c_{1}$ to third order.

The resulting calibration equation is: \\

\noindent ${\rm [Fe/H]}= -2.19-1.02(b-y)+7.34m_{1}-0.27c_{1}\\
+5.86(b-y)^{2}-43.74m_{1}^{2}-0.14c_{1}^{2}+25.03(b-y)m_{1}\\
+5.29(b-y)c_{1}+25.95m_{1}c_{1}-31.10(b-y)^{3}+46.19m_{1}^{3}\\
-3.86c_{1}^{3}+4.54(b-y)^{2}m_{1}+19.31(b-y)^{2}c_{1}-17.46m_{1}^{2}(b-y)\\
+18.36m_{1}^{2}c_{1}-9.99c_{1}^{2}(b-y)+6.60c_{1}^{2}m_{1}-59.65(b-
y)m_{1}c_{1}
$\\ 

\noindent This new calibration is used for all stars with 0.30 $<$ {\it b-y} $<$ 
0.46. Together with the blue and red calibrations already derived in the GCS, 
this completes the final metallicity calibration of the present paper. All 
results discussed in the following are based on it.

\begin{figure}[htbp] 
\resizebox{\hsize}{!}{\includegraphics[angle=0]{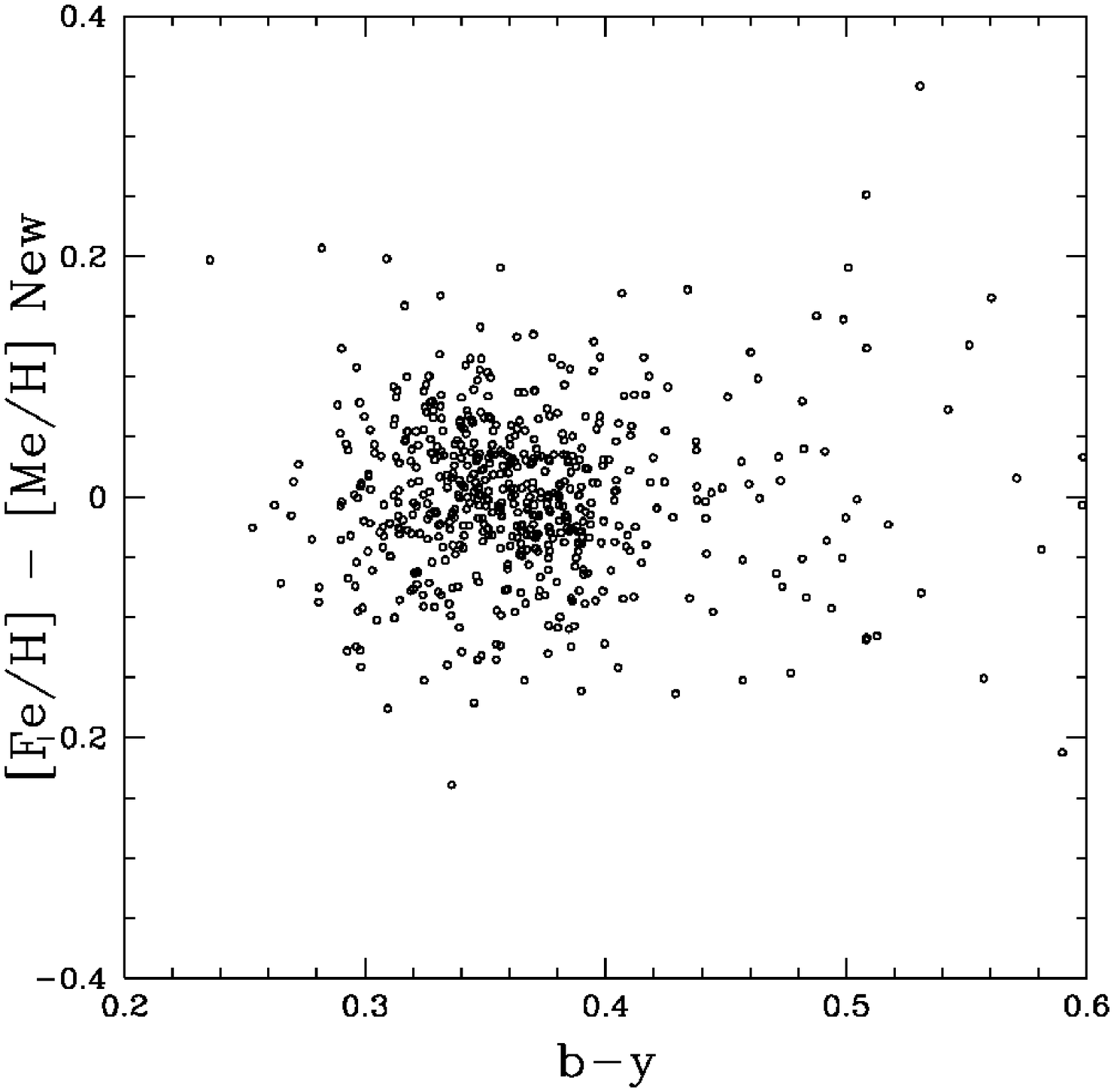}
                      \includegraphics[angle=0]{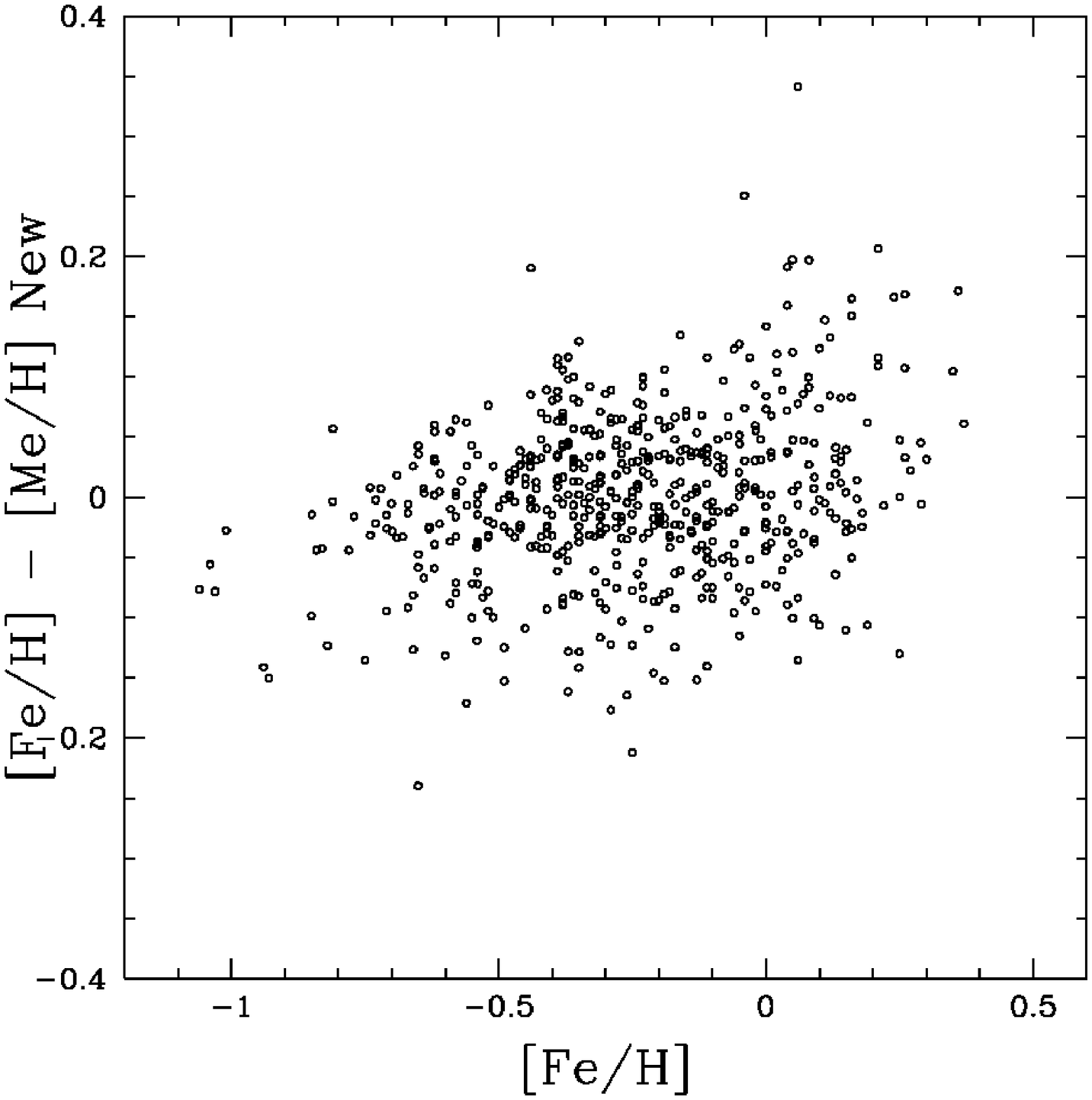}}
\caption{Same as Fig. \ref{fehdiff}, but using the new metallicity 
 calibration derived in this paper.} 
\label{fehdiffb} 
\end{figure}

\begin{figure}[htbp] 
\resizebox{\hsize}{!}{\includegraphics[angle=0]{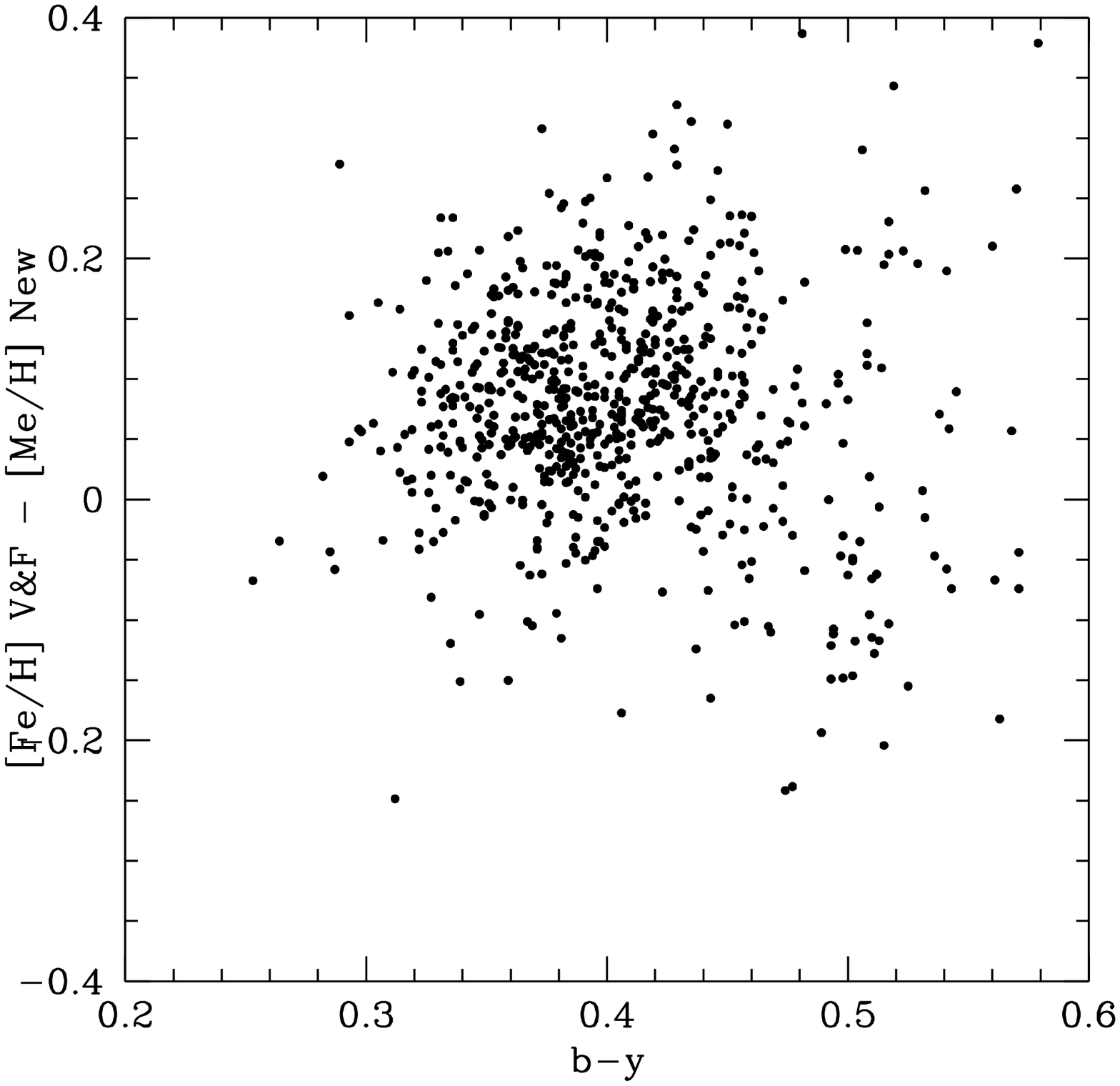}
                      \includegraphics[angle=0]{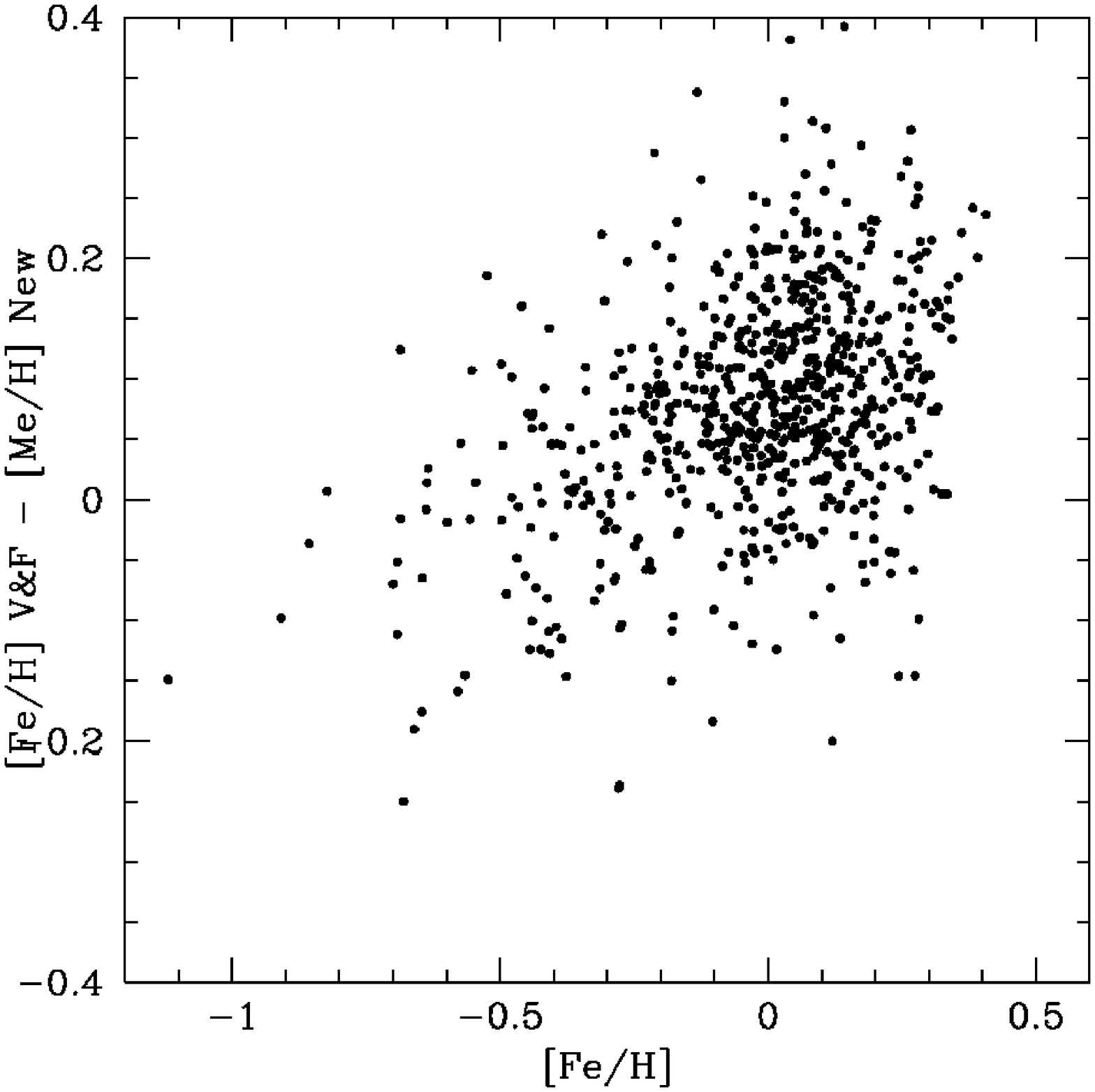}} 
\caption{Spectroscopic [Fe/H] data from VF05 
vs. our new photometric [Me/H] for the GCS stars. } 
\label{fehvf} 
\end{figure}

Fig. \ref{mehdiff} shows 
the differences between the GCS [Fe/H] values and those derived with the new 
calibration; note that, in the left panel of Fig. \ref{mehdiff}, the heavy 
sequence at zero difference is formed by the stars that already then used the 
blue calibration.

The consistency of the photometric metallicities from this calibration with 
the spectroscopic reference values is shown in Fig.~\ref{fehdiffb}. The 
dispersion around the mean is 0.07 dex -- the same as between two different 
spectroscopic 
measurements. The calibration derived above is also in good accordance with 
the two relations derived in the GCS. Compared to the blue relation, the mean 
difference is 0.02 (dispersion 0.04) in the common range 
0.3$<${\it b-y}$<$0.32. Compared to the red relation, the mean difference 
is 0.00 (dispersion 0.09) in the common range 0.44$<${\it b-y}$<$0.50. 

Finally, Fig.~\ref{fehvf} compares our new GCS metallicities with the 
spectroscopic [Fe/H] values from VF05. The mean difference 
VF05-GCS is 0.08 dex, with a standard deviation of 0.10 dex. The higher 
metallicities derived by VF05 are understandable, given the 
generally higher temperatures they adopt.

\begin{figure}[htbp] 
\resizebox{\hsize}{!}{\includegraphics[angle=0]{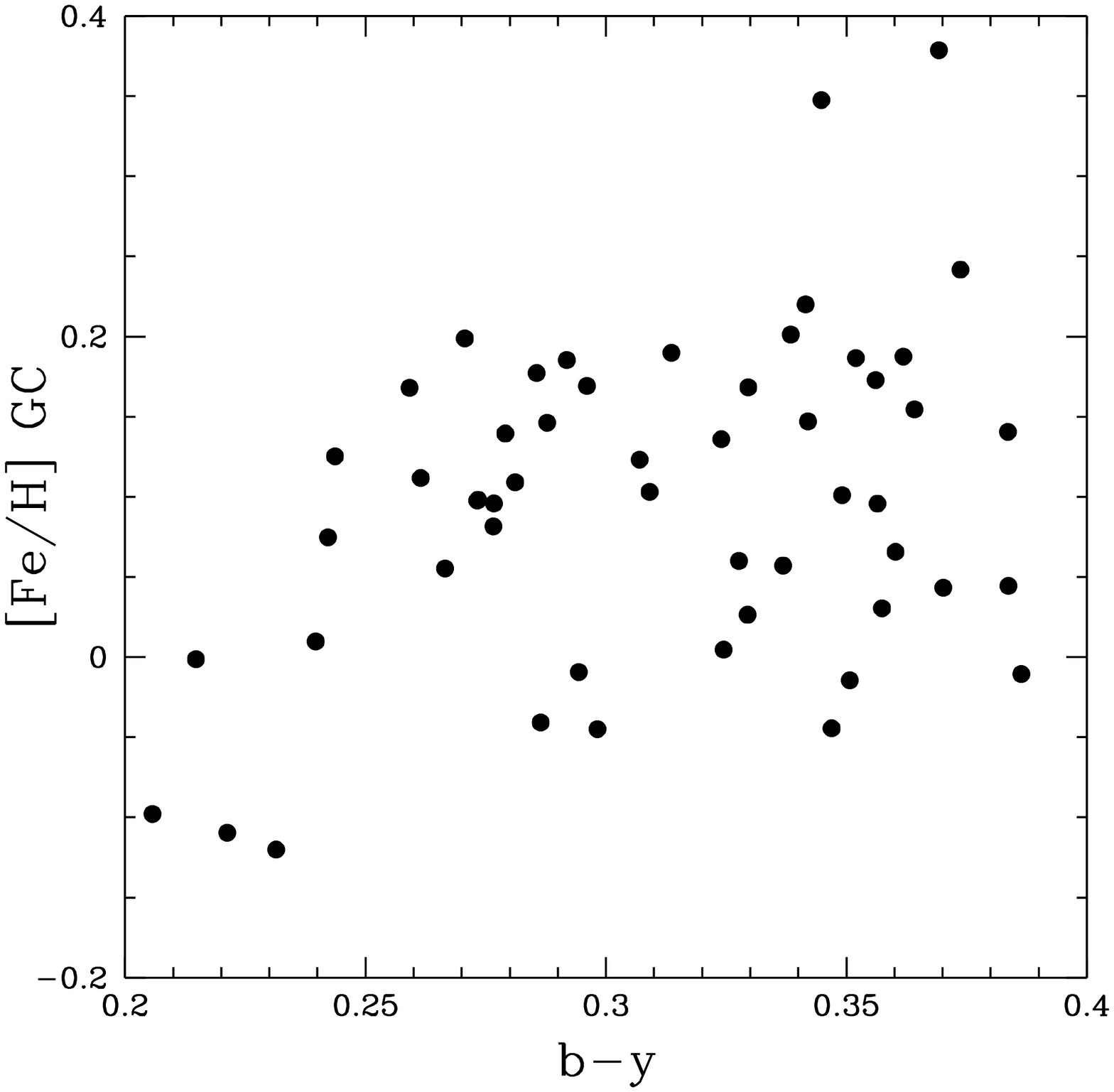}
                      \includegraphics[angle=0]{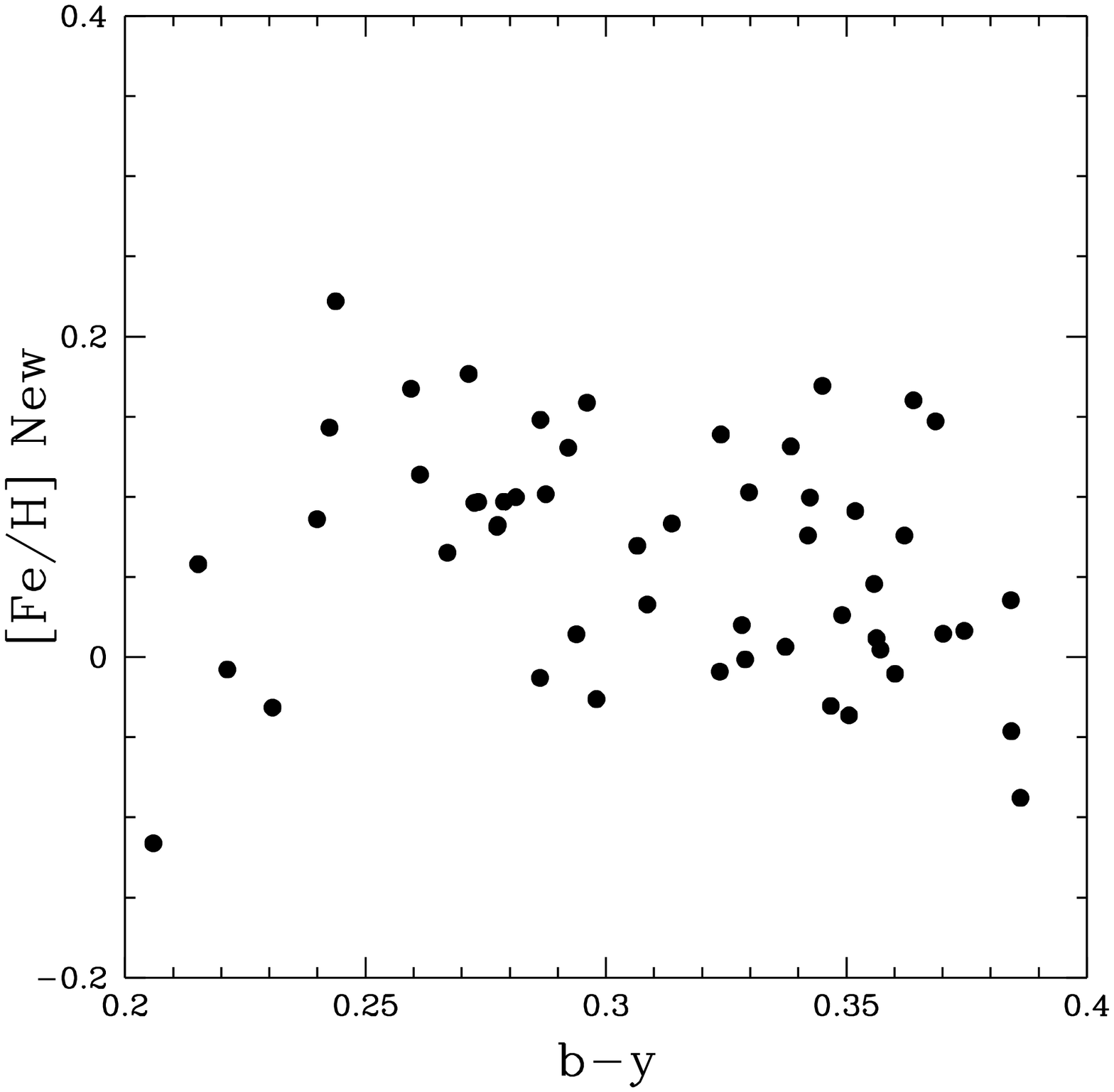}}
\resizebox{\hsize}{!}{\includegraphics[angle=0]{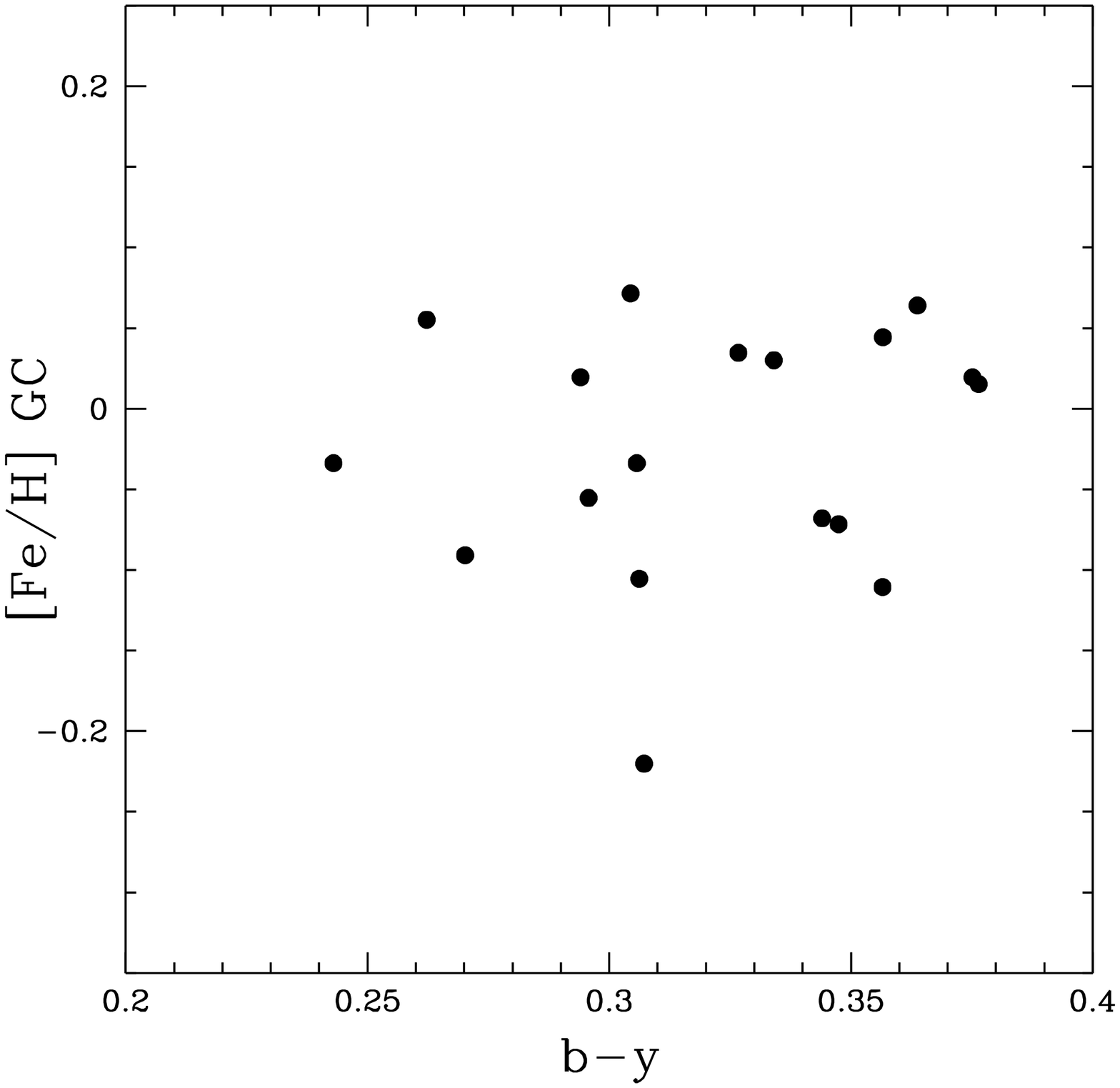}
                      \includegraphics[angle=0]{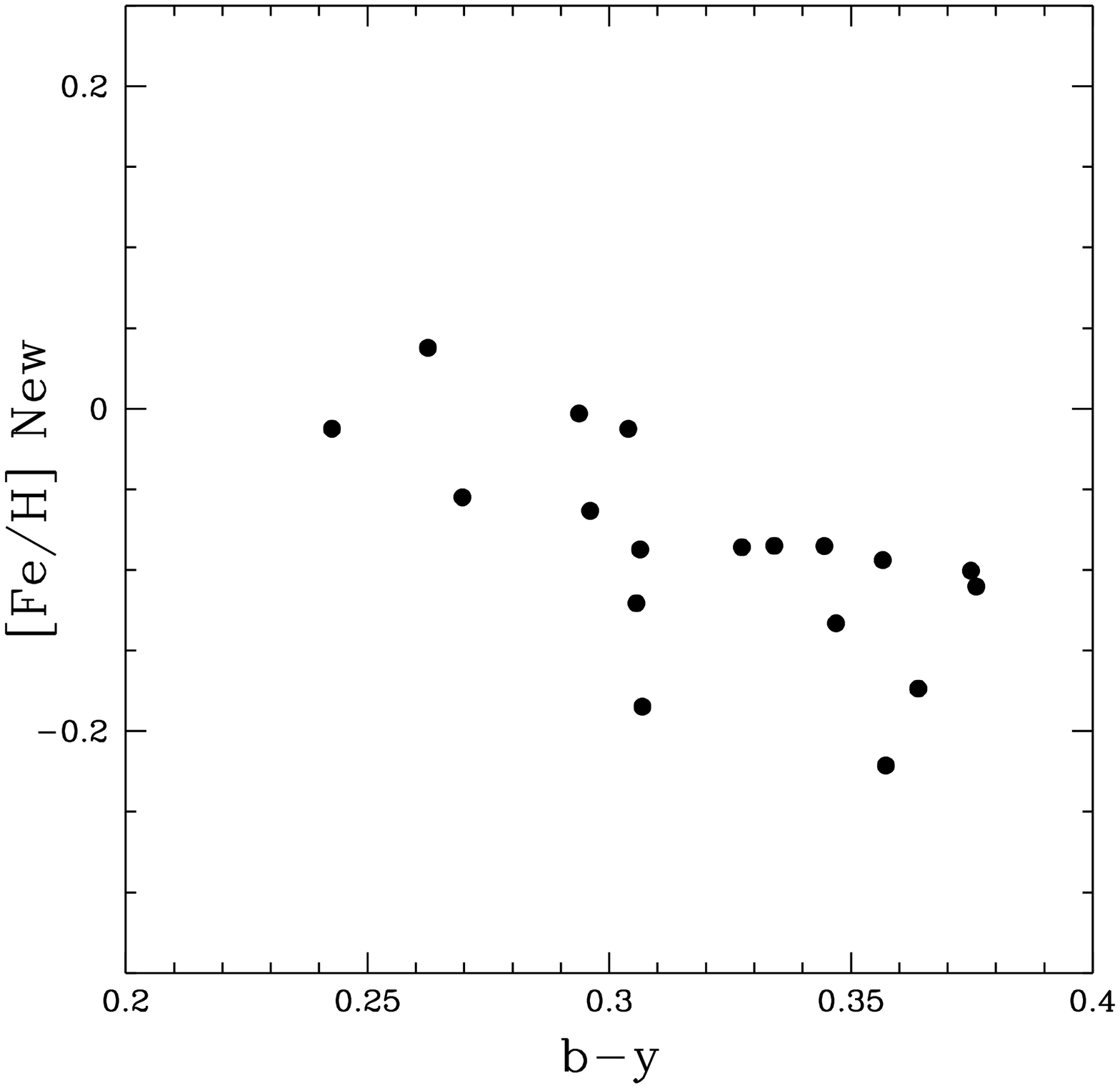}}
\caption{{\it Top left:} The GCS metallicity calibration applied to the
Crawford \& Perry (\cite{crawfordPerry}) {\it uvby}$\beta$ photometry 
of the Hyades. {\it Top right:} Using the new calibration from this paper 
instead. {\it Bottom left:} The GCS metallicity calibration applied to 
the Crawford \& Barnes  (\cite{crawfordBarnes}) {\it uvby}$\beta$ 
photometry of the Coma cluster. {\it Bottom right:} The Coma 
stars, using the new calibration.}
\label{clusters} 
\end{figure}

\subsection{The Hyades and Coma clusters}
One of the two open clusters included in the GCS, the Hyades, has been 
used by e.g. H06 to assess the zero-point of the GCS metallicity calibration. 
A comparison with the original {\it uvby}$\beta$ photometry of the Hyades 
and Coma clusters by Crawford \& Perry (\cite{crawfordPerry}) and Crawford 
\& Barnes (\cite{crawfordBarnes}) shows, however, that the photometry of 
the cluster stars that was listed in the GCS is not on the same system as 
the rest of the catalogue, perhaps because they were observed at high air 
mass from Chile. While unfortunate, this offset means that the Hyades data
will give misleading results if applied to the metallicities (and thus ages) 
of the whole GCS catalogue.

The result of using the photometry from the original sources combined with 
the earlier GCS calibrations is shown in Fig.~\ref{clusters}. For the Hyades, 
we then find $\rm [Fe/H]= 0.10\pm 0.10$; for Coma, $\rm [Fe/H]= -0.02\pm 0.08$ 
(mean and dispersion). The derived [Fe/H] for the Hyades corresponds well to 
the standard value, as expected because it is mostly based on the Schuster \& 
Nisssen calibration, which used the Hyades as a main anchor point. 

Using the new calibration instead yields $\rm [Fe/H]=0.06\pm 0.07$ for the 
Hyades and $\rm [Fe/H]= -0.09\pm 0.06$ for Coma. The dispersion among the 
cluster stars is no larger than in the field star calibration, but there is 
an offset of about 0.07 dex. 

A different photometric [Fe/H] for the Hyades is to be expected, considering 
the non-standard He/Fe ratio of this cluster (Vandenberg \& Clem \cite{clem03}), 
but Coma has a standard chemical mix, and the mean photometric metallicity 
is very near the accepted spectroscopic value. 

In summary, we conclude that it is inadvisable to use the Hyades for general 
calibrations of photometric metallicities or comparing metallicity scales. The 
same is true for the ages, especially for the unevolved solar-type stars for 
which isochrone ages are bound to be very uncertain by any method. 

Finally, we note that the revised data in Table 1 for Hyades and Coma stars
are based on the standard photometry and can be used with confidence.

\subsection{Solar analogs as a test of the calibrations}

Solar analogs offer yet another test of the temperature and metallicity
determinations. Such stars can be identified through either photometric or  
spectroscopic resemblance to the Sun -- preferably both. 18 Sco (HD 146233) 
is one of the best known examples, with {\it b-y}= 0.404, M$_{V}$= 4.77, as 
compared to the value {\it b-y}= 0.403 for the Sun by Holmberg et al. 
(\cite{holmberg06}), determined by comparison with a carefully selected 
sample of stars similar to the Sun.

[Fe/H] for 18 Sco is close to zero with both the old (+0.03) and new (-0.02) 
metallicity calibrations, whereas $\rm T_{eff}$= 5766K as derived from 
{\it b-y} and the new calibration is much closer to that of the Sun than the 
GCS value of 5689K, in good agreement with the general trend of 
Fig.~\ref{teffdiff}.

\section{Absolute magnitude/distance calibration}\label{distance}

In order to determine an improved absolute magnitude/distance calibration for 
{\it uvby}$\beta$ photometry, we selected a sample of 2451 stars with 
absolute 
magnitudes better than 0.10 mag from Hipparcos and no indication of binarity 
in the GCS. A relation using all combinations of {\it b-y}, $m_{1}$ and 
c$_{1}$ up to third order was then fitted to the absolute magnitudes of 
these stars. 

\begin{figure}[htbp] 
\resizebox{\hsize}{!}{\includegraphics[angle=0]{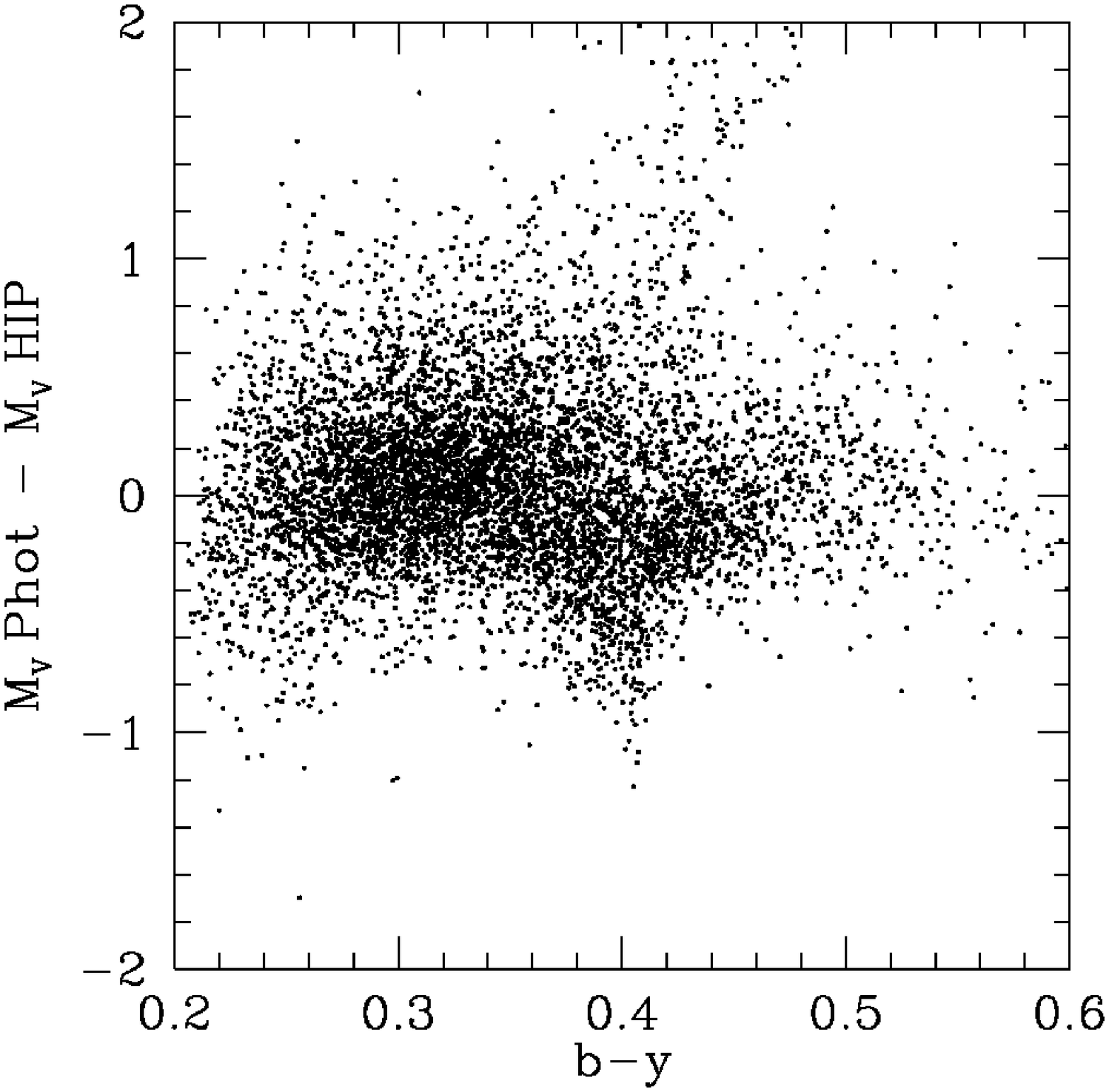}
                      \includegraphics[angle=0]{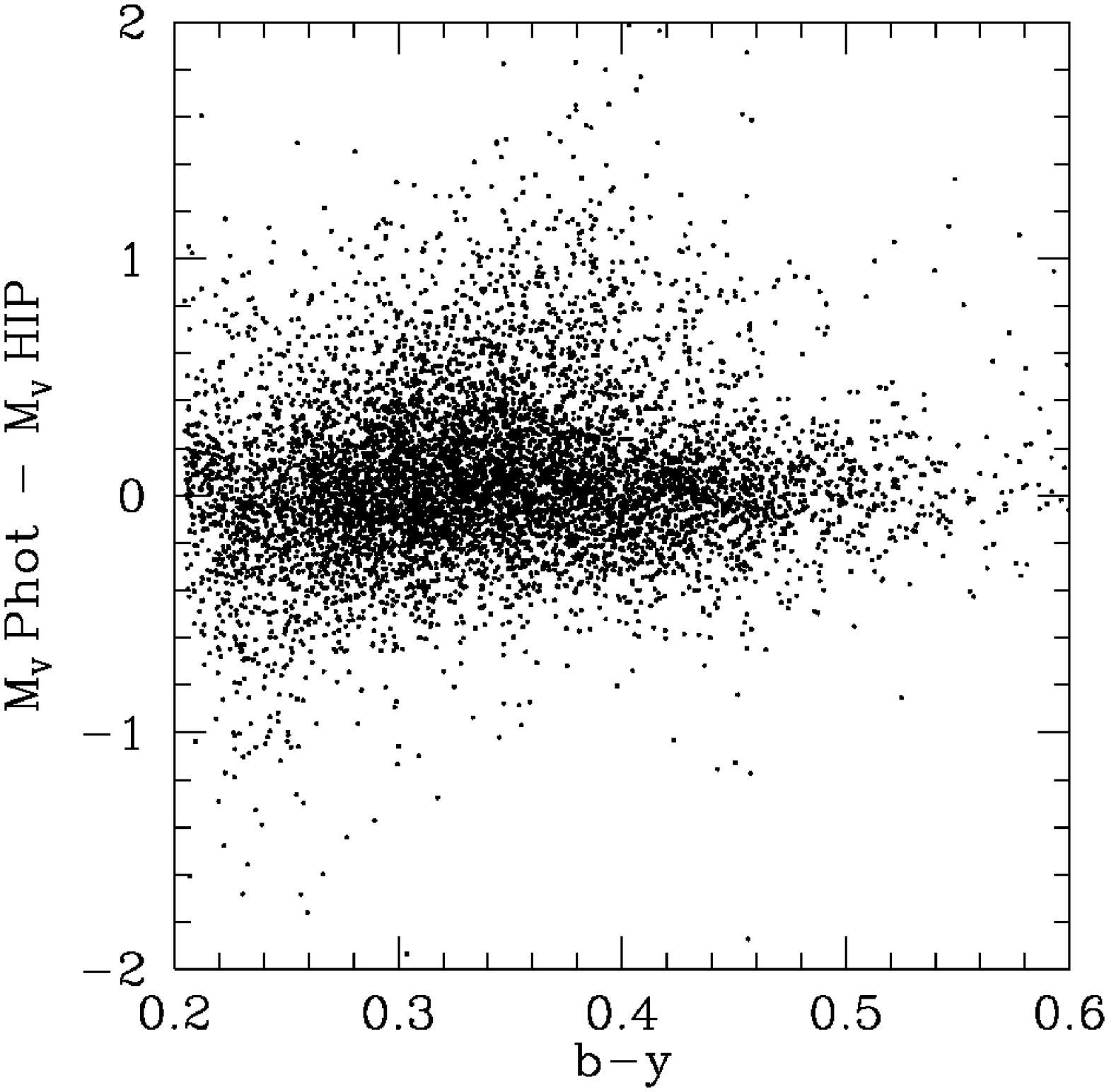}}
\caption{Differences in M$_V$ as determined from {\it uvby}$\beta$ 
photometry and from Hipparcos distances. {\it Left:} Original GCS 
photometric M$_V$. {\it Right:} M$_V$ from our improved calibration.}
\label{mvfit} 
\end{figure}

The difference between the calibration data and the fitted relation is shown 
in Fig.~\ref{mvfit}. The dispersion of the fit is 0.24 mag, but there is a 
clear excess of stars with large differences, probably due to still-undetected 
binaries. Repeating the fit after removing stars more than 2.5 $\sigma$ from 
the mean relation reduces the dispersion to 0.16 mag. The stars span the range:
 0.20 $\leq b-y \leq$ 0.60, 0.09 $\leq m_{1} \leq$ 0.64, 
0.16 $\leq c_{1} \leq$ 0.80, and -0.84 $\leq {\rm M_{V}} \leq$ 7.24.

The resulting calibration equation, valid within the above parameter ranges, 
is: \\

\noindent ${\rm M_{V}}= 2.99+2.68(b-y)+10.08m_{1}+3.14c_{1}+15.27(b-y)^{2}\\
-17.73m_{1}^{2}+26.28c_{1}^{2}-12.73(b-y)m_{1}-20.79(b-y)c_{1}\\
-38.86m_{1}c_{1}-17.47(b-y)^{3}-24.14m_{1}^{3}-8.16c_{1}^{3}\\
-56.98(b-y)^{2}m_{1}+27.86(b-y)^{2}c_{1}+103.32m_{1}^{2}(b-y)\\
-161.55m_{1}^{2}c_{1}-131.37c_{1}^{2}(b-y)+29.51c_{1}^{2}m_{1}\\
+257.37(b-y)m_{1}c_{1}$\\

${\sigma(\rm M_{V}})$ corresponds to distance errors of only 
11\% and 7\% (with and without 2.5-$\sigma$ outliers). Note that this is 
the total dispersion, 
including also the error in the Hipparcos distances. A further check of the 
new calibration based on the wide physical binaries in the sample 
confirms this estimate (see Sect. \ref{binaries}).

This relation is a clear improvement over the photometric distance 
calibration used in the GCS, which had an uncertainty of 13\% (0.28 mag). 
It can also be used to identify further 
binaries in the GCS catalogue, which only flagged stars with a difference 
between the Hipparcos and photometric M$_V$ larger than 3 $\sigma$, i.e. 
larger than 0.84 mag from the photometry alone. We have not, however, 
recomputed the statistics on binaries in the GCS, as other sources of 
uncertainty remain important.

As already expected from Fig. 10 in the GCS, there is no systematic 
difference between the GCS distances and those computed with the above 
formula (see Fig. \ref{mvfit}). However, the uncertainty of the individual 
distances - which dominates that of the space motions - is reduced enough 
that we have recomputed the distances and UVW velocity components for 
all the GCS stars, correcting a minor error in the GCS space velocities 
at the same time. 

\section{Interstellar reddening}

\begin{figure}[htbp] 
\resizebox{\hsize}{!}{\includegraphics[angle=0]{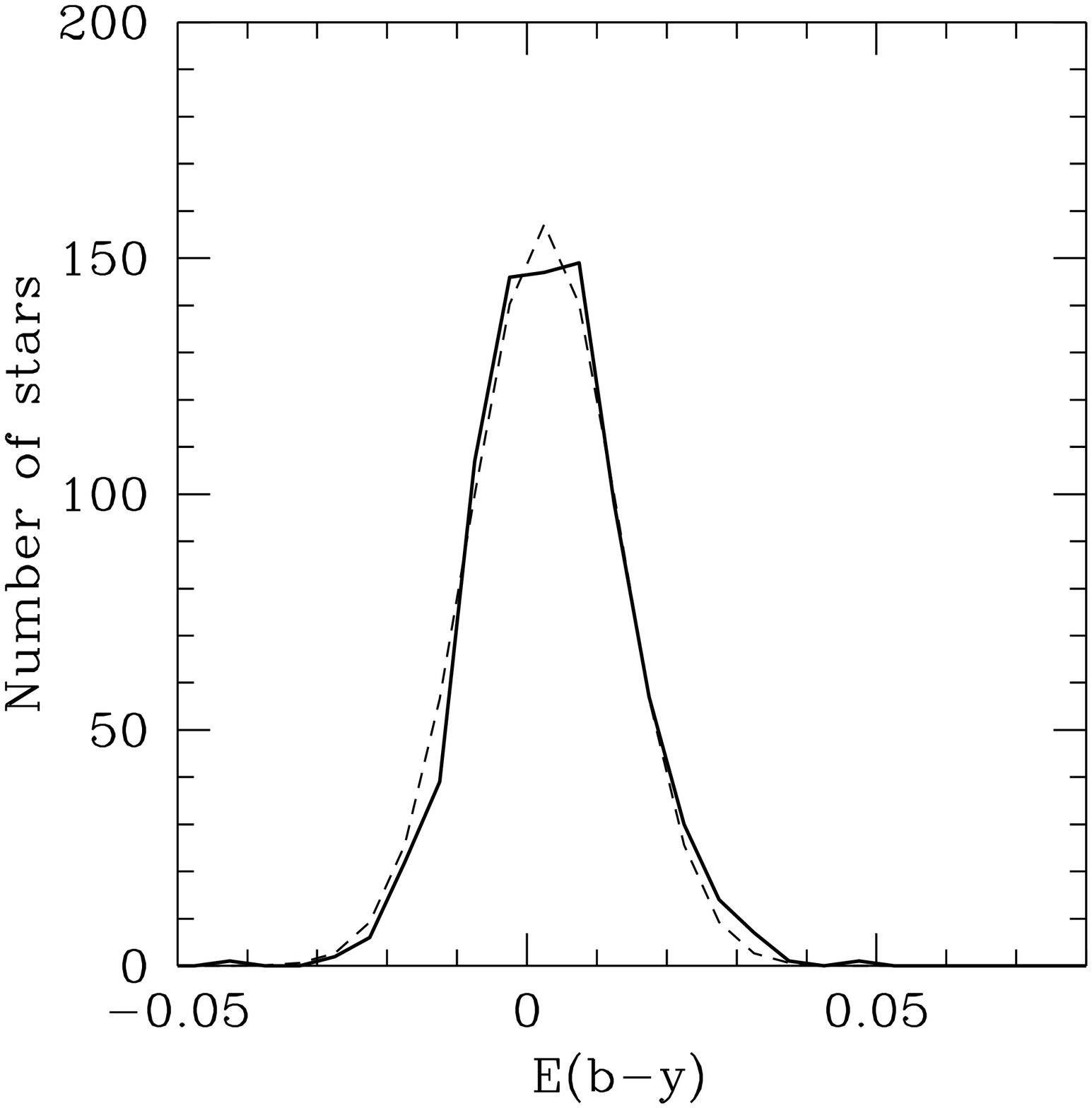}
                     \includegraphics[angle=0]{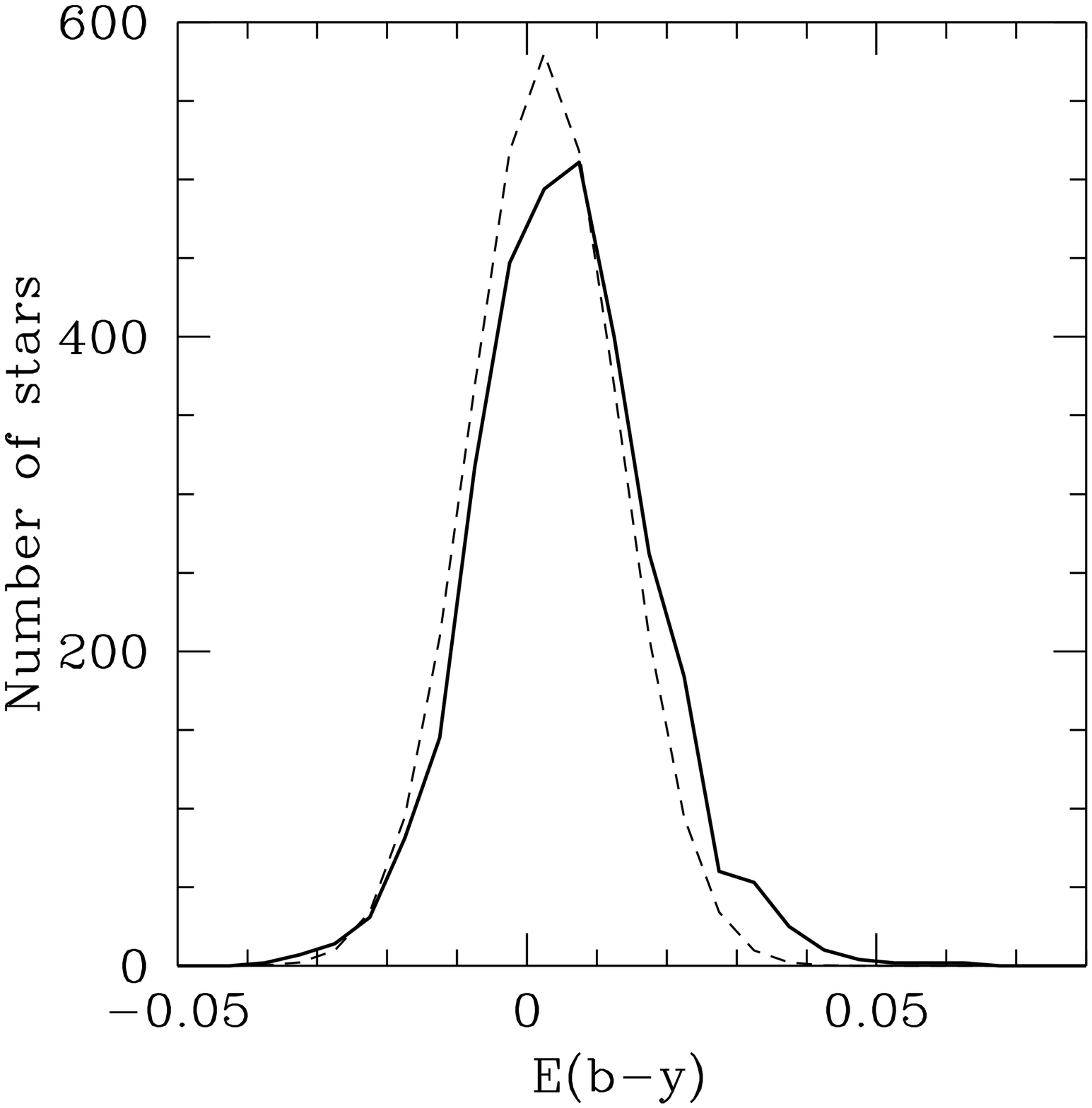}}
\caption{Distribution of {\it E(b-y)} values from the GCS.
{\it Left:} Stars within 40 pc (solid) and a Gaussian fit (dashed) for a mean 
reddening of 0.0025 mag and dispersion 0.0105 mag. {\it Right:} Stars with
distances 40 -- 70 pc and the same Gaussian, showing a significant excess 
of slightly reddened stars.}
\label{eby} 
\end{figure}

In the GCS, interstellar reddenings were derived from the intrinsic colour 
calibration by Olsen (\cite{eho88}), which yields reddening estimates with 
a stated precision of 0.009 mag. In the GCS there are 827 stars within 40 
pc which 
have {\it E(b-y)} estimates. Fig.~\ref{eby} (left) shows the histogram of 
these {\it E(b-y)} values; a best-fit pure Gaussian shows only a slight 
zero-point shift of 0.0025 and a dispersion of 0.0105 mag. 

The distribution is quite symmetric, showing that it is dominated by 
observational errors and that real reddening within 40 pc is negligible. 
This is no longer true for the distance range from 40 to 70 pc, where
the distribution is markedly skewed, with a clear excess of true positive 
extinctions (see Fig. \ref{eby}, right). Thus, contrary to the assertion 
by H06, this distance range contains a clear excess of stars with 
significant positive reddenings; the mean reddening is 0.0048 mag for this 
sample.

We conclude that the intrinsic colour calibration by Olsen (\cite{eho88}) 
remains valid with sufficient accuracy for the GCS sample, so we do 
not recompute {\it E(b-y)} and {\it (b-y)}$_0$.

\section{Stellar ages}\label{stellarages}

\subsection{Review of stellar models}

\begin{figure}[htbp] 
\resizebox{\hsize}{!}{\includegraphics[angle=0]{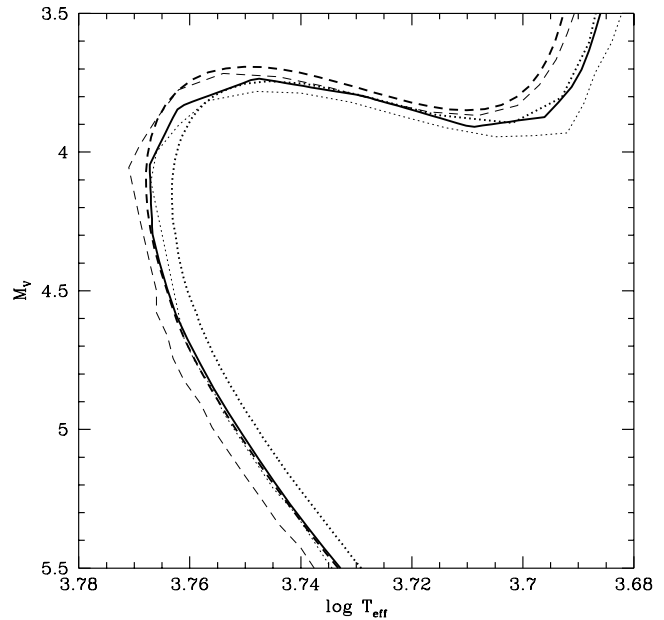}}
 \caption{Solar-metallicity 8-Gyr isochrones in the 
$\rm \log  T_{ eff}-M_{ V}$ diagram. {\it Thick lines:} Padova (full); 
Victoria-Regina (dashed); Yale-Yonsei (dotted). {\it Thin lines:} Geneva 
models:  basic (dashed) and ``best'' low-mass (dotted).}
\label{isochrones} 
\end{figure}

The determination of ages for individual field stars in the range 1-10 Gyr is 
a field with rich opportunities for large systematic as well as random 
errors. Some of these arise from the calibrations of the observational 
data, as reviewed in the preceding chapters. Others are due to the various 
approximations used in stellar model calculations, especially as regards 
the treatment of convective core overshooting in this regime of small to 
vanishing convective cores, and of the stellar atmospheres that are used 
to compute the observable properties of a model, notably log $g$ and 
$\rm T_{eff}$. 

The density and sampling of the grid of models and/or isochrones are 
also important in this range, where a major change in isochrone 
morphology near the turnoff occurs for ages of several Gyr. Other errors 
again may result from the techniques used to interpolate in the published 
isochrones and compute masses, ages, and error estimates from the observed 
data. 

In the GCS, great effort was devoted to the derivation of ages and masses 
and their uncertainties for the stars of the survey, using the sophisticated
interpolation method and Bayesian computational techniques of J{\o}rgensen \& 
Lindegren (\cite{jorgensen05}) and accounting for the average $\alpha$-element 
enhancement of metal-poor stars. Exactly the same method was adopted in the 
very recent paper by Takeda et al. (\cite{takeda}). The theoretical isochrones 
used in the GCS computations were the latest set from the Padova group 
(Girardi et al. \cite{girardi}, Salasnich et al. \cite{Salasnich}), while 
Takeda et al. (\cite{takeda}) computed an extensive set of new tracks using 
the Yale-Yonsei stellar evolution code (Demarque et al. \cite{demarque}).

In order to assess the effect of different model prescriptions, we have 
compared the Padova isochrones to both the Yale-Yonsei (Demarque et al. 
\cite{demarque}) and Victoria-Regina models (VandenBerg et al.
\cite{vandenberg}) as well as to two different Geneva model series (the 
``basic'' and ``best low-mass'' models, Lejeune \& Schaerer 
\cite{lejeune01}). 

Fig.~\ref{isochrones} shows that the differences between the various models 
are in fact quite small. The difference in age for a star located on the 
three isochrones in the middle, if derived instead from the two outlying 
ones, is below 1 Gyr at the turnoff and about 2 Gyr at $\rm M_{V}$= 5, 
respectively. For the 5 Gyr isochrones, the difference is about 1 Gyr at both 
points; the larger percentage age spread at the turnoff is related to the 
slightly different treatment of convective overshooting in the different models. 
The situation is similar at e.g. [Fe/H]= --0.71, the differences between the 
isochrones being even somewhat smaller. 

Accordingly, in the following we use the same Padova isochrones and 
$\alpha$-enhancement corrections as in the GCS itself and do not repeat the 
full age computations 
with all the different sets of models. This will highlight the effect of 
changing the calibrations as discussed above, while comparing with the ages 
computed by VF05 and Takeda et al. (\cite{takeda}) will illustrate 
the effects of photometric vs. spectroscopic input parameter determinations.

\begin{figure*}[bhtp] 
\resizebox{\hsize}{!}{\includegraphics[angle=0]{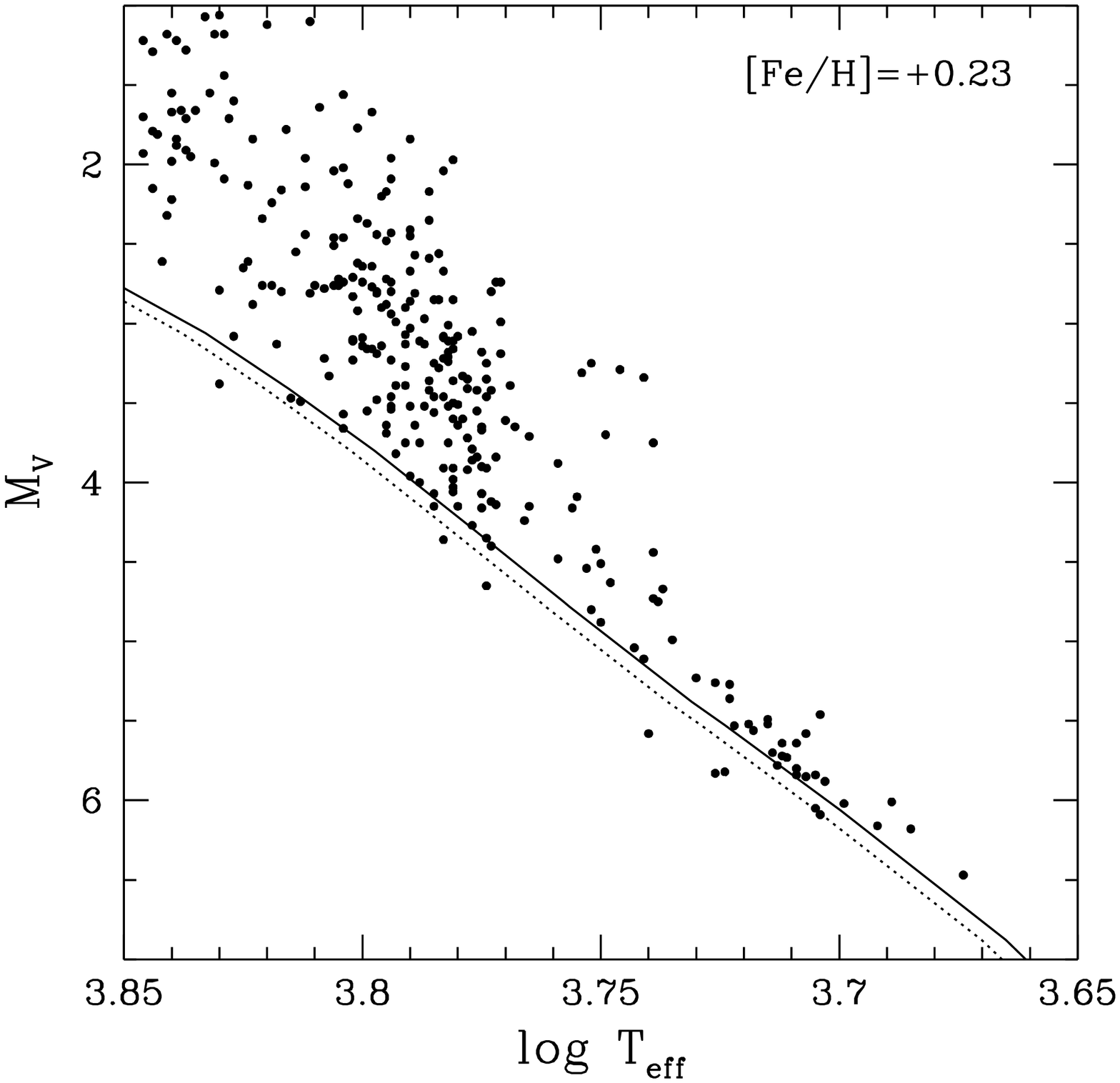}
                      \includegraphics[angle=0]{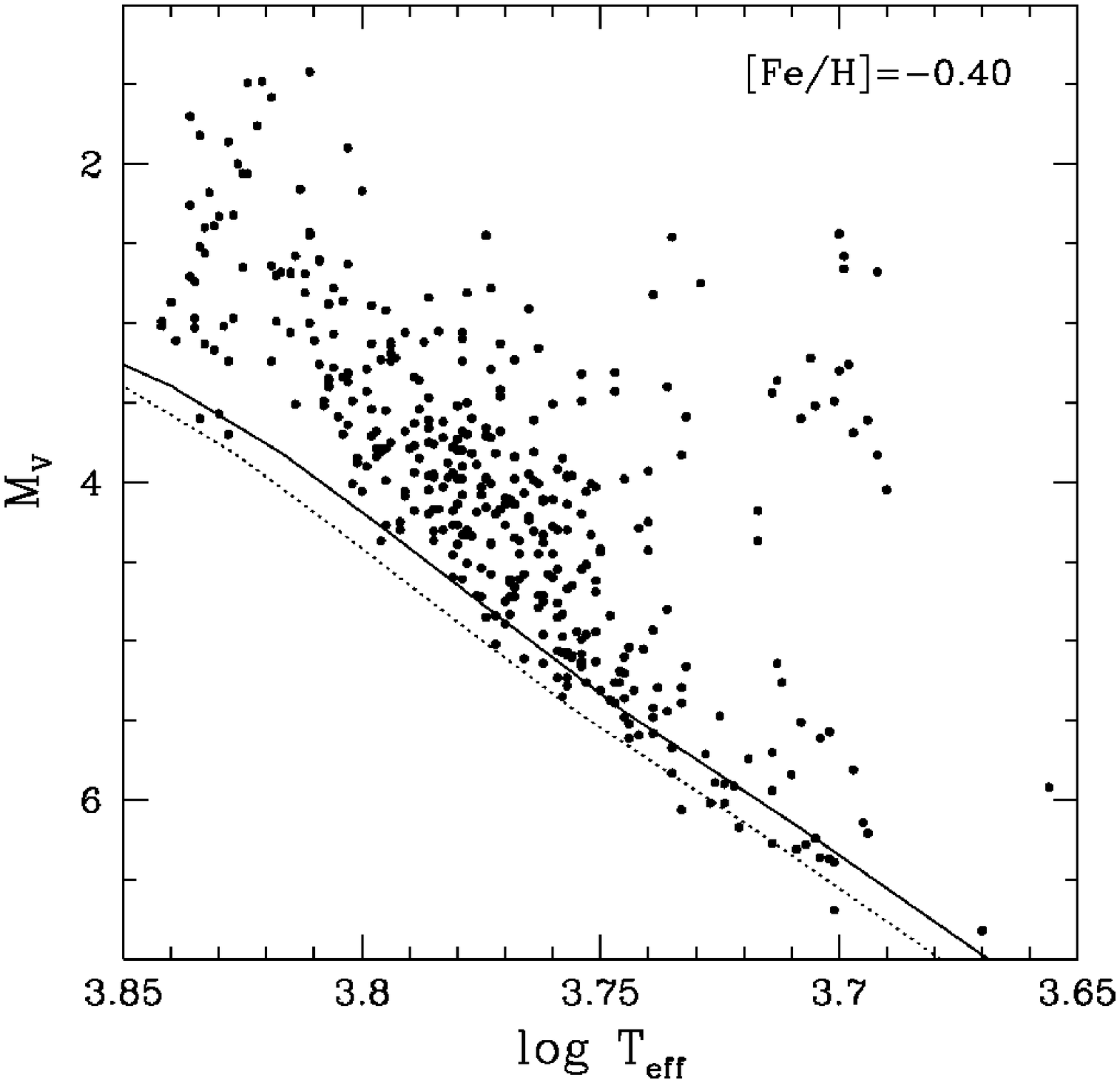}}
\resizebox{\hsize}{!}{\includegraphics[angle=0]{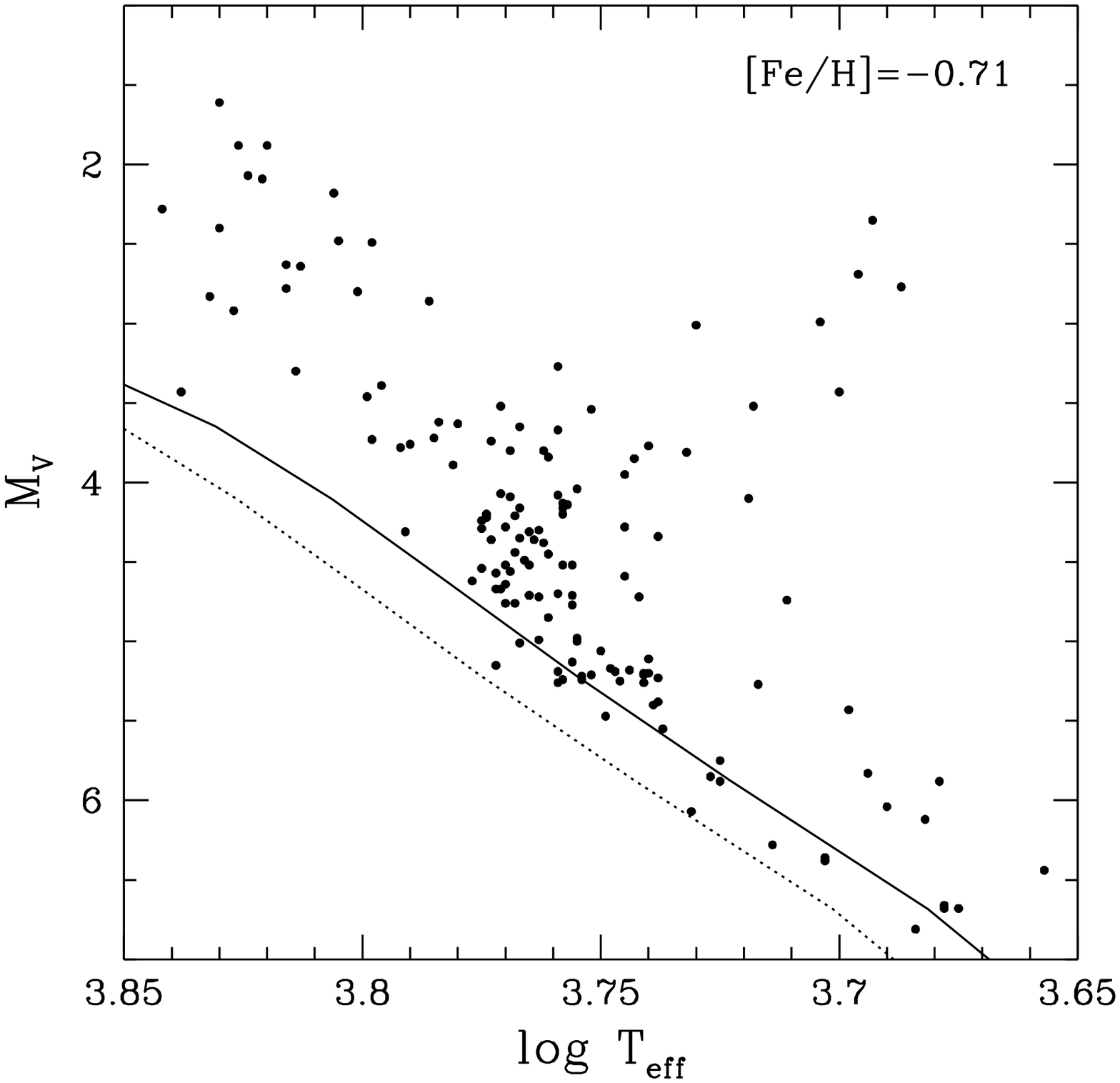}
                      \includegraphics[angle=0]{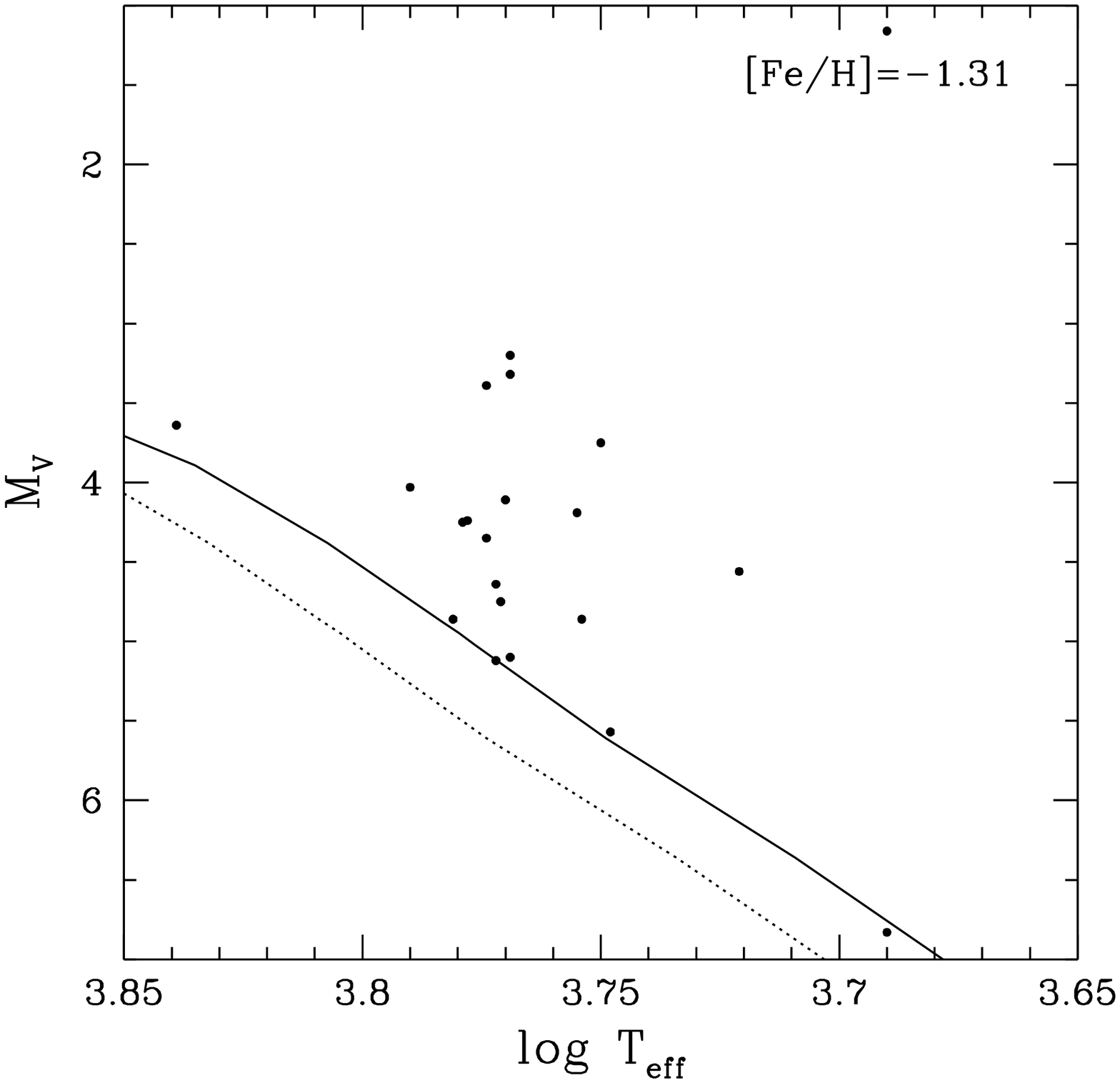}}
\caption{Comparison between the observed single GCS stars with [Fe/H]= 
+0.23$\pm$0.05, -0.40$\pm$0.05, -0.71$\pm$0.20, and -1.31 $\pm$0.30, after 
allowing for $\alpha$-enhancement, and ZAMS Padova isochrones with (solid) 
and without (dotted) temperature corrections.}
\label{isoshift} 
\end{figure*}

\begin{figure*}[hbtp] 
\resizebox{\hsize}{!}{\includegraphics[angle=0]{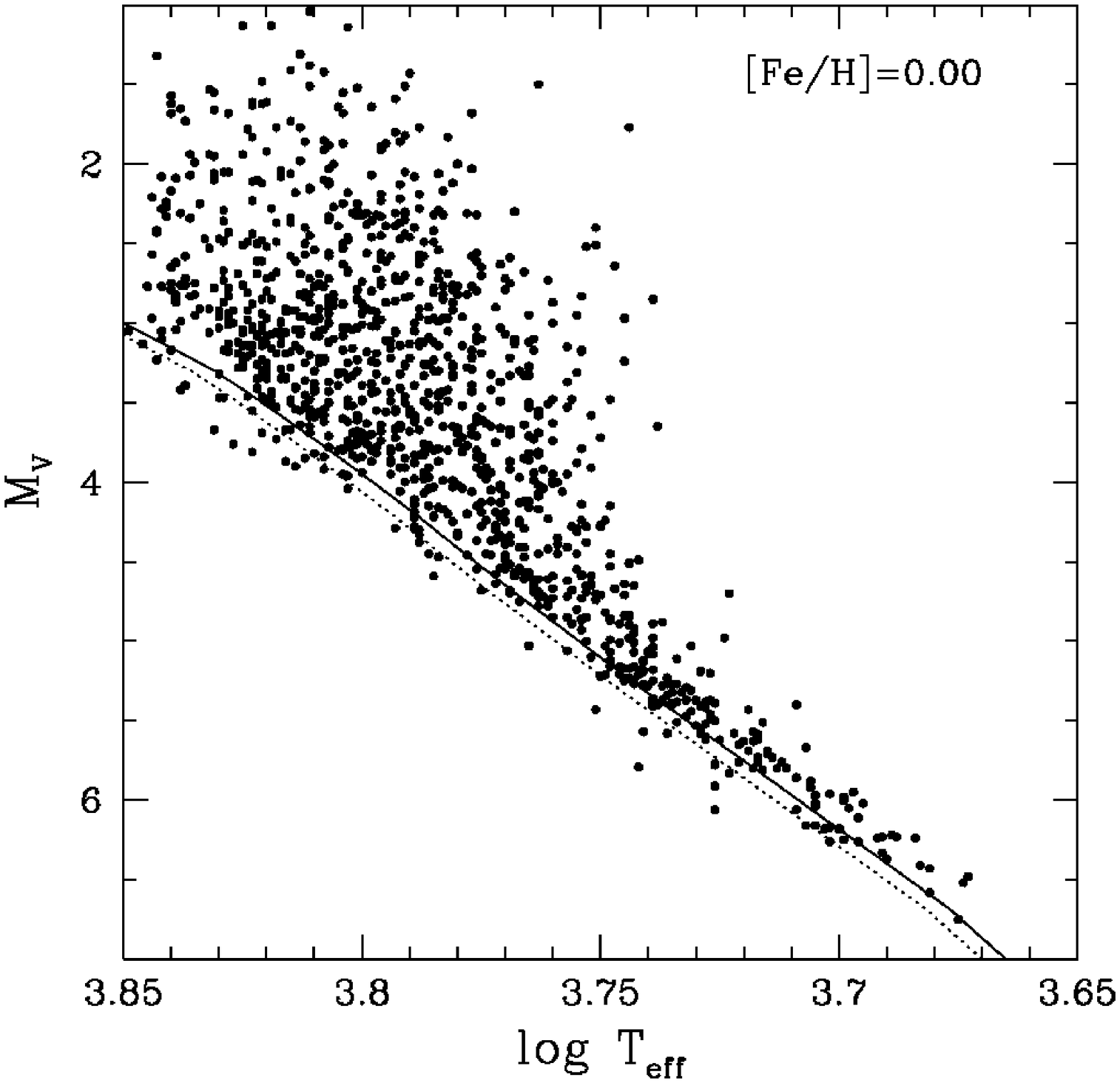}
                      \includegraphics[angle=0]{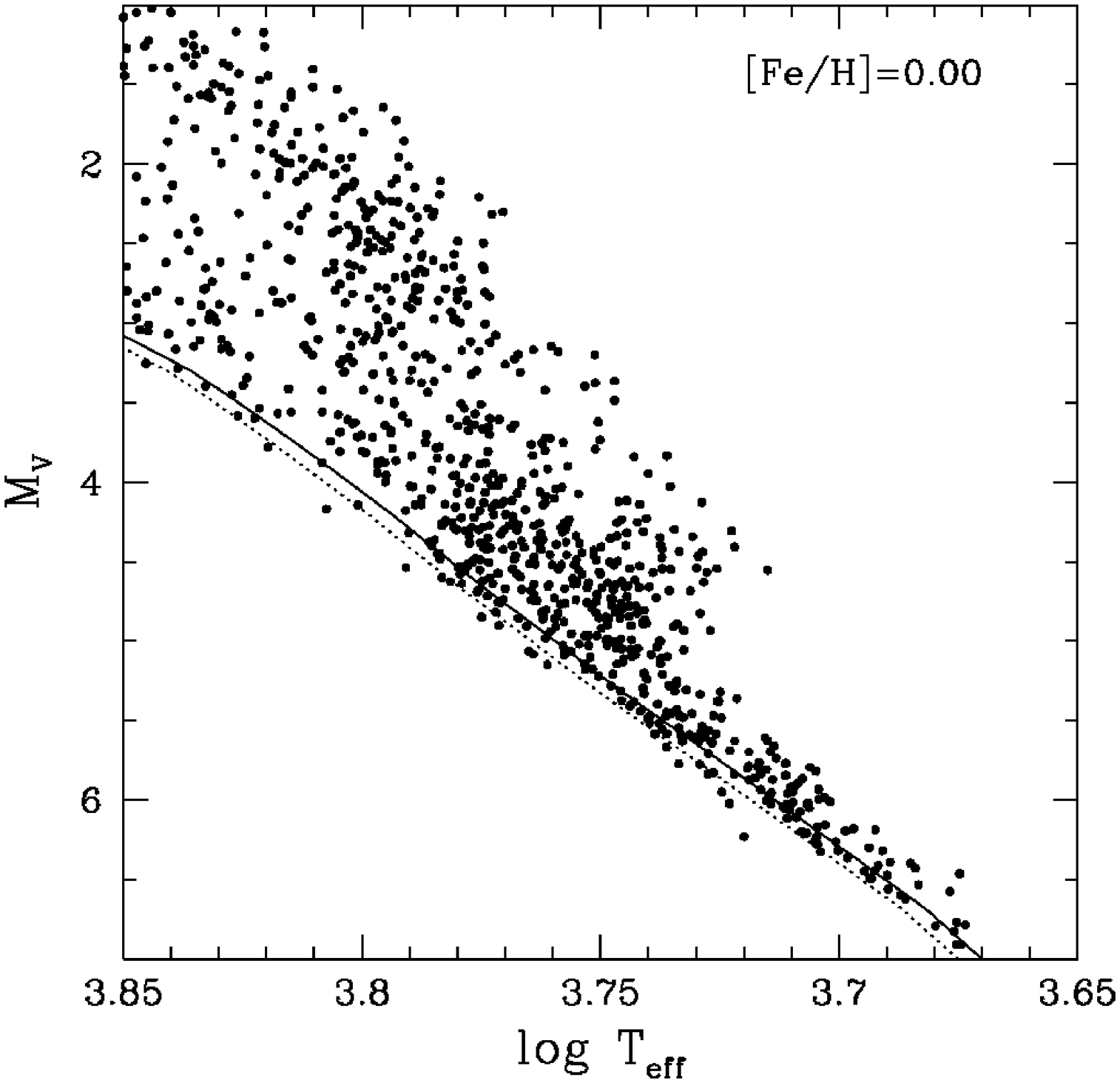}}
\caption{{\it Left:} The observed single GCS stars with [Fe/H]= 0.00 $\pm$ 
0.03 and the model ZAMS with the new temperature corrections (solid) and 
without (dotted). {\it Right:} Same, for the simulated catalogue (see Sect. 
\ref{syncat} and Fig. \ref{volcomp} on the sample differences).
}
\label{isocomp} 
\end{figure*}

\subsubsection{The model temperature scale}

In the GCS, close attention was paid to verifying the consistency between the 
computed lower main sequences and the observed unevolved stars. An appreciable 
offset was found, and substantial, metallicity-dependent, negative temperature 
corrections were applied to the Padova models to avoid deriving spuriously 
large ages for the low-mass stars. No similar offset seems to have been found 
necessary by Takeda et al. (\cite{takeda}), perhaps because the Yale-Yonsei 
models (Demarque et al. \cite{demarque}) themselves appear to be somewhat cooler 
than the Padova models (see Fig.~\ref{isochrones}). The offset is also masked 
by the hotter temperatures and higher metallicities they adopt for their stars.

With the new temperature and metallicity calibrations derived in this paper, 
the temperature corrections applied to the Padova isochrones in the GCS 
also need revision. We find corrections of $\rm \delta \log T_{eff}$ = 
-0.005 for [Fe/H] $\geq$ 0, -0.01 at [Fe/H]= -0.40, -0.02 at [Fe/H]= -0.70, 
and -0.025 at [Fe/H] $\leq$ -1.30 to be needed for the models to match the 
observed ZAMS; see Figs. \ref{isocomp} and \ref{isoshift}.

As a further check, Fig.~\ref{isocomp} compares the temperature-corrected 
solar-metallicity model main sequences with the distribution of stars in the 
observed catalogue as well as in a simulated sample (see Sect. \ref{syncat}) 
with errors in $\rm \log T_{eff}$, [Fe/H], and $\rm M_{V}$ as observed, both 
in the range [Fe/H]= 0.00$\pm$0.03. Note that the main purpose of the figure 
is to illustrate the scatter of stars around the ZAMS due to measurement 
errors; the model has not been designed to account for the effect of the 
age-dependent sample volume of the GCS on the distribution of stars in the HR 
diagram (cf. Sect. \ref{syncat}). 

\begin{figure}[thbp] 
\resizebox{\hsize}{!}{\includegraphics[angle=-90]{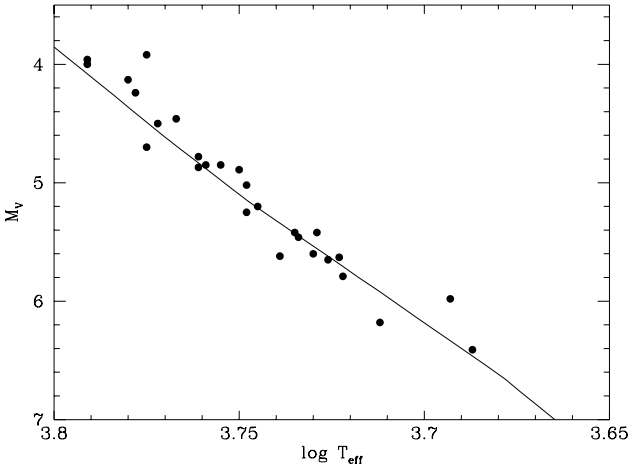}}
\caption{GCS stars with chromospheric ages below 1 Gyr and 
$\rm -0.1<[Fe/H]<0.1$, compared to a corrected 1-Gyr solar-metallicity 
isochrone.}
\label{chromo} 
\end{figure}

Another way to check the temperature corrections is to consider active 
stars with low chromospheric ages. Fig.~\ref{chromo} compares stars with  
-0.1$<$[Fe/H]$<$0.1 and chromospheric ages below 1 Gyr from the catalogue 
of Wright et al. (\cite{wright}) to a solar-metallicity 1-Gyr isochrone, 
corrected as above. As seen, the agreement is again quite satisfactory.

\begin{figure*}[hbtp] 
\resizebox{\hsize}{!}{\includegraphics[angle=0]{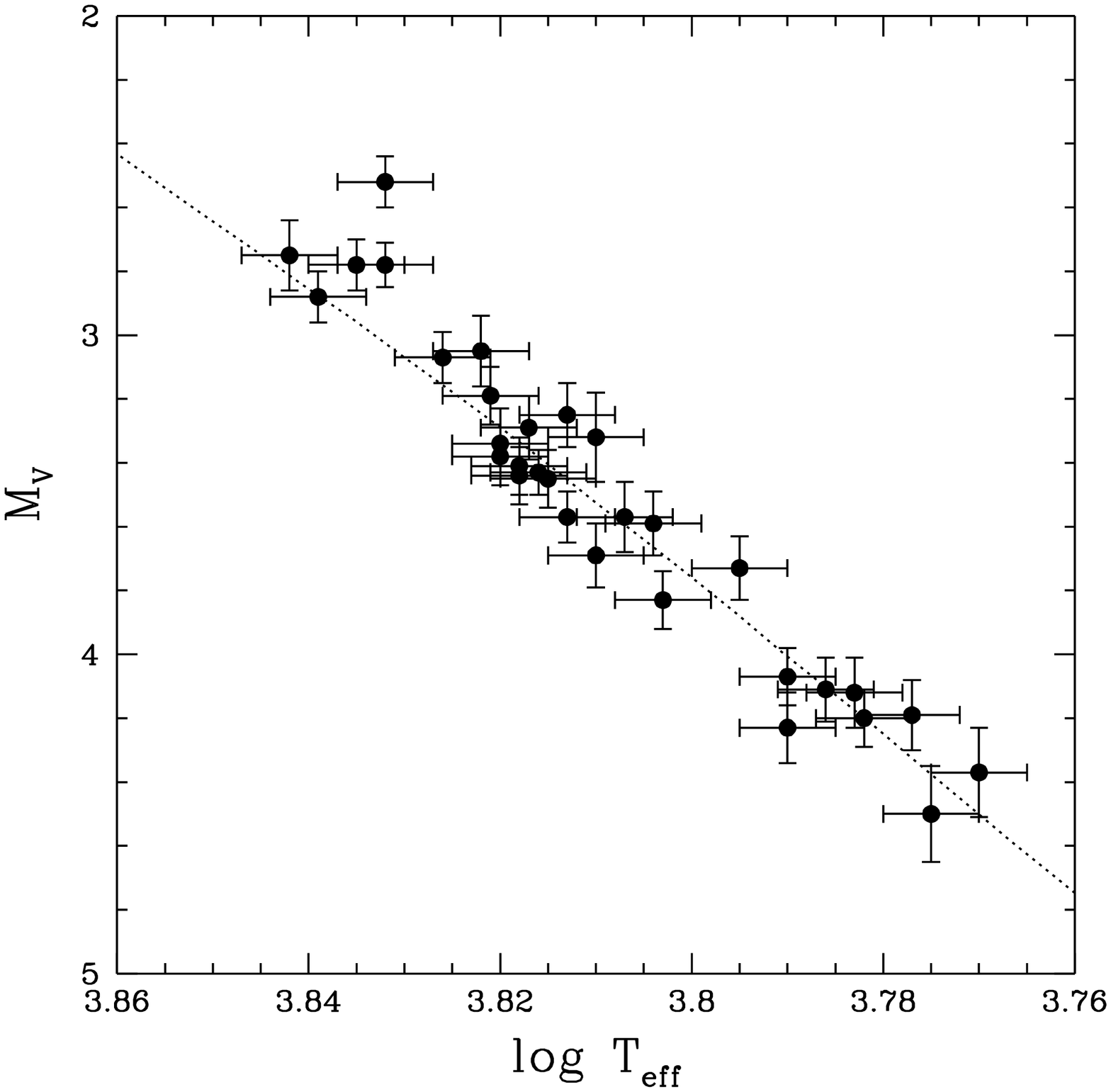}
                      \includegraphics[angle=0]{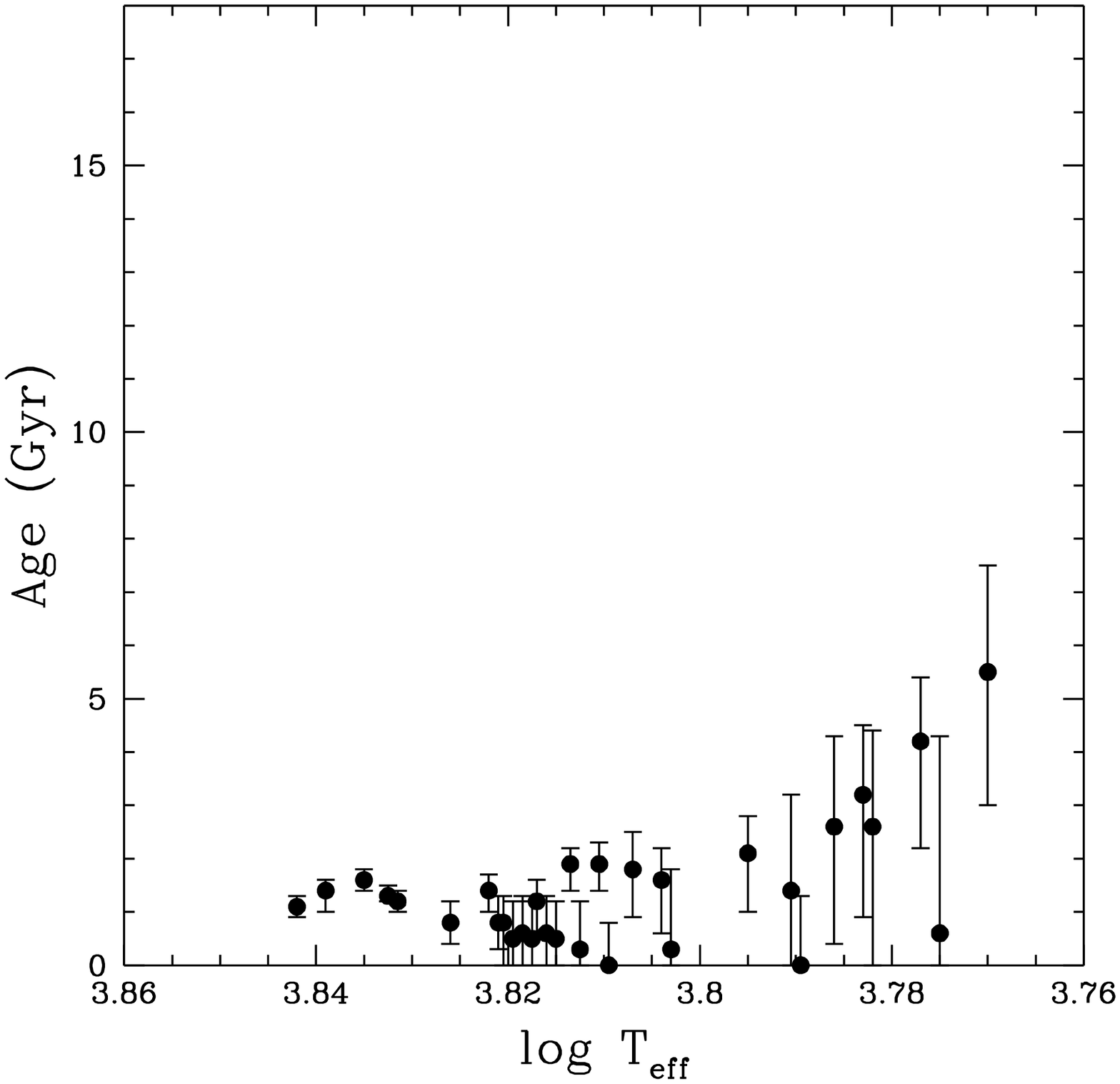}
                      \includegraphics[angle=0]{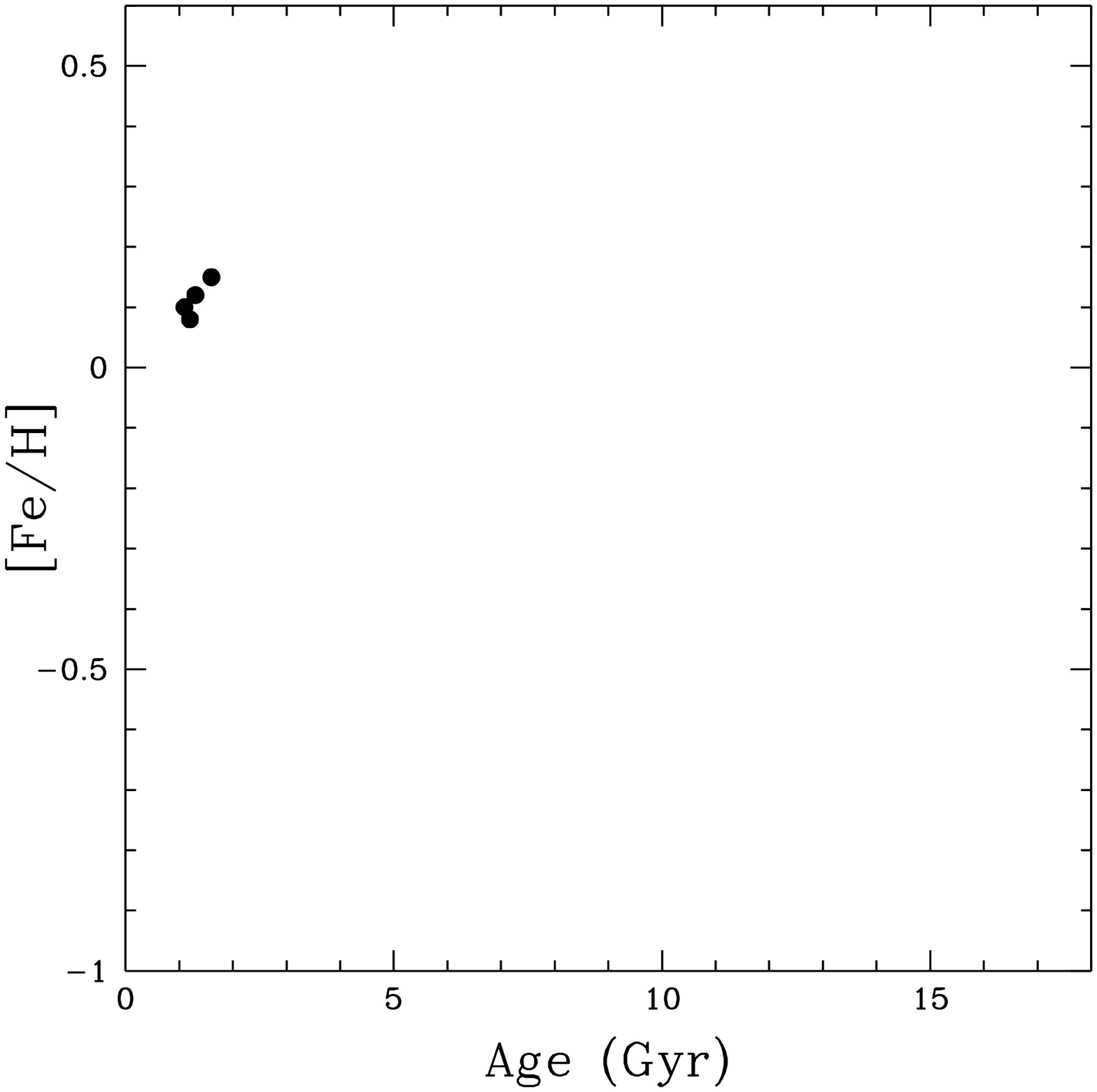}}
\caption{{\it Left:} Single Hyades stars compared to a corrected 0.7-Gyr 
isochrone for [Fe/H]= +0.15. {\it Middle:} Estimated ages and 
uncertainties; the scale offset is due to the non-standard Y/Z ratio of 
the Hyades. 
{\it Right:} Age-metallicity relation for the Hyades computed as for the 
whole catalogue, including only stars with age errors below 25\%. 
}
\label{hyadesage1} 
\end{figure*}

Finally, a test can be made using single Hyades stars with photometry from 
Crawford \& Perry (1966). Fig.~\ref{hyadesage1} compares these stars to a 
0.7-Gyr corrected isochrone and shows the estimated ages and their 
uncertainties. The scale offset in the ages is due to the use of a standard
Y/Z ratio in the models, which is incorrect for the Hyades (Vandenberg \& Clem 
\cite{clem03}).

\subsection{Consistency checks with wide binaries}\label{binaries}

\begin{figure}[htbp] 
\resizebox{\hsize}{!}{\includegraphics[angle=0]{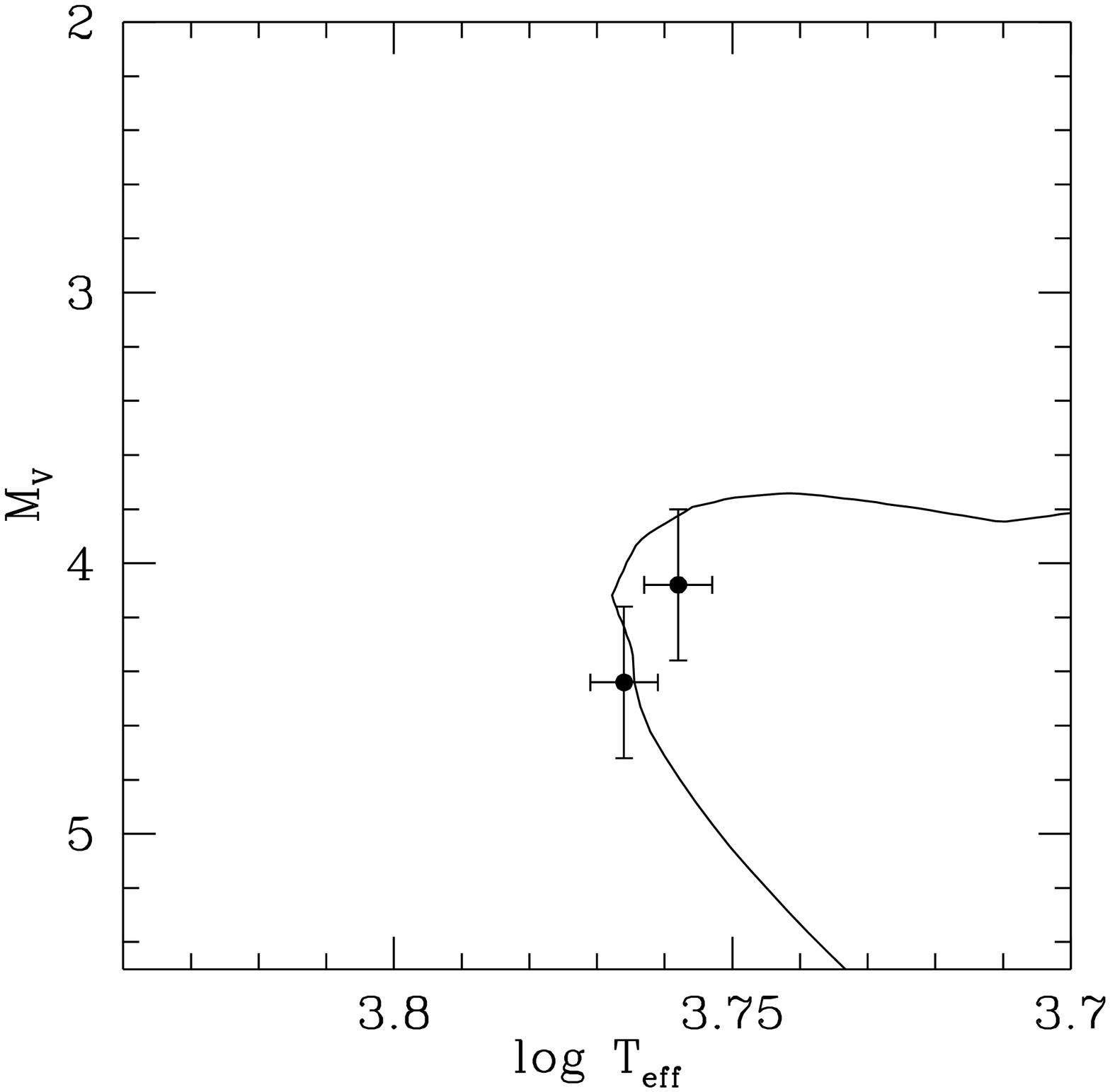}
                      \includegraphics[angle=0]{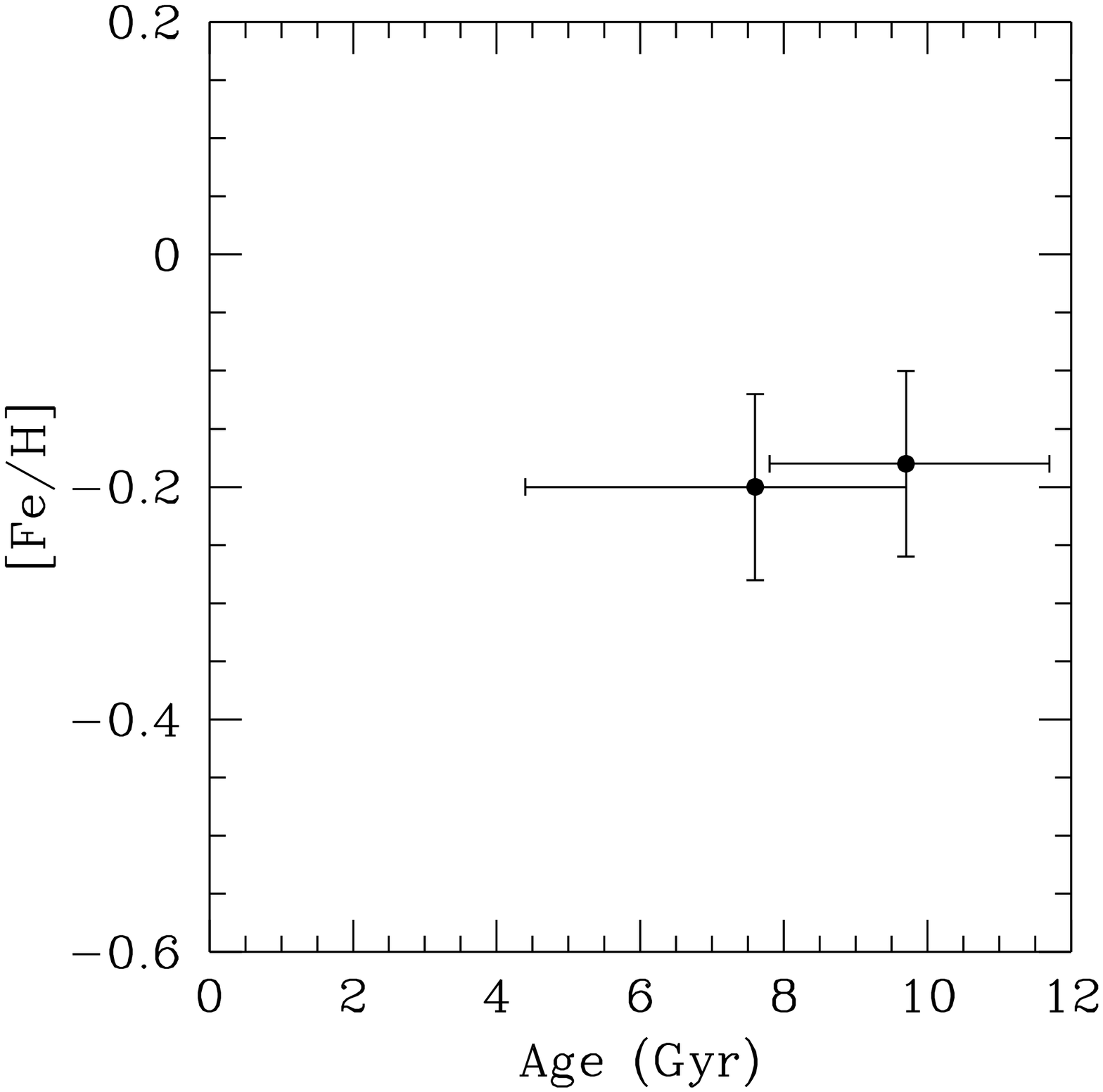}}

\caption{ ${\rm M_V}$, [Fe/H], and ages for the components 
of the physical pair HD 130140/1. The isochrone shown is for the best 
mean age of 9 Gyr. $\Delta Age$/$\sigma Age$= 0.74 for this pair. 
} 
\label{cpm} 
\end{figure}

Wide physical binaries as confirmed by proper motions and radial velocities
provide another check of the accuracy of our metallicity, distance, and age 
determinations. We have selected 18 such pairs from the GCS, with separate 
measurements of all parameters and no indication of further multiplicity. 
The pairs have a wide distribution in [Fe/H], from -0.4 to +0.1, and ages 
in the range 0--10 Gyr. Fig.~\ref{cpm} shows a typical example. 

With the GCS calibrations, the rms difference in [Fe/H] between the two 
components is 0.11dex; using the new calibrations reduces it to 0.06 dex,
in excellent agreement with our other error estimates. Similarly, using the 
new photometric distance calibration (Sect. \ref{distance}) instead of the 
older ones used in the GCS reduces the average difference in distance 
between the binary components from 16$\pm$3\% to 12$\pm$3\% (s.e. of mean). 

The accuracy of the age determination can be estimated from the ratio of 
the age difference between the two binary components, $\Delta Age$, and 
the combined 1$\sigma$ uncertainty of the two ages, $\sigma Age$. 
For the GCS ages we find $<\Delta Age$/$\sigma Age>$= 0.87.
Recomputing the ages and age uncertainties for the binary components, 
using the new calibrations and (lower) estimated errors for [Fe/H], $\rm 
T_{eff}$ from {\it b-y}, and photometric $\rm M_{V}$ as well as the newly 
corrected isochrones, we find $<\Delta Age$/$\sigma Age>$= 0.76. In 
other words, the age estimates become more consistent as well as more 
precise with the new calibrations.

\subsection{New ages vs. the GCS ages}

\begin{figure}[htbp] 
\resizebox{\hsize}{!}{\includegraphics[angle=0]{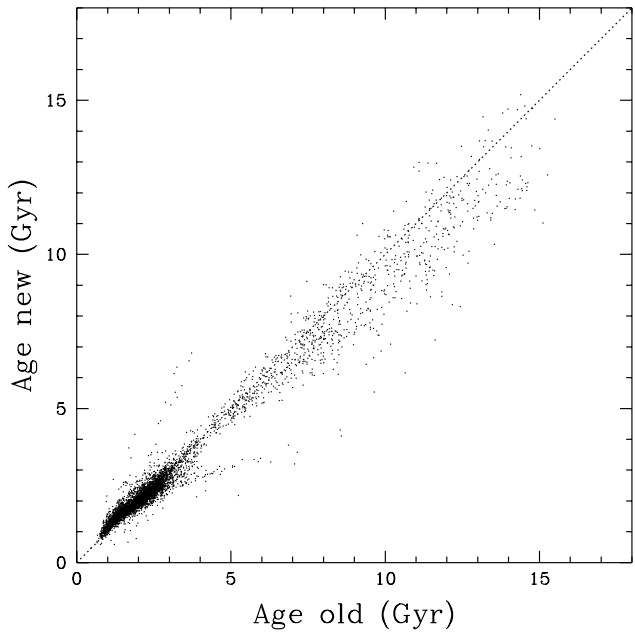}}
\caption{Ages based on the new metallicity and temperature calibrations 
and model corrections vs. the original ages GCS, using only single stars 
with ages better than 25\% in both sets. The dotted line shows the 
1:1 relation.} 
\label{agecomp1} 
\end{figure}

Fig.~\ref{agecomp1} compares the GCS ages, computed with the old [Fe/H]
and $\rm T_{eff}$ calibrations, with the ages computed using the new 
calibrations and model corrections from this paper. Overall, the differences 
are insignificant, much smaller than the estimated individual uncertainties. 
A linear fit gives the following mean relation between the new ages and 
those in the GCS: $\rm Age_{New} = 0.16 + 0.90 \times Age_{GCS}$ (all ages 
in Gyr). Thus, on average, the largest ages decrease by $\sim$10\%. This 
is similar to the differences seen when using different stellar models and 
has negligible impact on the interpretation.  

The only noteworthy deviations occur on the two ``branches'' that can 
be seen in Fig.~\ref{agecomp1}. This small group of stars is located in 
the ``hook'' region in the HR diagram, where an observed point is matched 
by two different isochrones, 
one placing the star on the main-sequence turnoff, the other on the early 
subgiant branch. As explained e.g. in J{\o}rgensen \& Lindegren 
(\cite{jorgensen05}), this leads to a two-peaked G-function structure. 
Small changes in the assumed [Fe/H] and $\rm T_{eff}$ may then change
the relative height of the peaks, and thus the most probable age of the 
star. For such double-peaked G-functions, one could use a weighted mean 
of the two maximum values rather than a simple fit to the highest peak, 
but the improvement would be cosmetic rather than real.

\begin{table*}[ht]
\caption[]{Sample listing of the recomputed parameters for the first 
five stars in the GCS catalogue. $Age_{low}$ and $Age_{up}$ are the 
lower and upper 1-$\sigma$ confidence limits on the computed age, respectively. 
The full table is available in electronic form from the CDS (see reference 
on title page).\label{table1}}
\begin{tabular}{|r@{\hspace{1mm}}l@{\hspace{0mm}}l@{\hspace{0mm}}|@{\hspace{2mm}
}r@{\hspace{1mm}}|@{\hspace{1.5mm}}r@{\hspace{1.0mm}}|r@{\hspace{1mm}}r@{\hspace
{1mm}}r@{\hspace{1mm}}r@{\hspace{1mm}}|@{\hspace{1mm}}r@{\hspace{1mm}}r@{\hspace
{1mm}}r@{\hspace{2mm}}|@{\hspace{0mm}}r@{\hspace{1mm}}r@{\hspace{1mm}}r|}
\hline  HIP&Name&Comp&RA ICRF{\hspace*{0mm}}&Dec ICRF{\hspace*{-1mm}}&$\rm 
logT_{e}$&[Fe/H]&d&$M_{v}$&Age&$Age_{low}$&$Age_{up}$& U&V&W\\
&&&h{\hspace{1.5mm}}m{\hspace{2.5mm}}s{\hspace*{3mm}}&$^{o}${\hspace{2.5mm}}$
^{\prime}${\hspace{2mm}}$^{\prime\prime}${\hspace*{2mm}}&&&pc&mag&Gy&Gy&Gy&
~km s$^{-1}$&~km s$^{-1}$&~km s$^{-1}$ \\
1&2&3&4{\hspace*{5.5mm}}&5{\hspace*{5.5mm}}&6&7&8&9&10&11&12&13&14&15\\
\hline
   437&HD 15      &    &00 05 17.8&+48 28 37&     &     &   &     &    &    &    
&&&\\
   431&HD 16      &    &00 05 12.4&+36 18 13&&&   &     &    &    &    &&&\\
   420&HD 23      &    &00 05 07.4&--52 09 06&3.776&-0.17& 42& 4.44& 3.7& 
0.3&6.3&40&-22&-16\\
   425&HD 24      &    &00 05 09.7&--62 50 42&3.768&-0.33& 70& 3.91& 8.5& 
7.5&9.6&-31&7&14\\
      &HD 25      &    &00 05 22.3&+49 46 11&3.824&-0.30& 79& 3.09& 2.0& 1.8& 
2.2&17&0&-22\\
  \hline
\end{tabular}
\end{table*}

Finally, we compare our results to two recent sets of ages, both computed
from the spectroscopic results by VF05. One set is given in that paper itself, 
using a relatively crude method, while the more recent one by Takeda et al. 
(\cite{takeda}) was computed with the same Bayesian technique as used in the 
GCS, based on a dense grid of stellar evolution tracks computed with the 
Yale-Yonsei code (Demarque et al. \cite{demarque}). In order to avoid confusion 
by including stars with highly uncertain ages (see, e.g. Fig. 8 of Reid et al. 
\cite{reid07}), we select only single stars with ages better than 25\% as given 
by both sources.

\begin{figure}[htbp] 
\resizebox{\hsize}{!}{\includegraphics[angle=0]{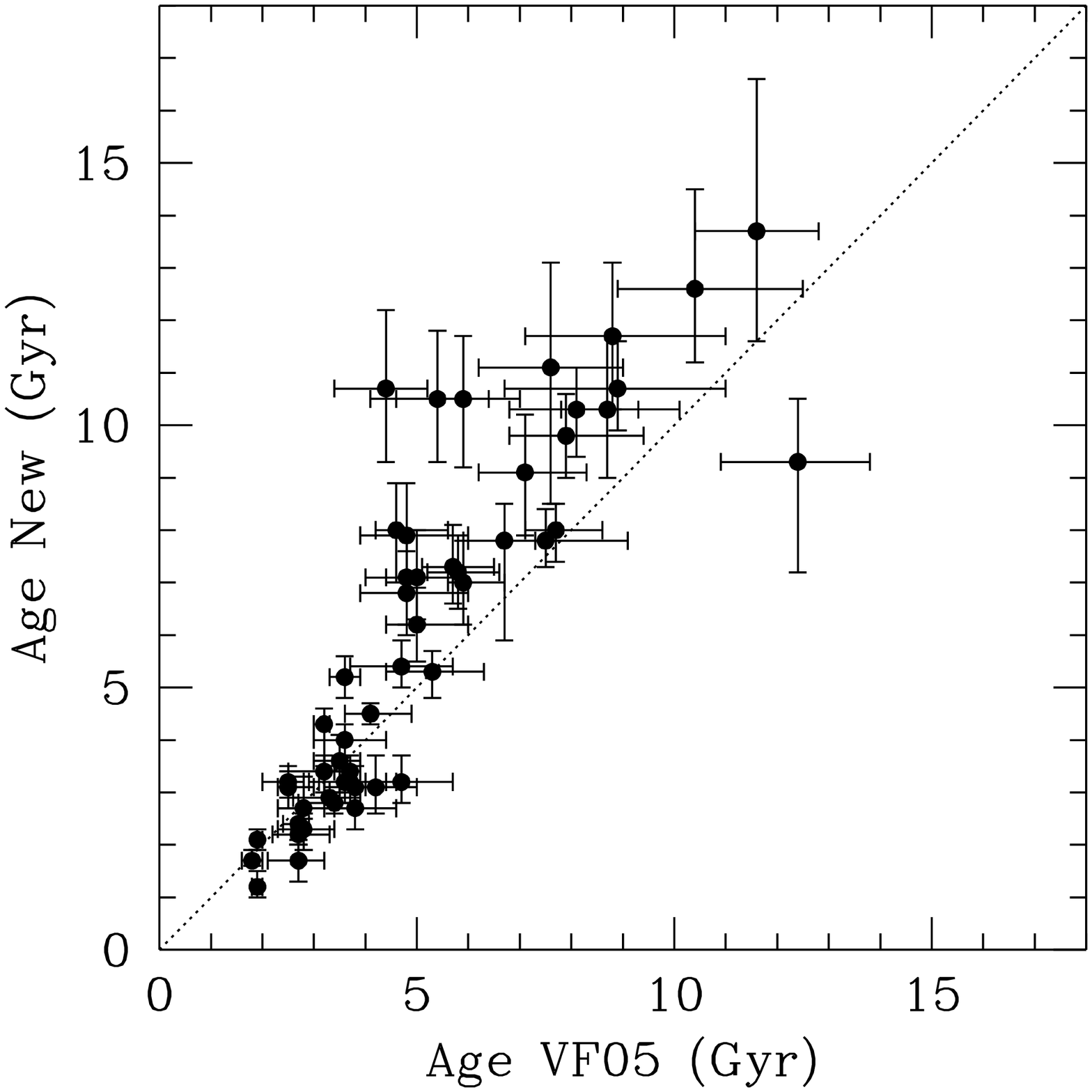}
                      \includegraphics[angle=0]{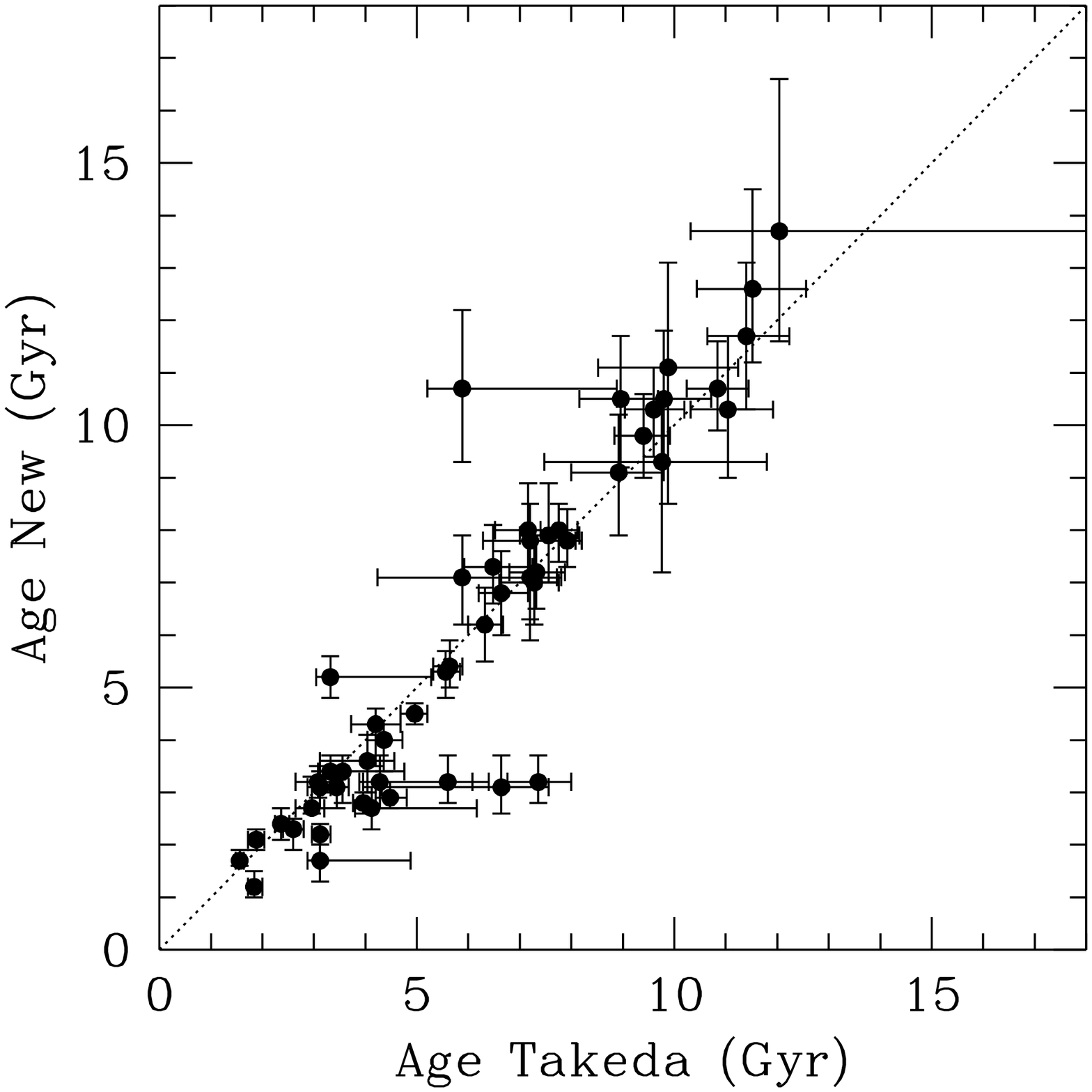}}

\caption{{\it Left:} New ages vs. ages from VF05. The sample consists 
of single GCS stars with both new and VF05 ages better than 25\%; the 1:1 
relation is shown. {\it Right:} The same comparison with ages from Takeda 
et al. (\cite{takeda}).} 
\label{agecompvft} 
\end{figure}

Fig.~\ref{agecompvft} shows the remarkable result of this comparison. While 
the VF05 ages are on average $\sim$10\% lower 
than ours, those by Takeda et al. (\cite{takeda}) are in essentially perfect 
agreement with ours. Considering that these results are based on totally 
different observational data and analysis techniques, calibrations, and 
stellar models, the agreement is extraordinarily close and inspires a solid 
confidence in age determinations from isochrones. Note that the differences 
between Takeda et al. (\cite{takeda}) and VF05
are due to different computational techniques applied to the same data, while 
the differences between the old and new GCS ages are due to improved 
calibrations, the computational method remaining the same. That the 
systematic differences seen in both cases are only of the order of 
$\sim$10\% gives further confidence in the robustness of the method.

For future reference we recall that, temperature and 
metallicity calibrations aside, the age estimates computed by H06 
were based on an unrealistically small value for the uncertainty of $\rm 
T_{eff}$. Further, the computations ignored the uncertainty in $M_V$ as well 
as the temperature offset of the models, the $\alpha$-enhancement of the 
metal-poor stars, and the presence of binaries in the sample. Moreover, the 
bright limit of $M_V$=2 excluded many young stars for which good ages can be 
readily determined (see GCS Fig. 11a), while at the same time including 
low-mass giants for which meaningful ages cannot be derived. Finally, the 
uncertainty of the resulting ages was not discussed.

\section{Results and sample characteristics}\label{sample} 

Based on the new calibrations and model corrections discussed above, we have 
computed new $\rm T_{eff}$, [Fe/H], distances, ages and age uncertainties,
and space motions for all the stars in the GCS. Table \ref{table1} gives 
the results; the full version is available in electronic form at the CDS. 
We have not recomputed Galactic orbital elements for the stars, as 
uncertainties in the assumed (smooth, axisymmetric) potential are more 
important than minor revisions of the space motions. 
In the following, we discuss the implications of the new data for our 
understanding of the evolution of the Milky Way disk. 

In such discussions, it is crucial to not only employ the best possible 
calibrations from observed to astrophysical parameters, but also to 
understand to what extent the criteria used to select the stellar sample 
may influence the conclusions drawn. This is especially important when 
studying subsamples selected on the basis of having one or more of the 
derived parameters available, perhaps within a certain level of precision, 
rather than the entire GCS. Age is the most striking example: The samples 
of stars having ages better than 25\%, or ``well-defined'' ages in the 
GCS sense, or any ages at all, are all very different from the full GCS 
sample in very non-random and non-intuitive ways. 

Much effort was spent checking these issues in preparation for the GCS. 
Only overall results were mentioned in the paper itself, details being 
left to the present paper. These tests are the subject of the following 
sections.

\subsection{Simulating the sample selection effects}\label{syncat}
To estimate the interplay between the sample selection criteria and any 
errors in the determination of the derived parameters in the GCS, we have
performed an extensive set of numerical simulations based on a synthetic 
catalogue. The synthetic catalogue assumes a constant density of stars in a 
spherical volume of $\sim$250 pc radius, a mixture of 90\% thin-disk and 10\% 
thick-disk stars with an even distribution in stellar age and evolutionary 
stage. Several distributions in metallicity and kinematics were imposed on the 
thin disk in order to explore the possible systematic effects; for the thick 
disk we assumed an asymmetric drift in $V$ of 65 km~s${-1}$ 
and ($\sigma_U$,$\sigma_V$,$\sigma_W$) = (72,43,36) km~s$^{-1}$. 

The star density was normalised such that, after applying the appropriate 
cutoffs, the resulting sample would contain about 15\,000 stars total, with 
about 97\% thin-disk and 3\% thick-disk stars, as observed. Each component has a 
distribution in metallicity and kinematics similar to the observed values, and 
the distribution of stellar parameters and measurement errors is also like the 
real catalogue. In order to better test the age estimation process at faster 
evolutionary stages, the stars are more uniformly distributed along the 
isochrones than what would result from a standard IMF. 

By applying the various calibrations, computations, and parameter cuts to 
the simulated catalogue in parallel with the observed sample, we can 
ascertain in quantitative terms which systematic effects may be introduced 
at each stage and whether they have any significant influence on the 
conclusions. 

\begin{figure}[htbp] 
\resizebox{\hsize}{!}{\includegraphics[angle=-90]{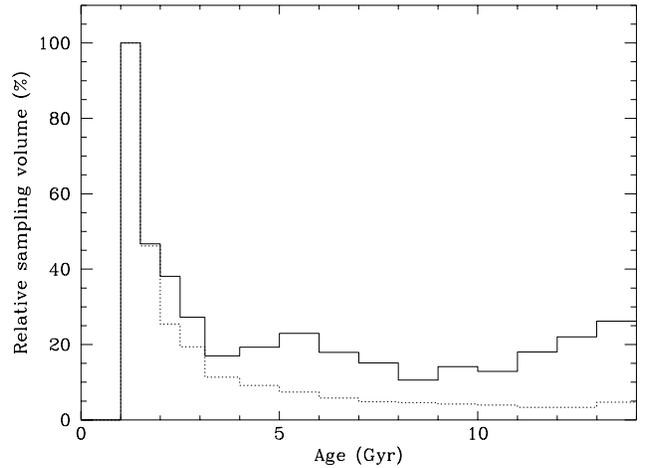}}
\caption{
Average volume occupied by single stars with ``well-determined'' ages in the 
real GCS (full line) and in the simulated catalogue with a fixed apparent 
magnitude limit of $V$= 8 (dotted).
}
\label{volcomp} 
\end{figure}

As a first example, we show how the volume of space sampled by the GCS varies 
with age. Because older stars are, on average, fainter (and cooler) than younger 
stars, and the GCS sample is limited by apparent magnitude, the volume sampled 
by the GCS decreases with stellar age. GCS Fig. 24 showed this effect as a 
function of colour; here we illustrate the dependence on age directly by 
imposing a simple apparent-magnitude cutoff at $V$= 8 and computing the average 
volume occupied by the stars in successive 1-Gyr age bins, normalised to that 
occupied by the youngest stars (=100\%). 

Fig. \ref{volcomp} compares the result for the simulated sample (dotted curve) 
with the same computation for the real GCS (full line). Only single stars with 
``well-determined'' ages are included. The GCS adopts fainter limiting 
magnitudes for the redder stars in order to achieve volume completeness to 40 
pc, so the fraction of older stars is higher than for the simplified simulation. 

However, the origin of the preponderance of young stars in, e.g., Figs. 
\ref{agecomp2} and \ref{avrnew} is clear: They are included from a volume 
$\sim$5 times larger than the stars older than $\sim$3 Gyr. Accounting properly 
for these sampling volume differences is, of course, crucial when deriving the 
star formation history of the solar neighbourhood from the data -- a study we 
do not undertake here.

More specific simulations have been carried out to test the most significant 
conclusions of the GCS; they are described in the following as appropriate in 
each context.

\section{The metallicity distribution function}\label{dGproblem}

Models of star formation and chemical evolution in the galactic disk make 
predictions of the distribution of heavy-element abundances in stars that 
(could) have survived from all stages of the evolution. The classic 
failure of 'closed-box' models -- the so-called ``G-dwarf 
problem -- refers to the observed lack of those metal-poor dwarf stars that 
should have accompanied the high-mass stars that produced the heavy 
elements we do observe in the younger stars, assuming a constant IMF. One 
may ask whether the metallicity calibration or the way the data are 
compared with the models may cause a spurious difference. 

\begin{figure}[hbtp] 
\resizebox{\hsize}{!}{\includegraphics[angle=0]{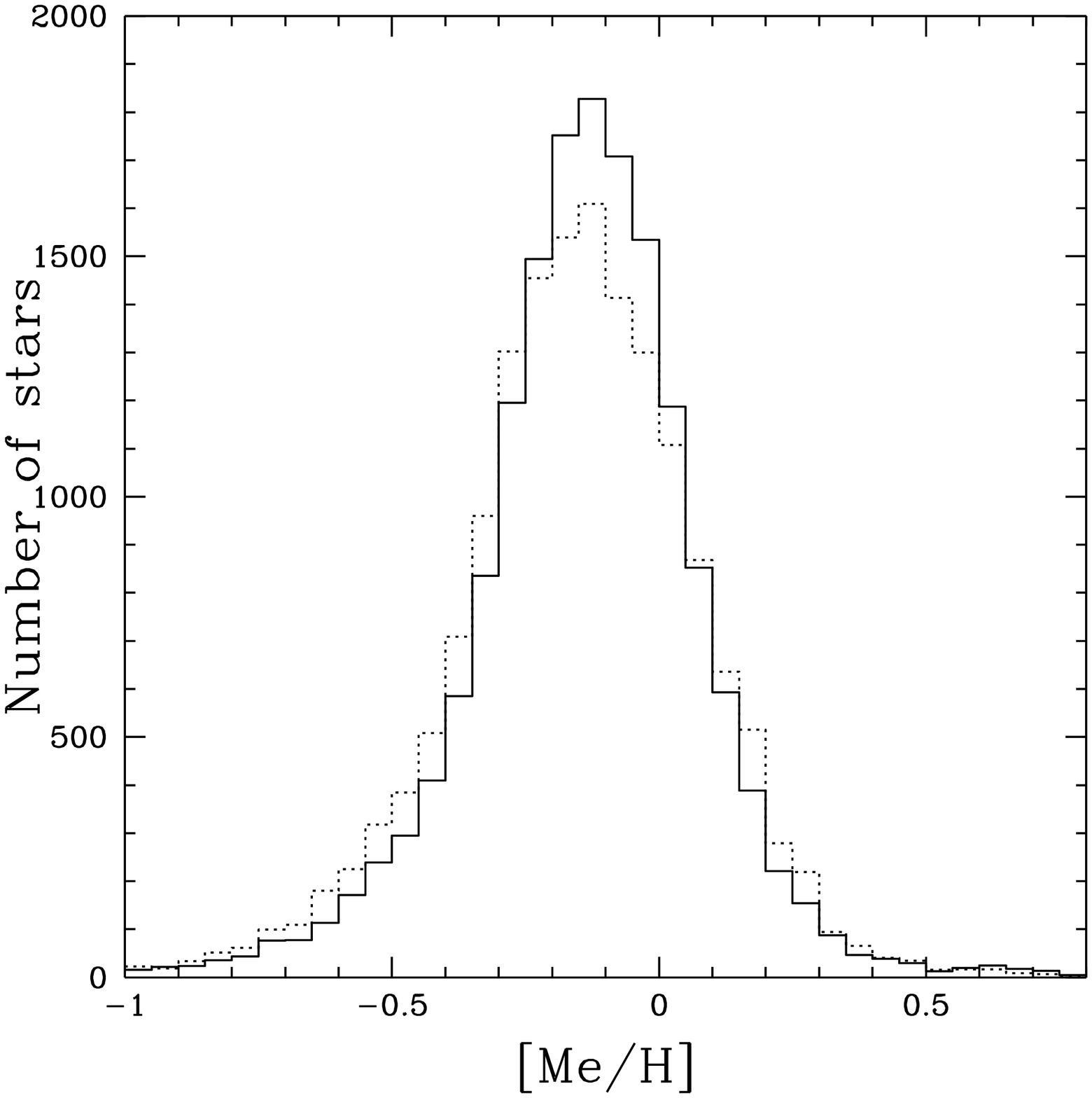}
                      \includegraphics[angle=0]{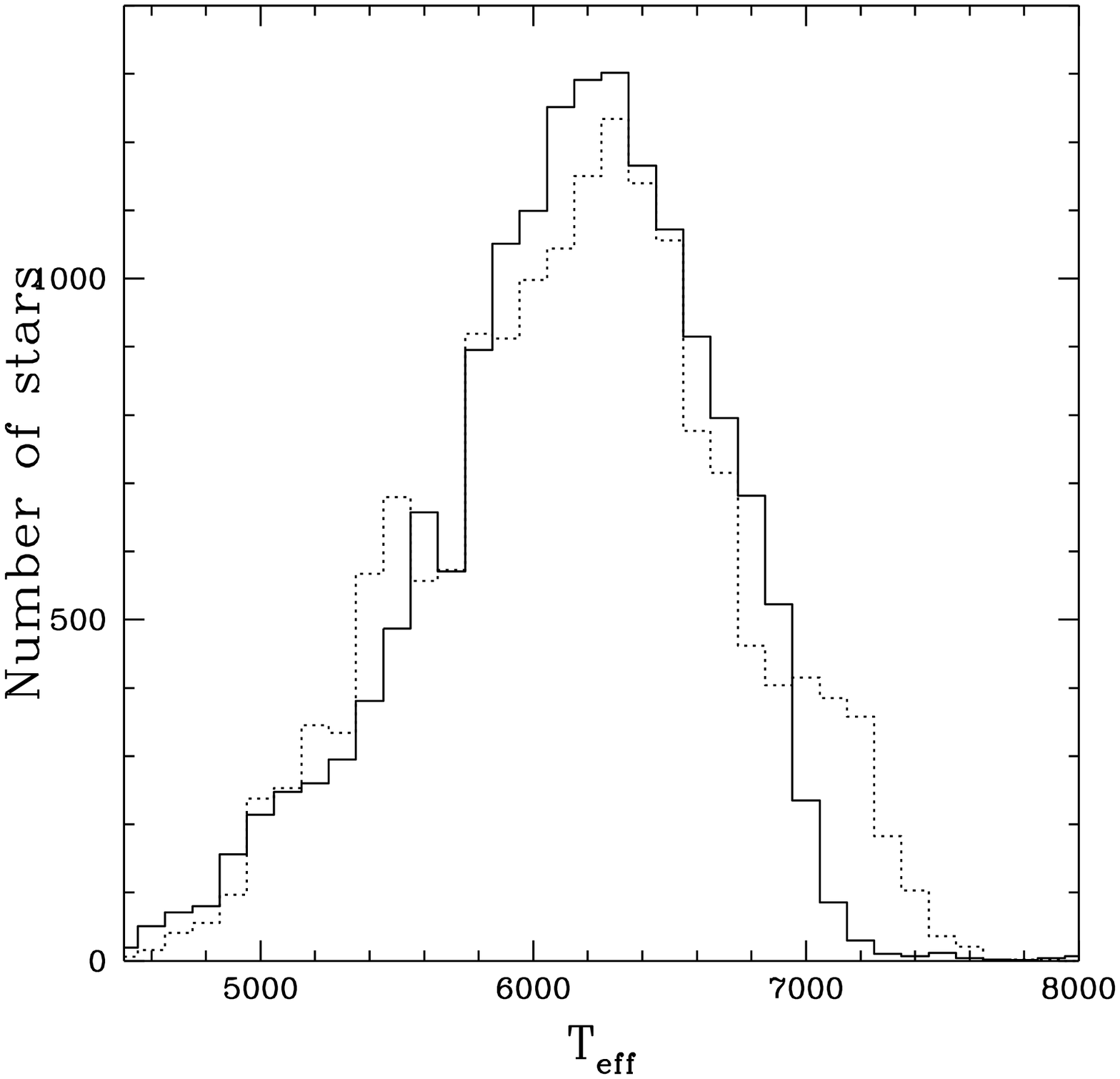}}
\caption{{\it Left:} Metallicity distribution function for all stars using 
the new calibration (solid) as well as the original GCS values (dotted). 
{\it Right:} The distribution of $\rm T_{eff}$ with the new and old 
calibrations (same symbols).}
\label{mehhist} 
\end{figure}

Fig. \ref{mehhist} compares the distributions of the GCS stars in [Fe/H] 
and $\rm T_{eff}$ as derived with the old and new calibrations. As seen, the 
new calibration leads to even fewer metal-poor stars in the sample than the 
old one, but the difference is marginal.

Galactic evolution models typically predict average properties of the stellar 
population in a vertical column through the disk at the position of the Sun, 
while observations typically -- as in the GCS -- refer to a roughly 
spherical volume centred on the Sun. Knowing the motions of the stars 
enables us to allow for this difference.

\begin{figure}[htbp]
\resizebox{\hsize}{!}{\includegraphics[angle=0]{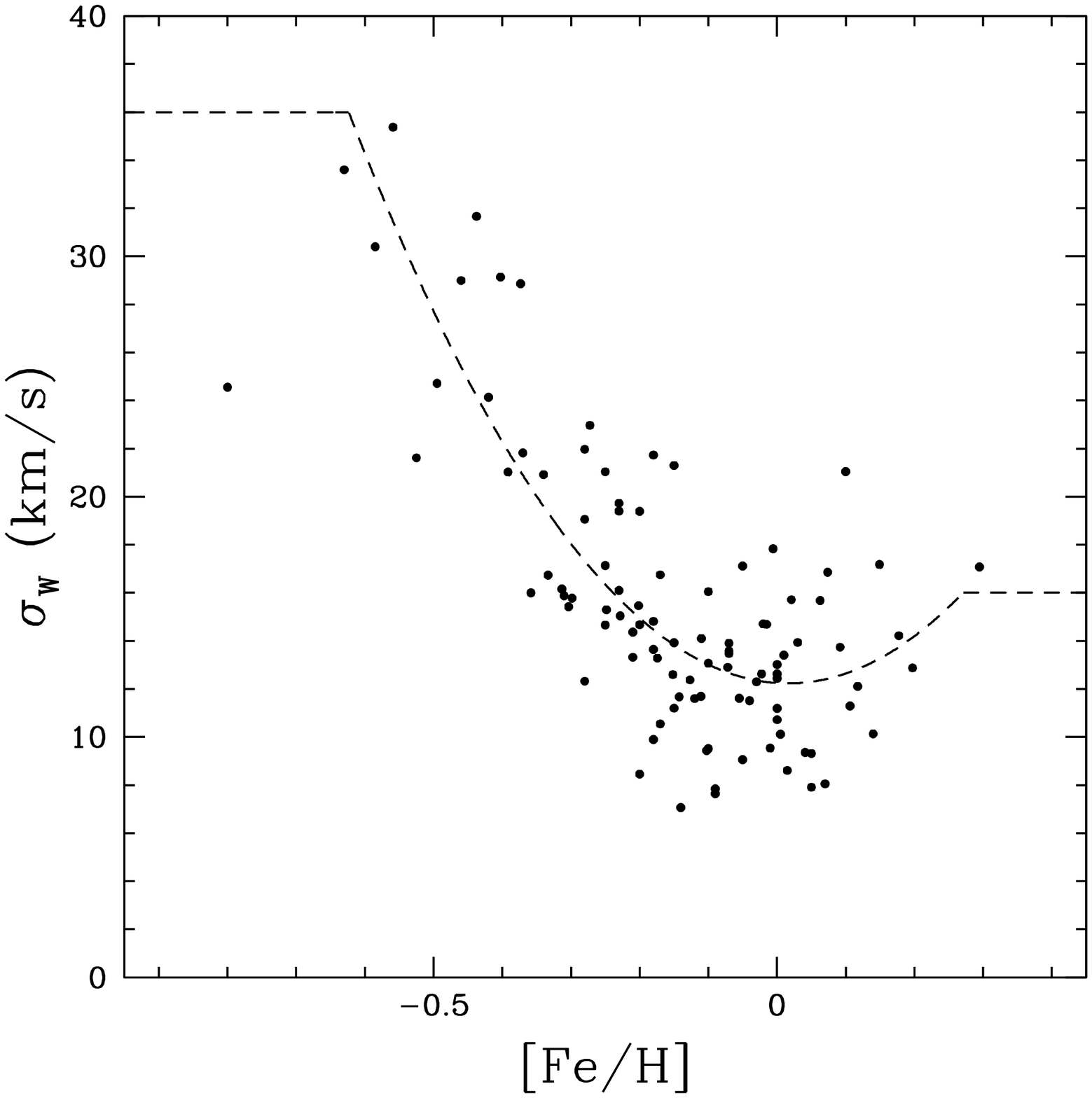}
                       \includegraphics[angle=0]{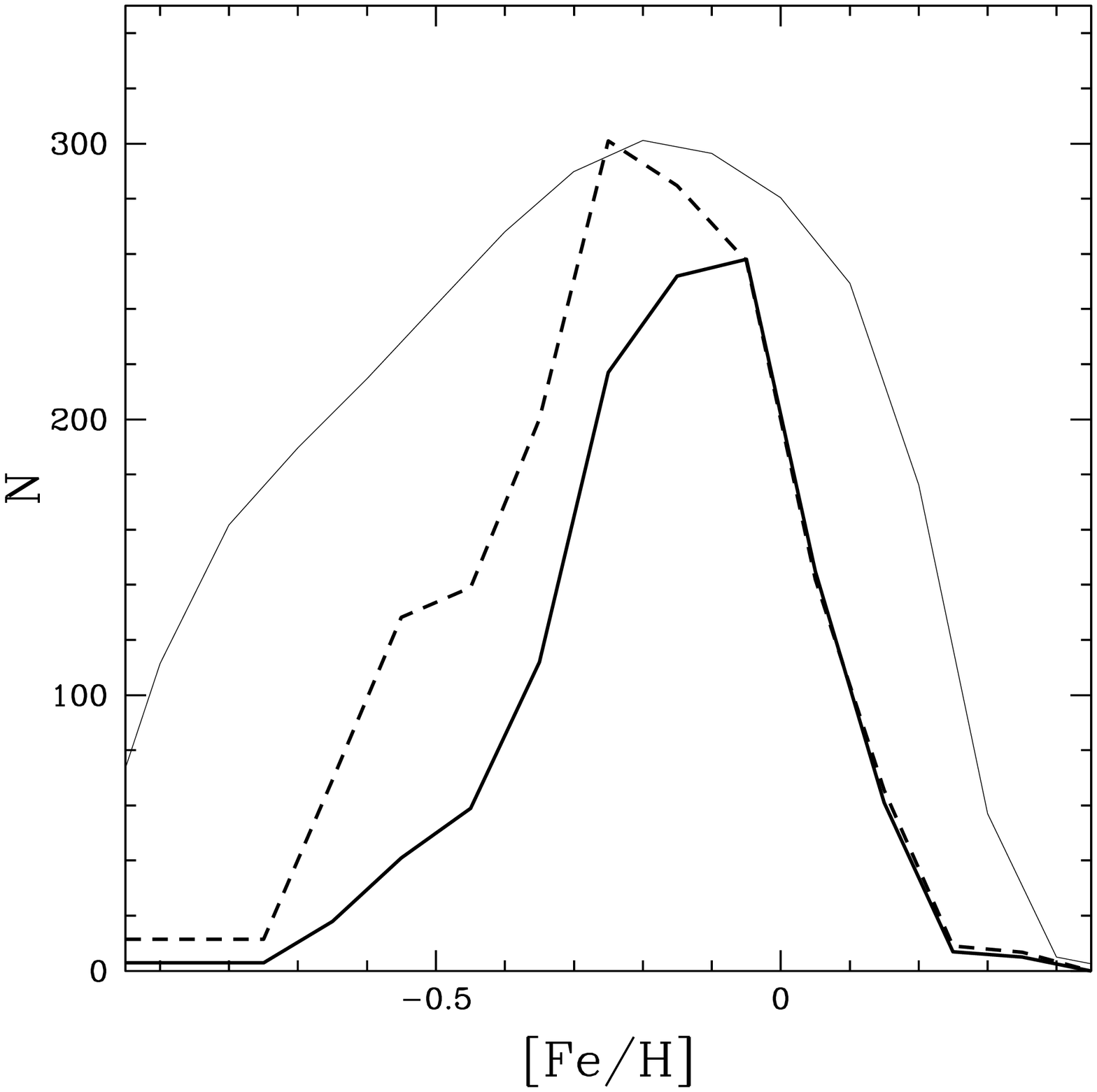}}
\caption{{\it Left:} The $W$ velocity dispersion as a function of [Fe/H] for 
the volume-complete sample (equal-size bins). {\it Right:} Metallicity 
distribution function for stars in the volume-complete sample (solid); 
in the column, using the fit in the left-hand panel (dashed); and in the 
closed-box model of Casuso \& Beckman (\cite{casuso04}, thin line).}
\label{cb}
\end{figure}

We have performed the correction from a volume-complete to a column-complete 
sample in a rigorous manner, using the self-consistent mass models of the 
disk derived by Holmberg \& Flynn (\cite{holmberg00}, \cite{holmberg04}; 
latest version in Flynn et al. \cite{holmberg06}). With this model, we derive 
the relation between volume and column density for each velocity dispersion.  
Fig.~\ref{cb} shows the vertical velocity dispersion of the GCS stars as a 
function of metallicity, together with a smooth polynomial fit. This fit is 
then used to transform the volume densities to corresponding column densities 
as functions of [Fe/H]. 

Fig.~\ref{cb} compares the volume and column metallicity distribution 
functions with each other and with a closed-box model without instantaneous 
recycling approximation (Casuso \& Beckman \cite{casuso04}), convolved with 
a Gaussian of $\sigma$= 0.08 to account for the observational error in
[Fe/H]. Clearly, neither the calibration nor the volume/column correction 
can account for the observed dramatic deficiency of metal-poor stars 
relative to the model.

We cannot comment on the reasons why H06 reached a different conclusion from 
the same data (albeit with a different metallicity calibration), since the 
closed-box model discussed by H06 was not described there. 

\begin{figure}[htbp] 
\resizebox{\hsize}{!}{\includegraphics[angle=-90]{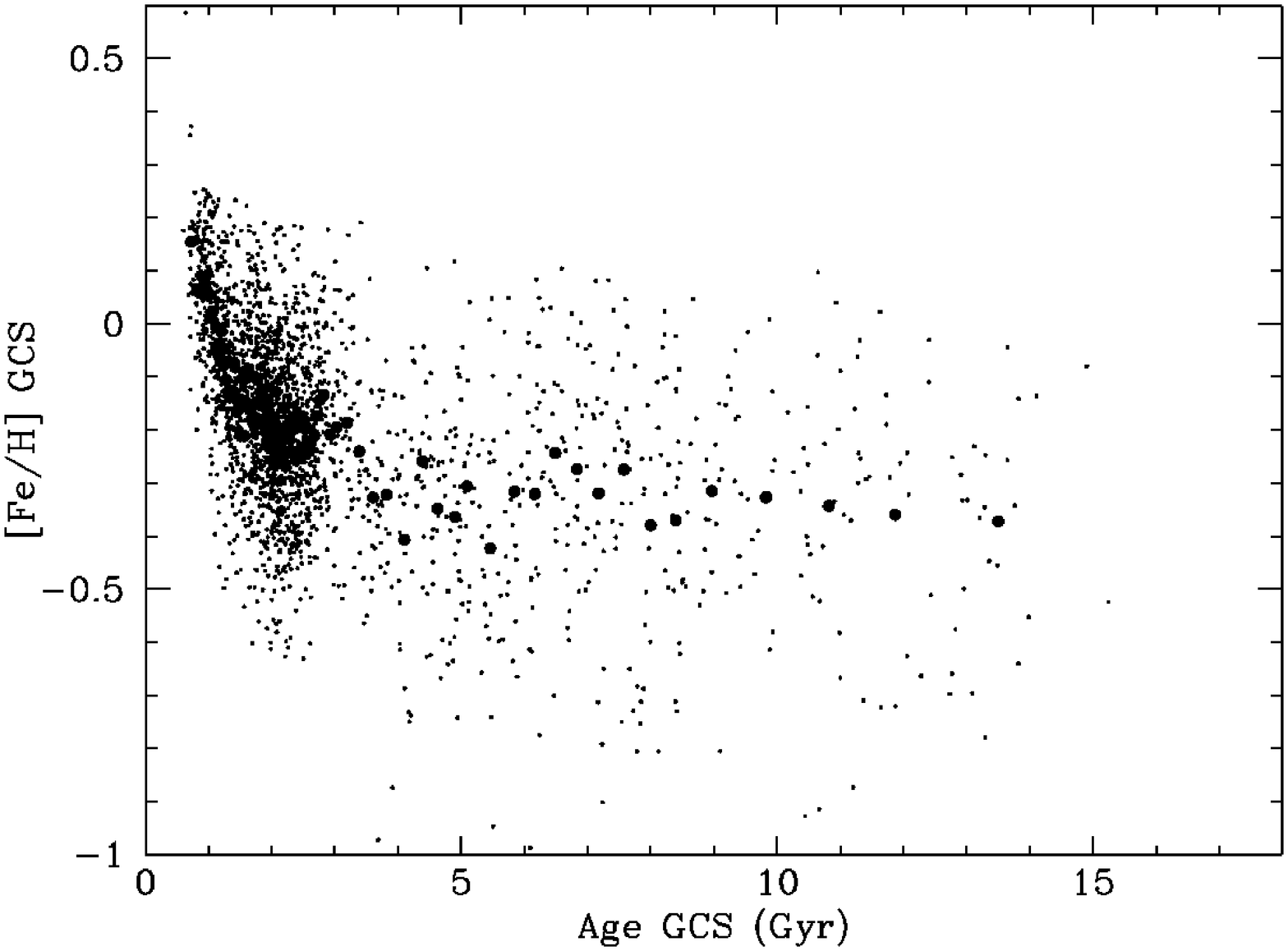}}
\resizebox{\hsize}{!}{\includegraphics[angle=-90]{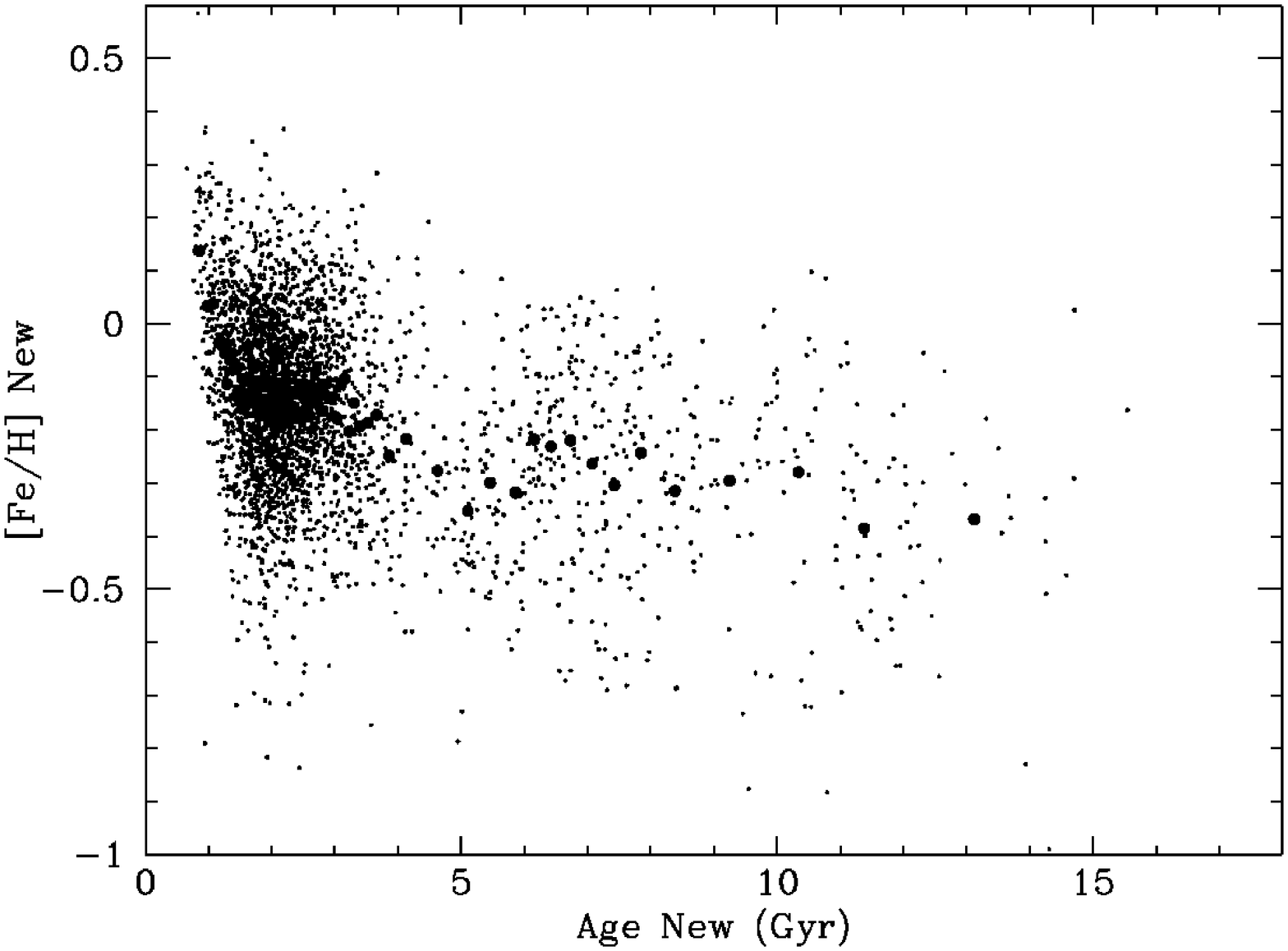}}
\caption{ {\it Top:} AMR from the original GCS, for single stars with ages 
better than 25\% and with {\it E(b-y)} $<$ 0.02 or $d<$ 40 pc. Large dots are
mean values in bins with equal numbers of stars. {\it Bottom:} Same, using 
the improved stellar parameters.} 
\label{agecomp2} 
\end{figure}

\begin{figure}[htbp] 
\resizebox{\hsize}{!}{\includegraphics[angle=-90]{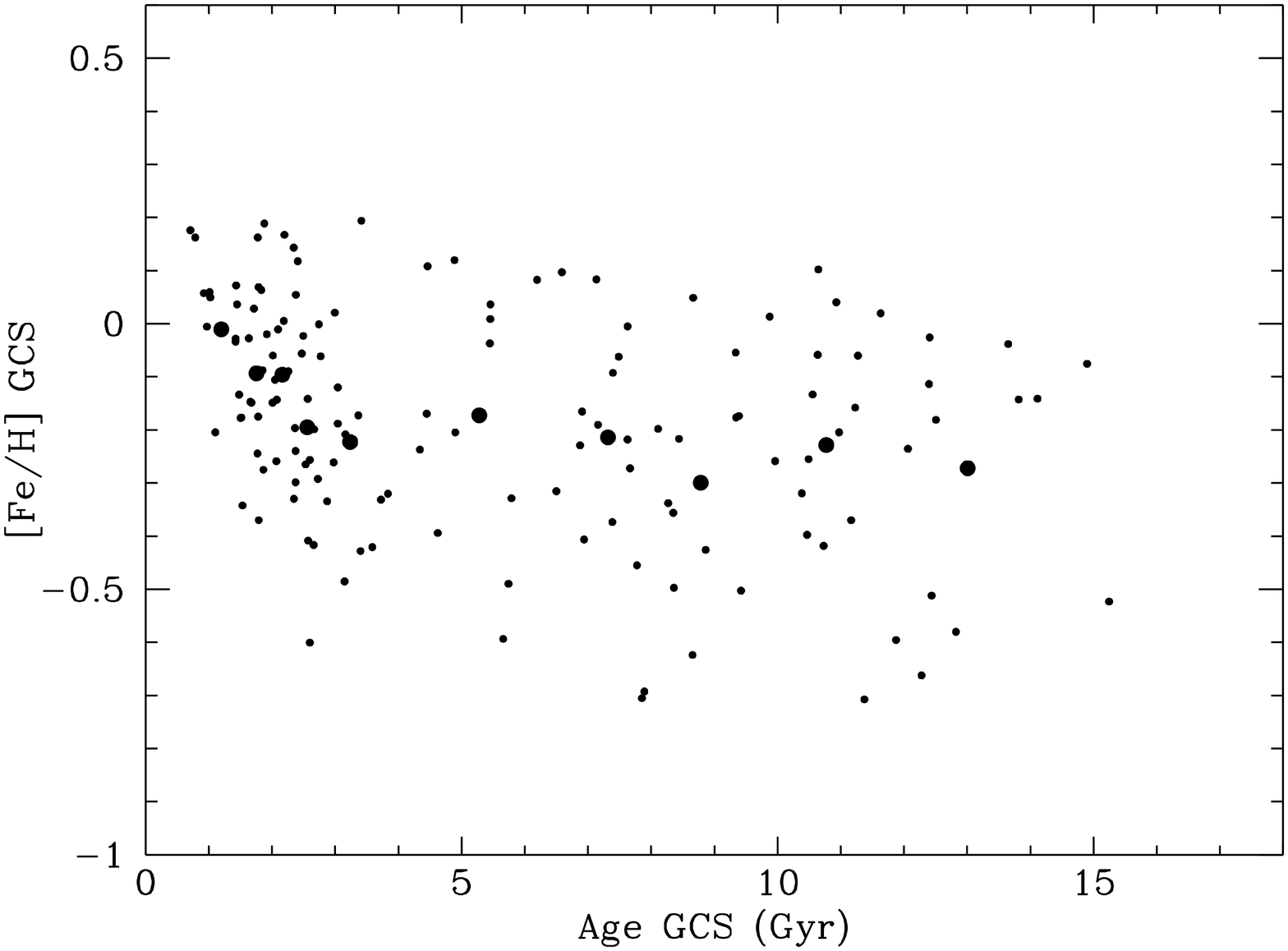}}
\resizebox{\hsize}{!}{\includegraphics[angle=-90]{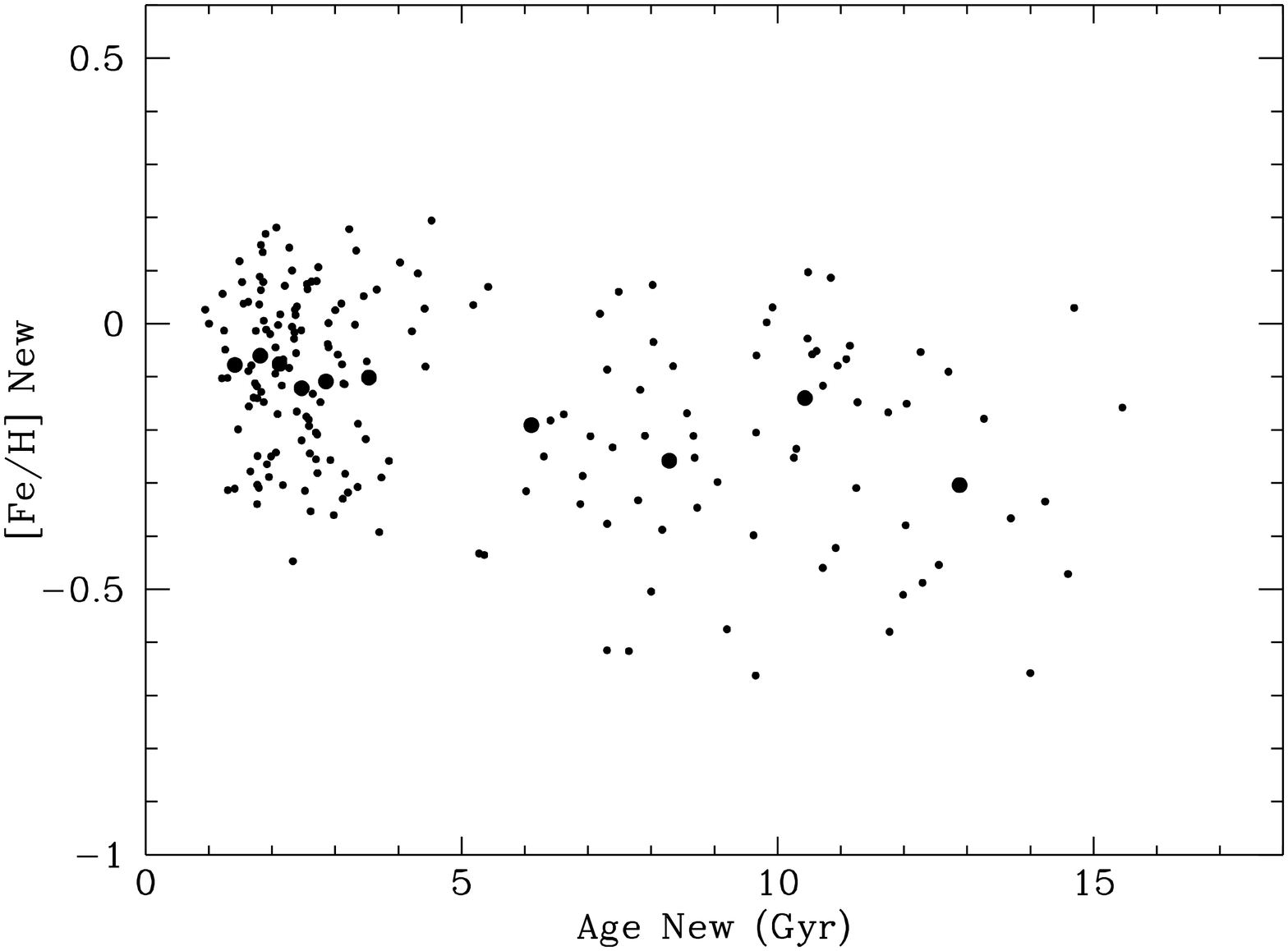}}
\caption{Same as Fig. \ref{agecomp2}, but only for the volume-limited 
sample with $d<$ 40 pc.} 
\label{agecomp3} 
\end{figure}

\section{The age-metallicity diagram}\label{AMRdiagram}

The relationship between average age and metallicity for stars in the solar 
neighbourhood -- the so-called age-metallicity relation (AMR) -- is probably 
the most popular diagnostic diagram for comparing galactic evolution models 
with the real Milky May. There is, however, no consensus on its shape or 
interpretation.
 
The discussion centres essentially on {\it (i)} the presence or absence 
of a general slope of the AMR (a gradual increase in mean [Fe/H] with time), 
and {\it (ii)} how much of the scatter in [Fe/H] at any given age is 
accounted for by observational errors and how much reflects the complexity 
in the evolution of a real galaxy as compared to most current models. 
It is thus crucial to ascertain whether the distribution of stars 
in the age-metallicity diagram (AMD) reflects primarily the intrinsic 
properties of the sample or primarily artefacts of the measurement and 
parameter calculation procedures. 

This can be investigated by means of our simulated sample.
First, we check whether the new calibrations have by themselves caused 
any significant change in the AMD, using the stars with the very best 
ages ($\sigma(Age) <$25\%). In order to eliminate 
the bright, distant early F stars in the GCS whose high [Fe/H] we 
suspected to be due to overestimated reddening corrections, we have limited 
the sample to single stars within 40 pc or with {\it E(b-y)} $<$ 0.02 mag. 
As seen, the revised calibrations cause hardly any difference in the AMD, 
neither for the full sample (Fig. \ref{agecomp2}) nor for the strictly 
volume-limited one (Fig. \ref{agecomp3}). 

Perhaps the most obvious structure in the observed AMD is the marked 
increase in mean metallicity and the absence of metal-poor stars for ages 
below 2 Gyr (see Fig. \ref{agecomp2} and GCS Figs. 27-28). It is important 
to understand if this is due, at least in part, to the selection criteria 
used to define the catalogue sample, or whether it is a genuine property 
of the solar neighbourhood. In the latter case, it could be interpreted 
as a cut-off in the resupply of fresh low-metallicity gas, followed by 
closed-box evolution. 

\begin{figure}[htbp] 
\resizebox{\hsize}{!}{\includegraphics[angle=0]{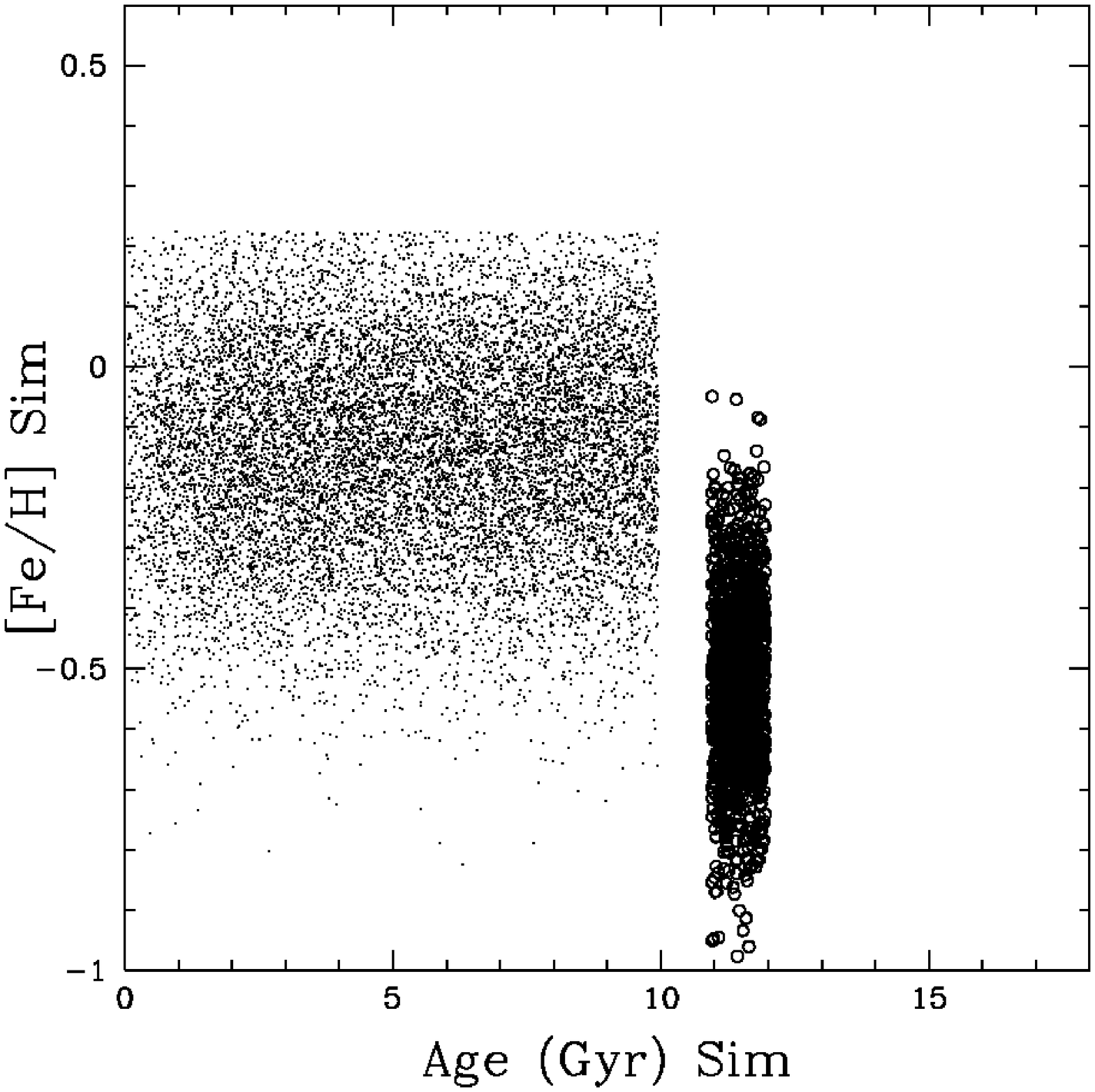}
                      \includegraphics[angle=0]{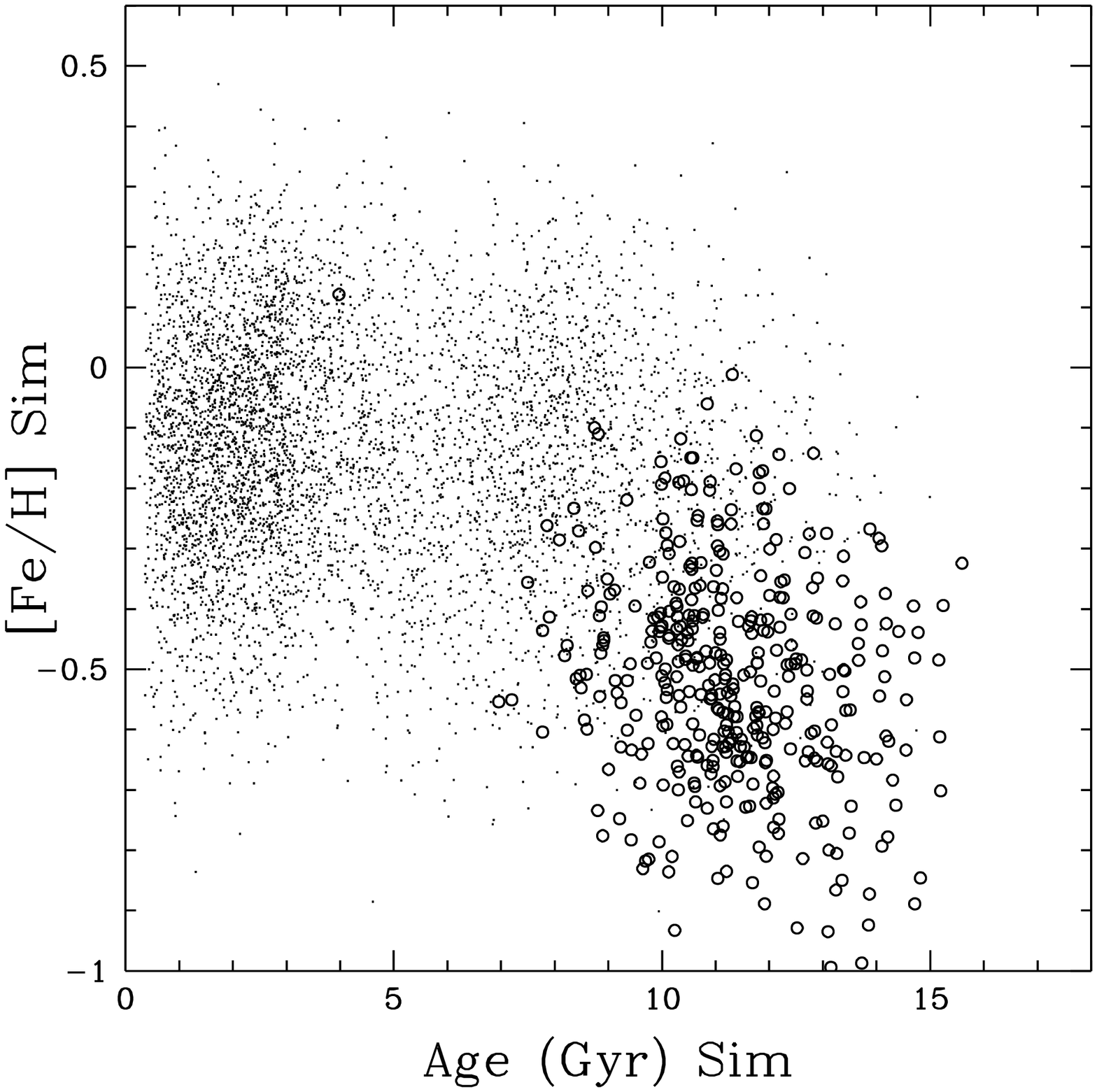}}
\resizebox{\hsize}{!}{\includegraphics[angle=0]{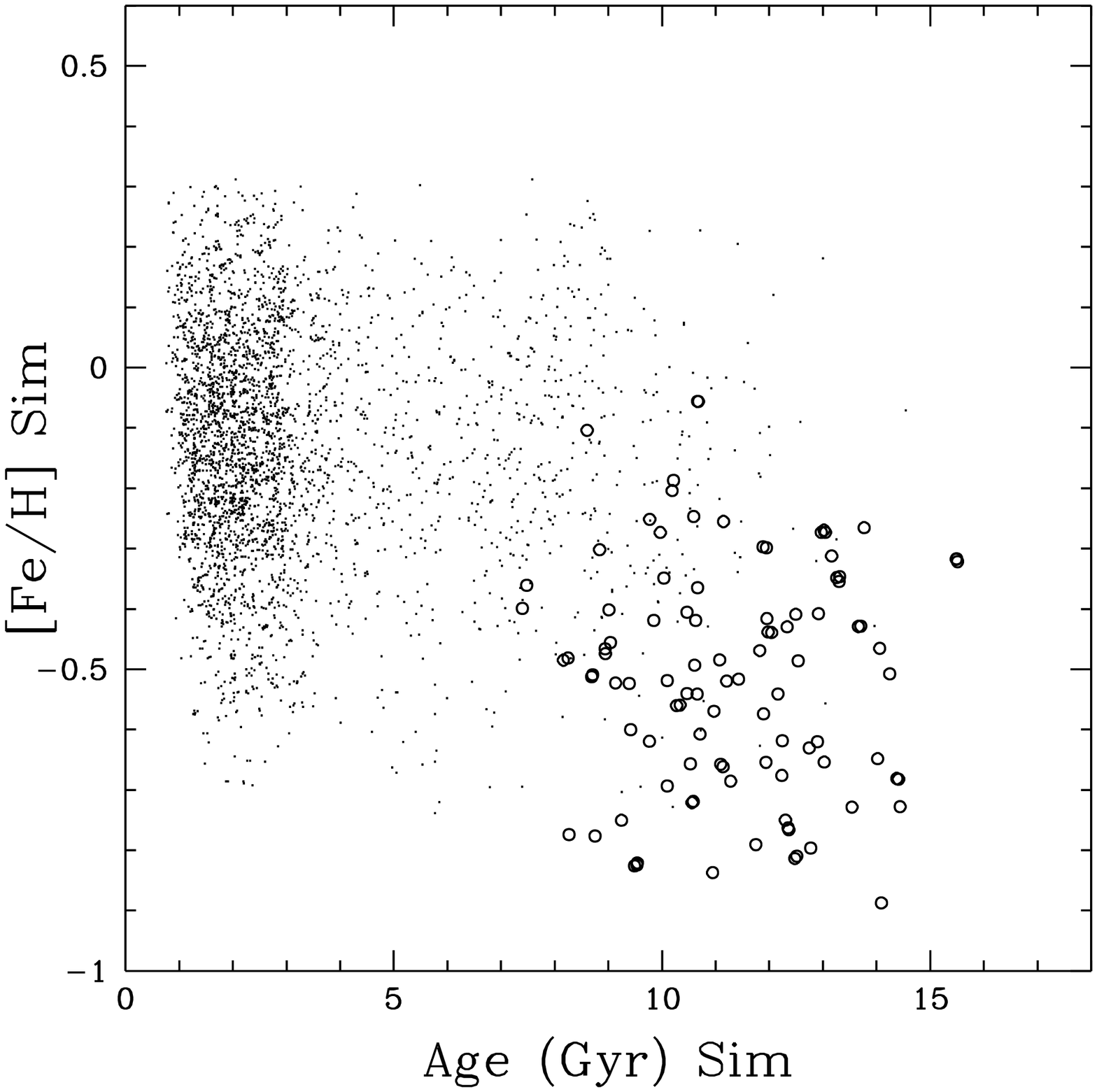}
                      \includegraphics[angle=0]{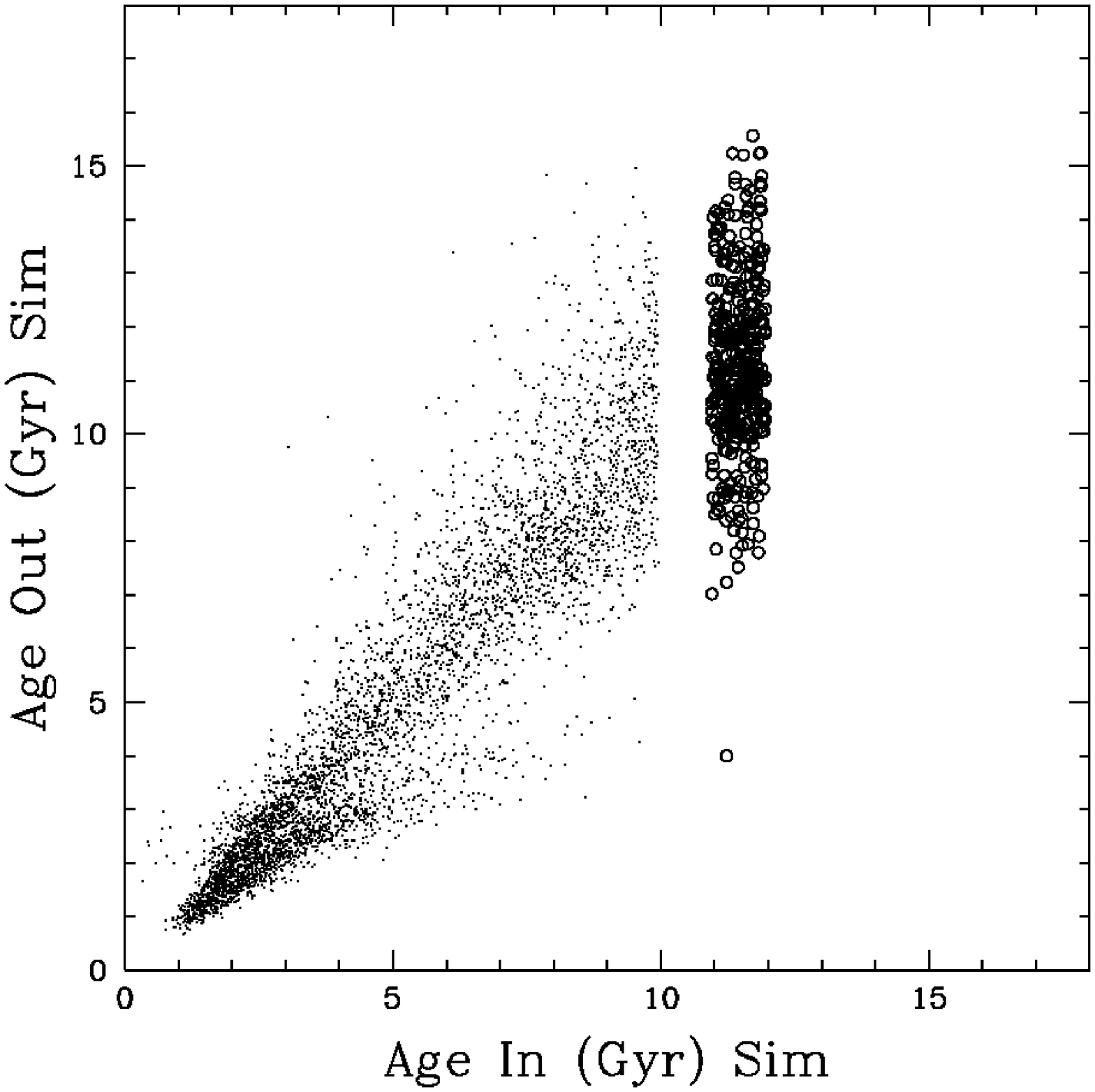}}
\caption{AMR simulations for the synthetic sample. {\it Top left:}  
Input AMR for a volume complete synthetic sample, flat within each of 
the thin and thick disks (dots and open circles, respectively).  
{\it Top right:} ``Observed'' AMD for the synthetic sample, using the 
new calibrations with the synthetic observations and retaining only 
stars with resulting ages better than 25\%. 
{\it Bottom left:} ``Observed'' AMD as above, but after imposing a blue 
colour cutoff ({\it b-y} $\geq$ 0.205) and apparent magnitude limit 
as in the GCS. 
{\it Bottom right:} Derived ages vs. the ``true'' input values.
}
\label{amrsim} 
\end{figure}

Fig.~\ref{amrsim} shows the results of computing the ``observed'' AMR 
from a simulated sample with a flat input AMR (first panel). 
The second panel shows the effect of recomputing the ages and metallicities 
from the synthetic data, using the new calibrations and restricting the sample 
to stars with ages better than 25\%. This illustrates the varying difficulty 
of determining precise ages over the HR diagram, especially in the region 
where isochrones overlap at 4-5 Gyrs. 

The third panel shows the AMR after applying the same blue colour cutoff 
({\it b-y} $\geq$ 0.205) and apparent magnitude limit as used to define the 
GCS sample. It exhibits the same preponderance of young metal-rich stars 
as the observed diagrams (Figs. \ref{agecomp2} - \ref{agecomp3}). Thus, 
the apparent lack of young metal-poor stars in the AMD is caused simply 
by the blue colour cutoff of the GCS. None of these selection effects 
seems to have been considered by H06 or by Reid et al. (\cite{reid07}).

In order to verify whether our calibrations or age computations could 
introduce (or remove) trends at higher ages in the AMD, we imposed tight 
AMRs (width 0.1 dex at all ages) on a subset of the simulated catalogue. 
In order to test all combinations of age and metallicity, we imposed both 
a linear {\it in}crease as well as a linear {\it de}crease of [Fe/H] with 
time. We then computed synthetic colours for the stars with random errors 
as observed, and processed these synthetic stars with the new calibrations 
in the same manner as in the GCS, including the colour cutoffs, and 
retaining only stars with ages better than 25\%. 

\begin{figure}[htbp] 
\resizebox{\hsize}{!}{\includegraphics[angle=0]{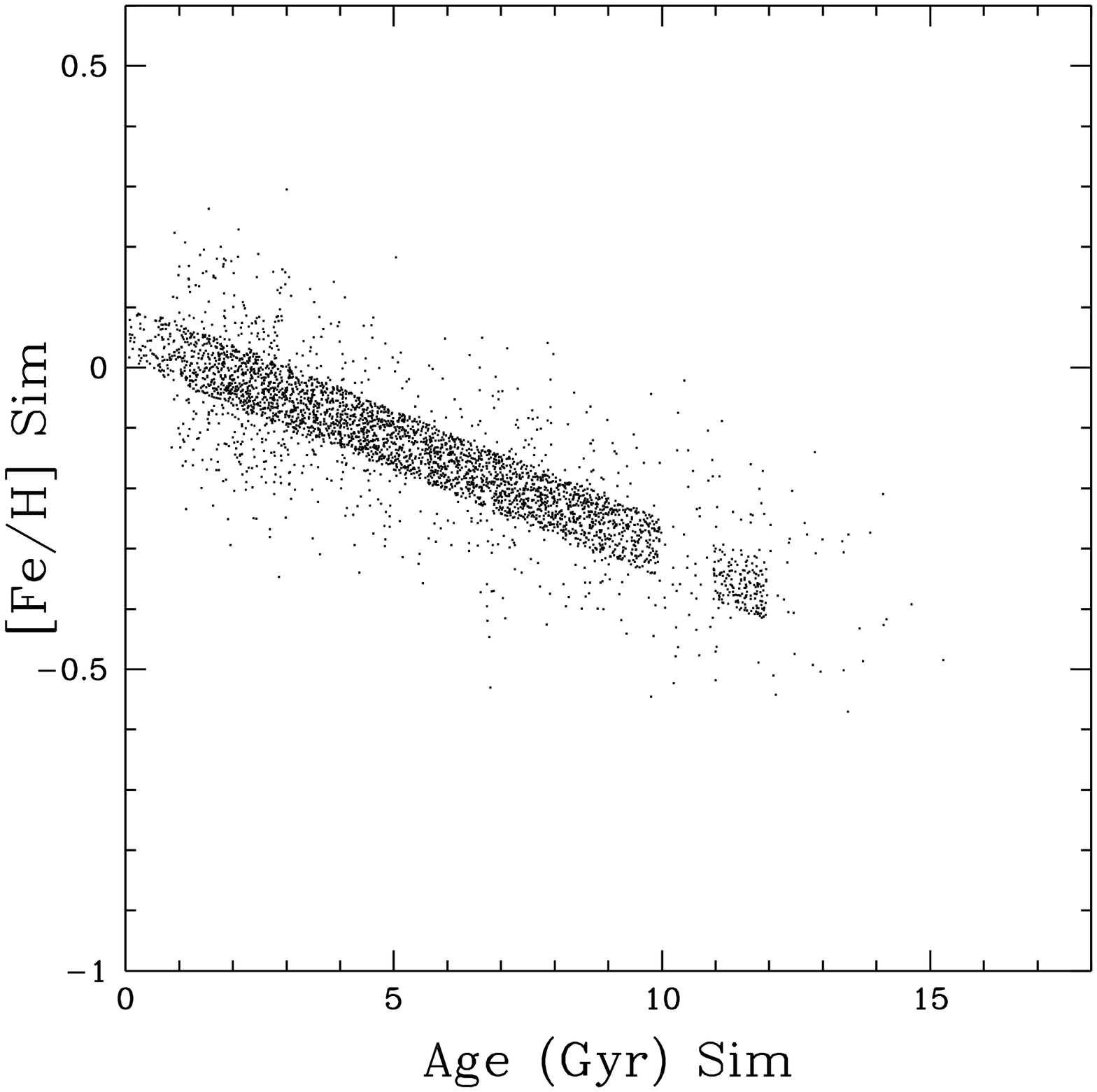}
                      \includegraphics[angle=0]{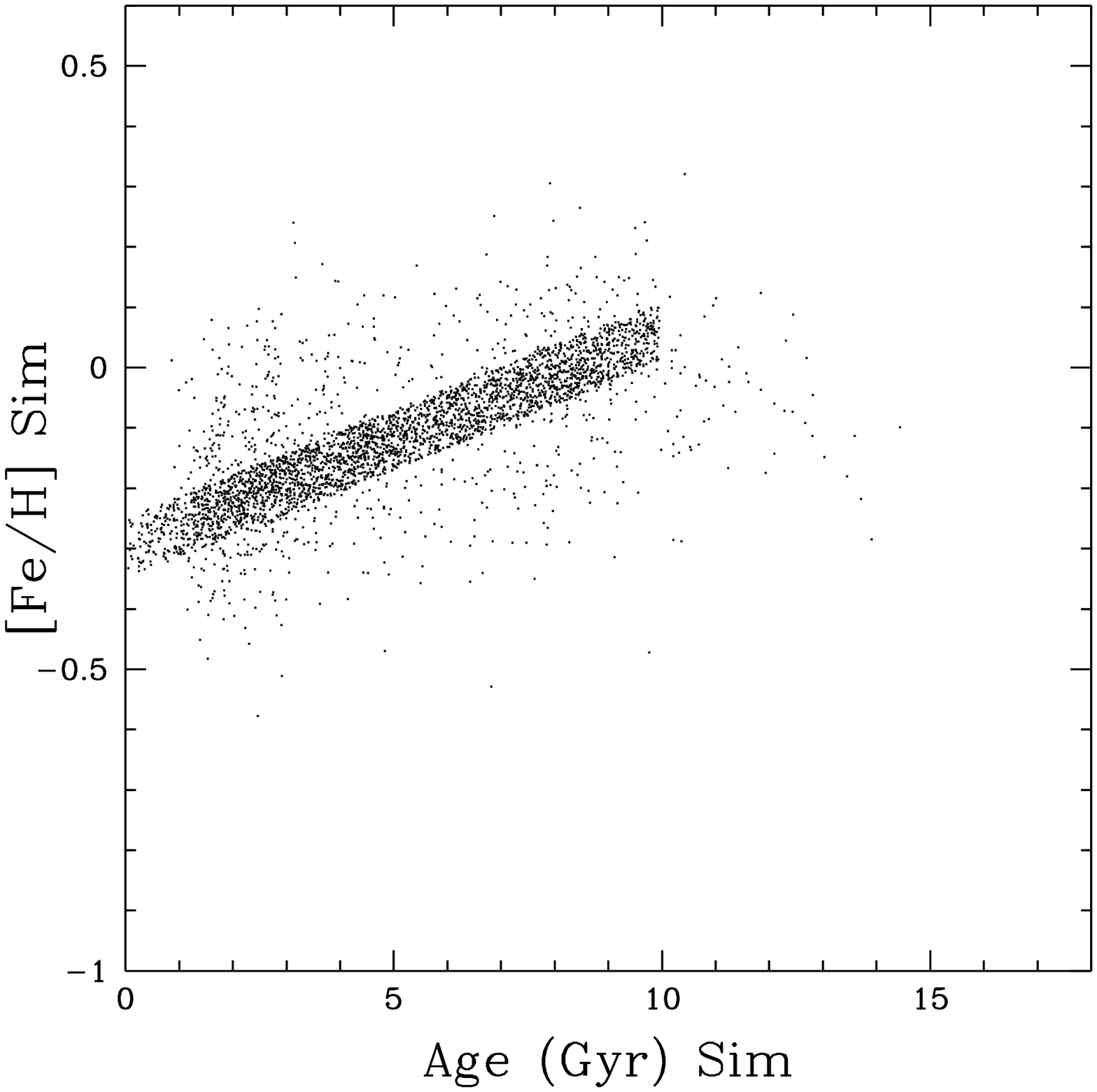}}
\caption{
Alternative input and ``observed'' AMRs for the synthetic catalogue 
(Fig. \ref{amrsim} showing a flat AMR). {\it Left:} Input AMR: A 0.10 dex 
wide band of slope -0.036 dex/Gyr. {\it Right:} Same, but with the opposite 
slope (+0.036 dex/Gyr).
}
\label{sim2} 
\end{figure}

Fig. \ref{sim2} compares the input and recovered AMD for both of these 
cases. As can be seen, the slope of the input AMR is faithfully 
reproduced in both cases, despite the inevitable scatter introduced by 
the observational errors. We thus conclude that the absence of a 
significant mean slope of the data points in Fig. \ref{agecomp3} reflects 
the true situation of the solar neighbourhood. 

Focusing on the stars older than 4 Gyr in Fig. \ref{agecomp3}, we find 
a mean [Fe/H]= -0.24 dex 
and standard deviation $\sigma$= 0.22 dex for the original GCS data 
(top panel, 70 stars), and a mean [Fe/H]=  -0.21 dex and 
$\sigma$= 0.21 dex using the new calibrations (bottom panel, 75 stars). 
Subtracting the estimated observational error in each case (0.10 and 
0.07 dex) yields an intrinsic (``cosmic'') scatter in [Fe/H] at a given 
age of 0.20 dex -- identical to the result by Edvardsson et al. 
(\cite{edv93}) from high-resolution spectroscopy. 

\begin{figure}[htbp] 
\resizebox{\hsize}{!}{\includegraphics[angle=0]{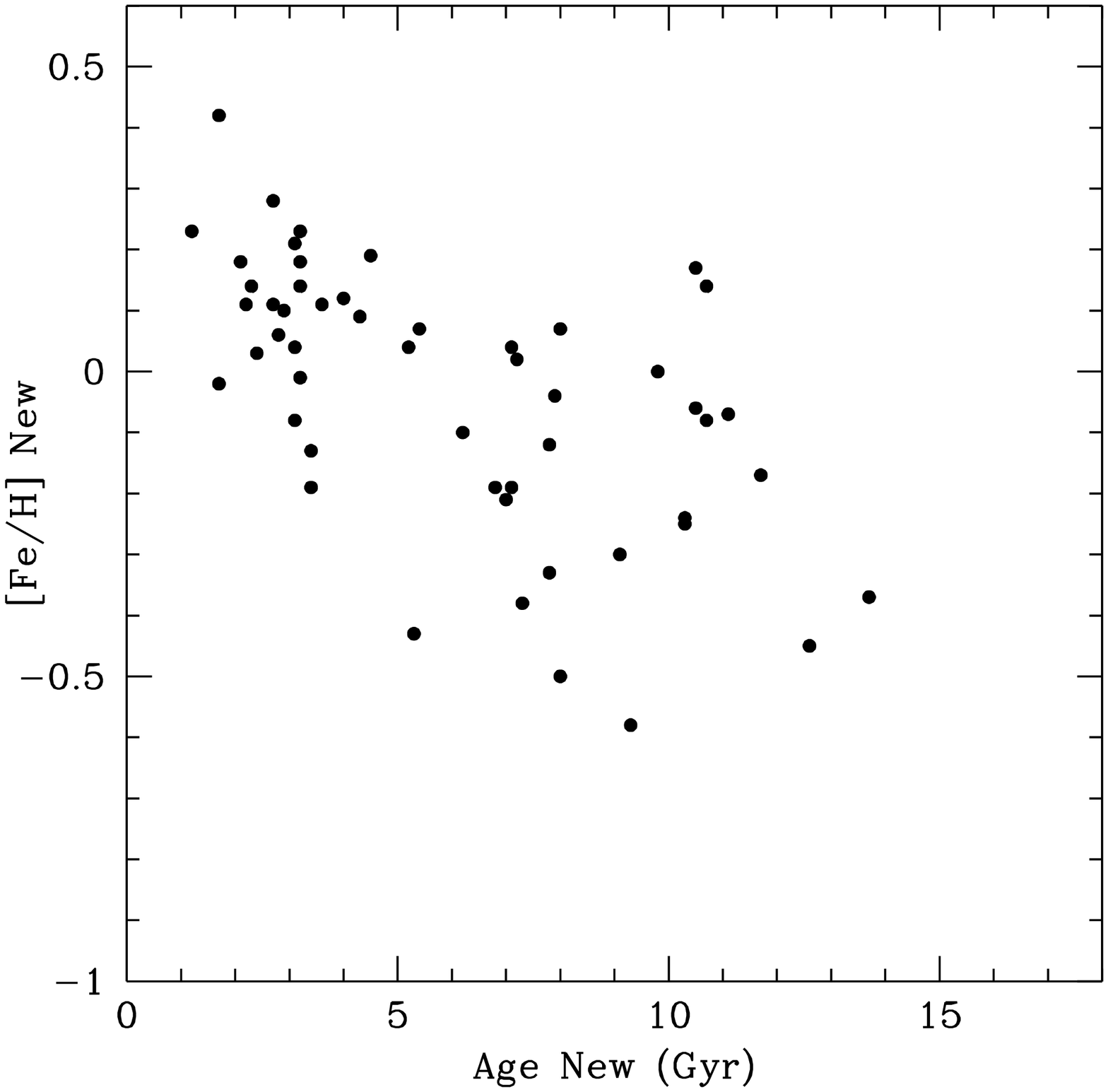}
                      \includegraphics[angle=0]{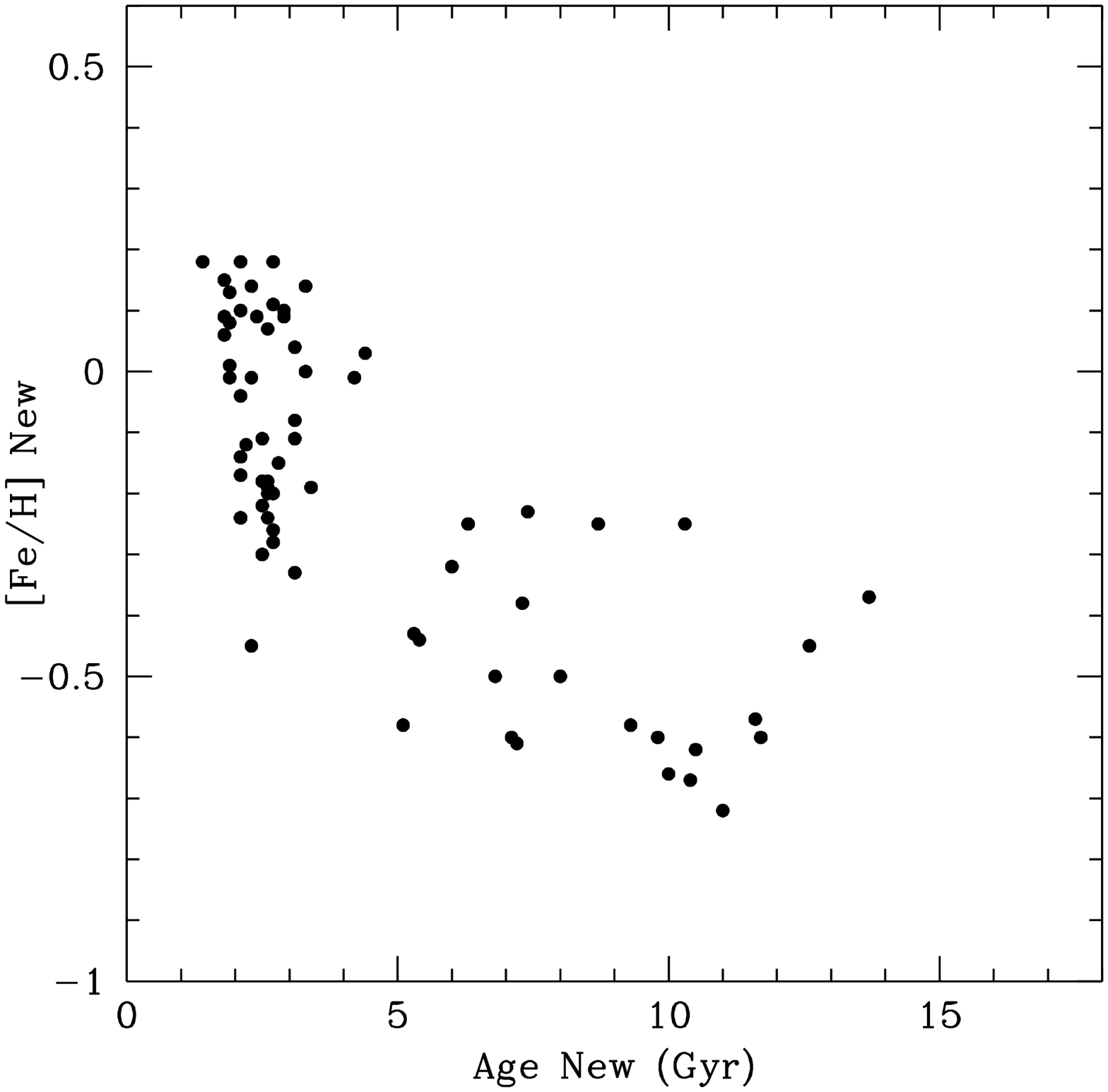}}
\resizebox{\hsize}{!}{\includegraphics[angle=0]{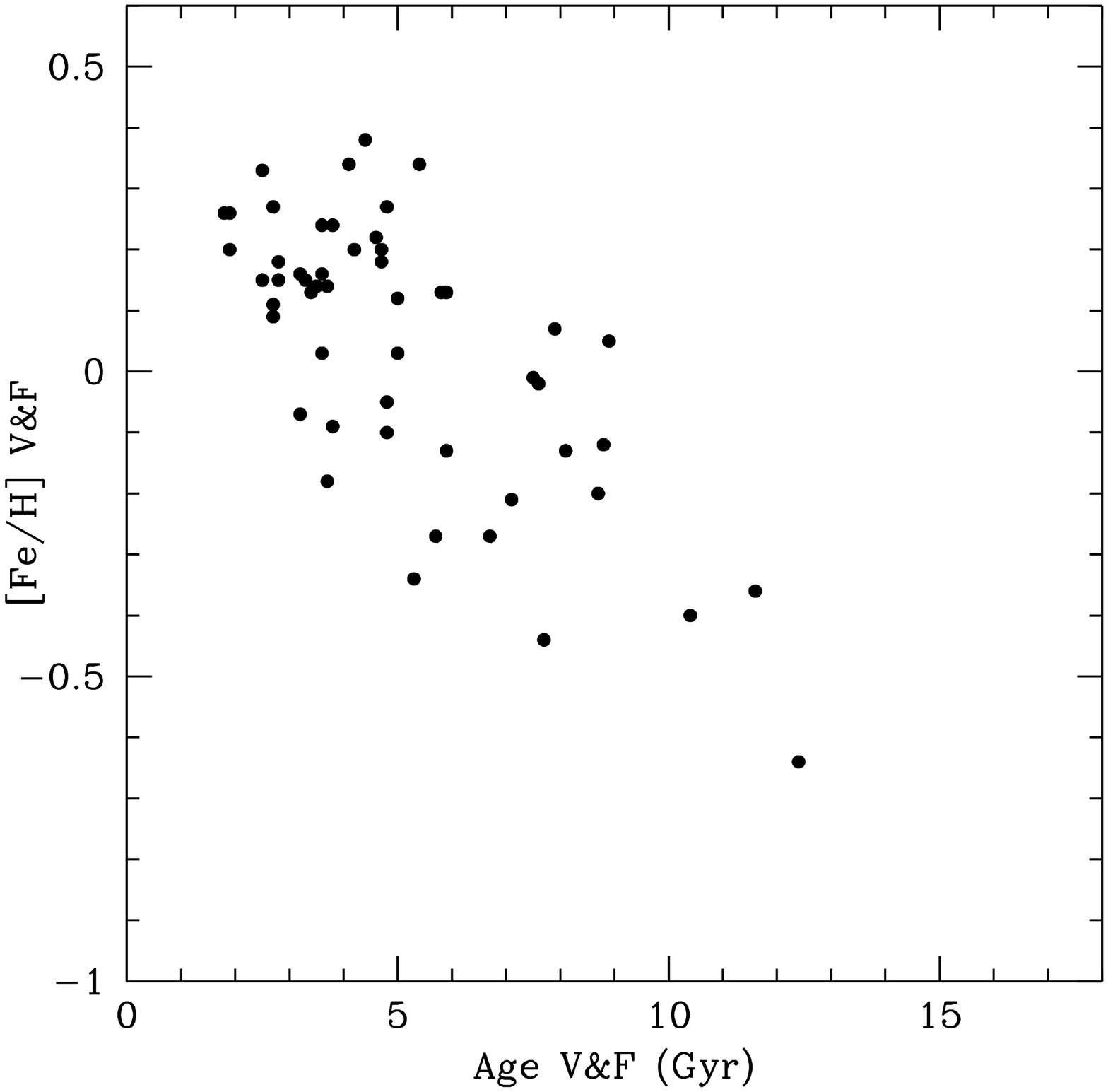}
                      \includegraphics[angle=0]{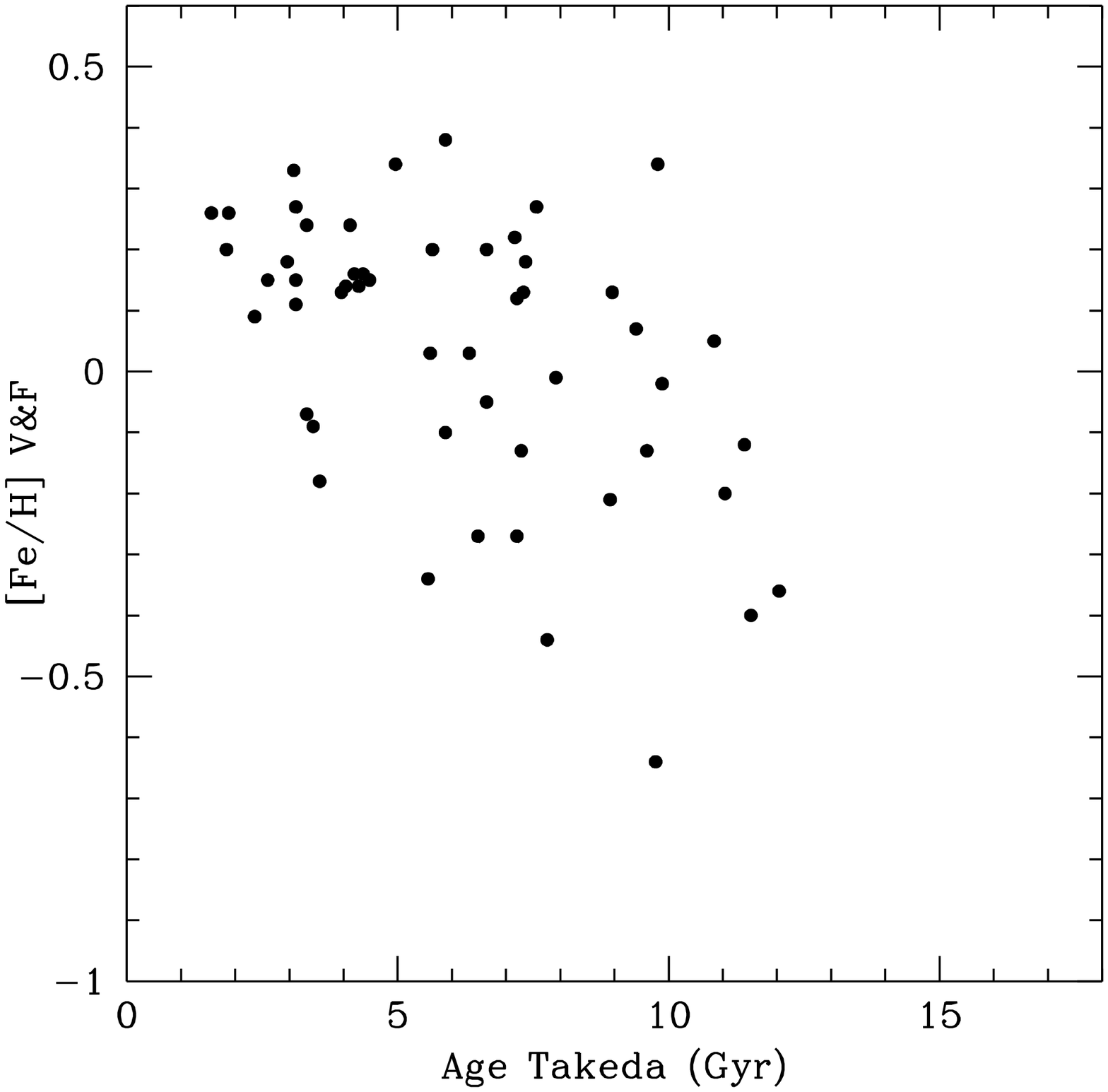}}
\caption{AMR for two pre-selected samples using different metallicity and
age determination methods. All ages have errors below 25\%. {\it Top left:}  
The VF05 sample, using our data. 
{\it Top right:}  The Edvardsson et al. sample, using our data. 
{\it Bottom left:} The VF05 sample, using their data. 
{\it Bottom right:} Same, using Takeda et al. ages.}
\label{amrvf} 
\end{figure}

We conclude the discussion by reiterating the importance of the sample 
selection on the resulting AMR. If certain types of stars are excluded {\it 
a priori}, the gross shape of the AMR may be predetermined whatever new 
metallicities or ages may be observed. Fig.~\ref{amrvf} illustrates this, 
using two popular sources, VF05 and Edvardsson et al. (\cite{edv93}). 
Both were selected for specific purposes, and both are prone to selection 
effects affecting the AMR. 

Edvardsson et al. selected {\it evolved} F-type dwarfs in order to be able 
to determine isochrone ages, excluding {\it a priori} any old, metal-rich 
and therefore redder (i.e. G) stars that might have populated the upper 
right-hand part of the AMR. VF05 studied stars used in planet searches and 
thusavoided stars with weak lines, i.e. hot metal-poor stars in the 
opposite corner of the AMR. These facts are clearly discussed in the 
papers, but make these and other inherently biased samples unsuitable for 
discussions of the general AMR of the solar neighbourhood.

\subsection{AMR slope vs. a radial metallicity gradient}

Recently, Rocha-Pinto et al. (\cite{rochapinto06}) interpreted an observed 
variation of mean metallicity with the difference between the mean 
orbital radius of the star (R$_{\rm m}$) and that of the Sun (R$_{\odot}$) 
as evidence for a tight age-metallicity relation in the thin disk. 
However, because their values of R$_{\rm m}$ range from 6 to 9 kpc, many of 
their stars must have large velocities and orbital eccentricities, usually 
associated with thick-disk stars. This was already noted by Edvardsson et al. 
(\cite{edv93}), although they did not use the explicit term ``thick disk''.

\begin{figure}[htbp] 
\resizebox{\hsize}{!}{\includegraphics[angle=-90]{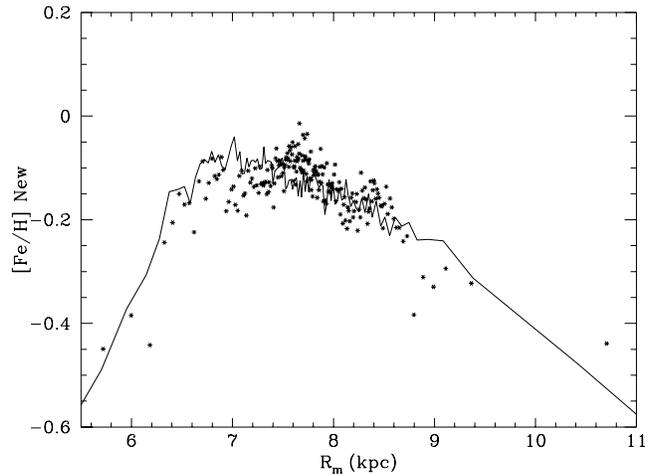}}
\caption{Radial variation in [Fe/H] for the single GCS stars (dots; new 
calibration), and for a simulated catalogue with the AMD of 
Fig.~\ref{amrsim} (line), i.e. with no correlation between age and metallicity 
in either the thin or the thick disk, but a radial metallicity gradient of 
-0.09 dex/kpc in the thin disk. We obtain a similar overall relation as 
Rocha-Pinto et al. (\cite{rochapinto06}), but without any AMR in either 
thin or thick disk.} 
\label{metrad} 
\end{figure}

A much more natural explanation of the apparent radial variation 
in metallicity is that the observed stellar sample is a mix of thin-disk 
stars dominating close to the solar circle, and thick-disk stars 
dominating at both high and especially low R$_{\rm m}$. To verify this, we 
have compared the observed radial metallicity distribution for single stars 
in the GCS (using the new calibrations) with that of our simulated catalogue, 
which has the flat age-metallicity distribution of Fig.~\ref{amrsim}, i.e. no 
correlation between age and metallicity in either the thin or the thick 
disk. However, we have added a radial metallicity gradient of -0.09dex/kpc 
for the thin-disk stars, close to the observed values (see, e.g., discussion 
in Sect. 6.2 of the GCS). Fig.~\ref{metrad} shows that this reproduces the 
observed radial metallicity variation without any variation of metallicity 
with age.

\begin{figure}[htbp] 
\resizebox{\hsize}{!}{\includegraphics[angle=0]{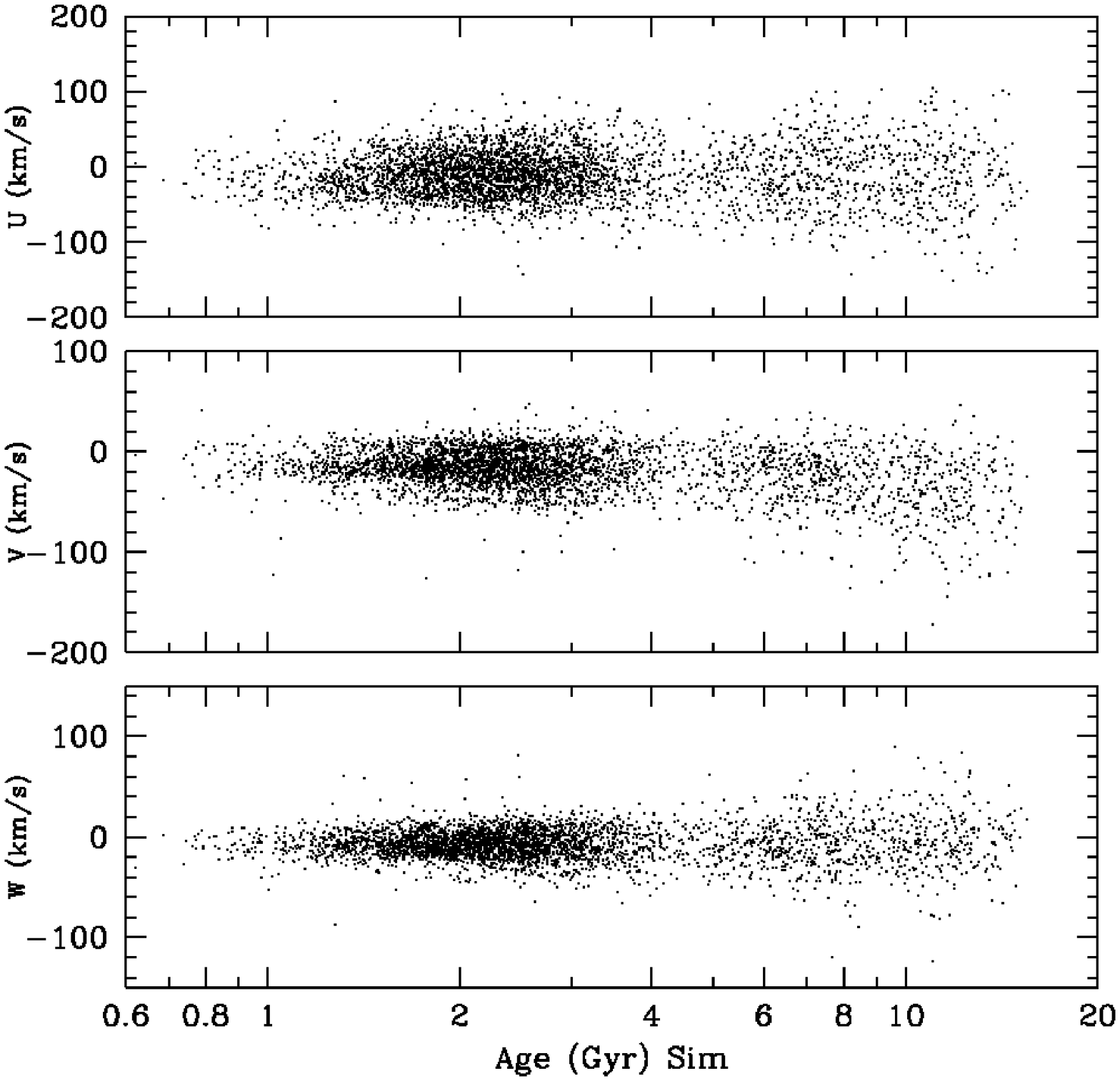}}
\caption{
U, V and W velocity vs. age for the 4065 single GCS stars 
with $\sigma(Age)<25$\%).} 
\label{avrnew}
\end{figure}

\begin{figure}[htbp] 
\resizebox{\hsize}{!}{\includegraphics[angle=0]{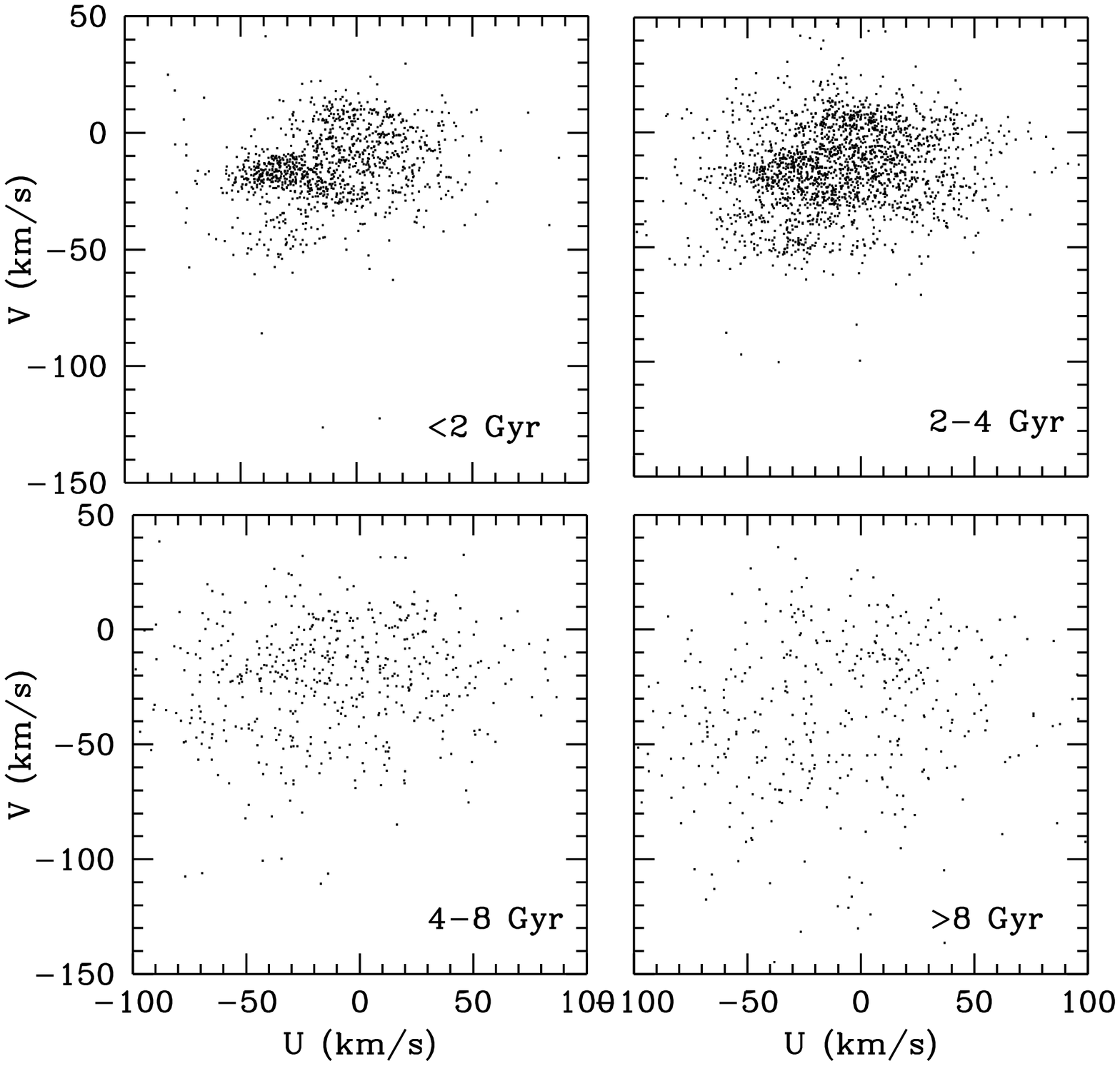}}
\caption{
U-V diagrams for the GCS subsample of Fig. \ref{avrnew}, separated 
into four age groups.} 
\label{uv}
\end{figure}

We have deliberately made no attempt to fit the detailed structure in the 
R$_{\rm m}$ - [Fe/H] distribution. This depends on the complex structure of
the distribution of the U and V velocities (see Figs. \ref{avrnew} - \ref{uv} 
and GCS Fig. 20), which exhibits a multitude of kinematic groups not accounted 
for in a simple diffusion picture. As an example, the peak in [Fe/H] at 
R$_{\rm m}\simeq$ 6.8 kpc is associated with the $\zeta$ Herculis stream 
discussed in the GCS and again by Famaey et al. (\cite{famaey05}) and Bensby 
et al. (\cite{bensby07}). 

\begin{figure}[htbp] 
\resizebox{\hsize}{!}{\includegraphics[angle=0]{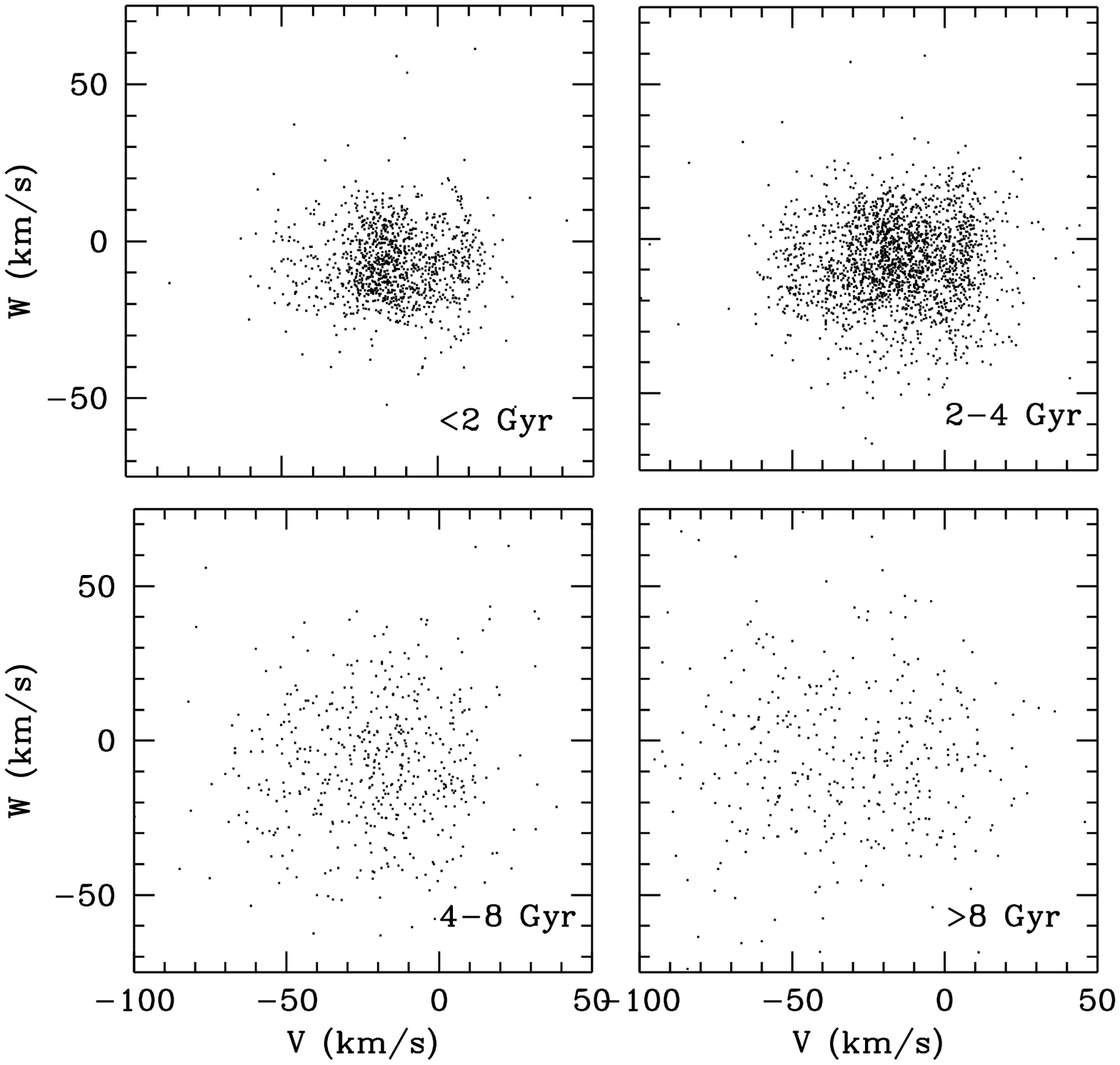}}
\caption{
V-W diagrams for the GCS subsample of Fig. \ref{avrnew}, separated 
into four age groups.} 
\label{vw}
\end{figure}

\begin{figure}[htbp] 
\resizebox{\hsize}{!}{\includegraphics[angle=0]{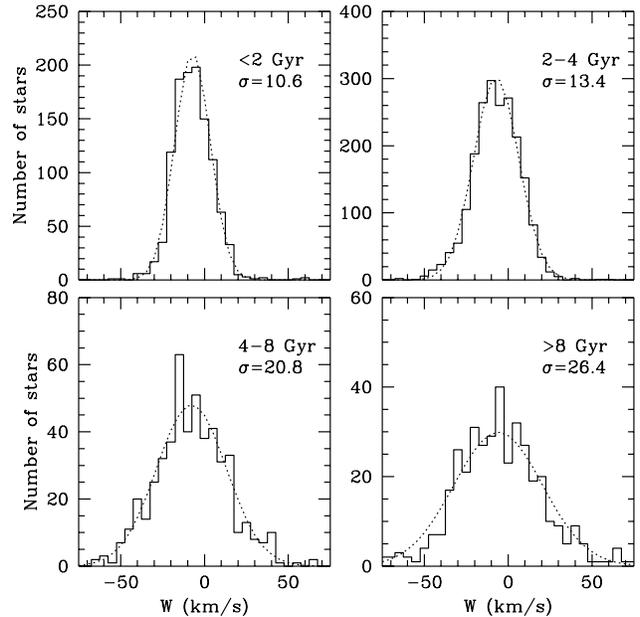}}
\caption{
W velocity distributions for the GCS subsample of Fig. \ref{avrnew}, 
separated into four age groups. No significant substructure is seen. }
\label{whist}
\end{figure}

\section{Age-velocity relation and disk heating}\label{diskheating}
 
As demonstrated earlier in this paper, our new 
calibrations cause no significant systematic changes in the velocities 
and ages of the GCS stars (apart from the overall $\sim$10\% age scale 
reduction). Thus, no major revision of the GCS results on the evolution 
of the kinematics of the disk with age is to be expected. However, the 
size of our sample allows us to examine that evolution in greater detail 
than was done in the GCS. 

For this task, we select the cleanest sample 
of GCS stars, i.e. the single stars with $\sigma(Age)<25$\% and with 
complete space velocity data. Note that, due to our improved age 
determination, there are now 4065 stars in this class as compared to 
2852 in the original GCS -- an increase by almost 50\%.

\begin{figure*}[bhtp] 
\resizebox{\hsize}{!}{\includegraphics[angle=0]{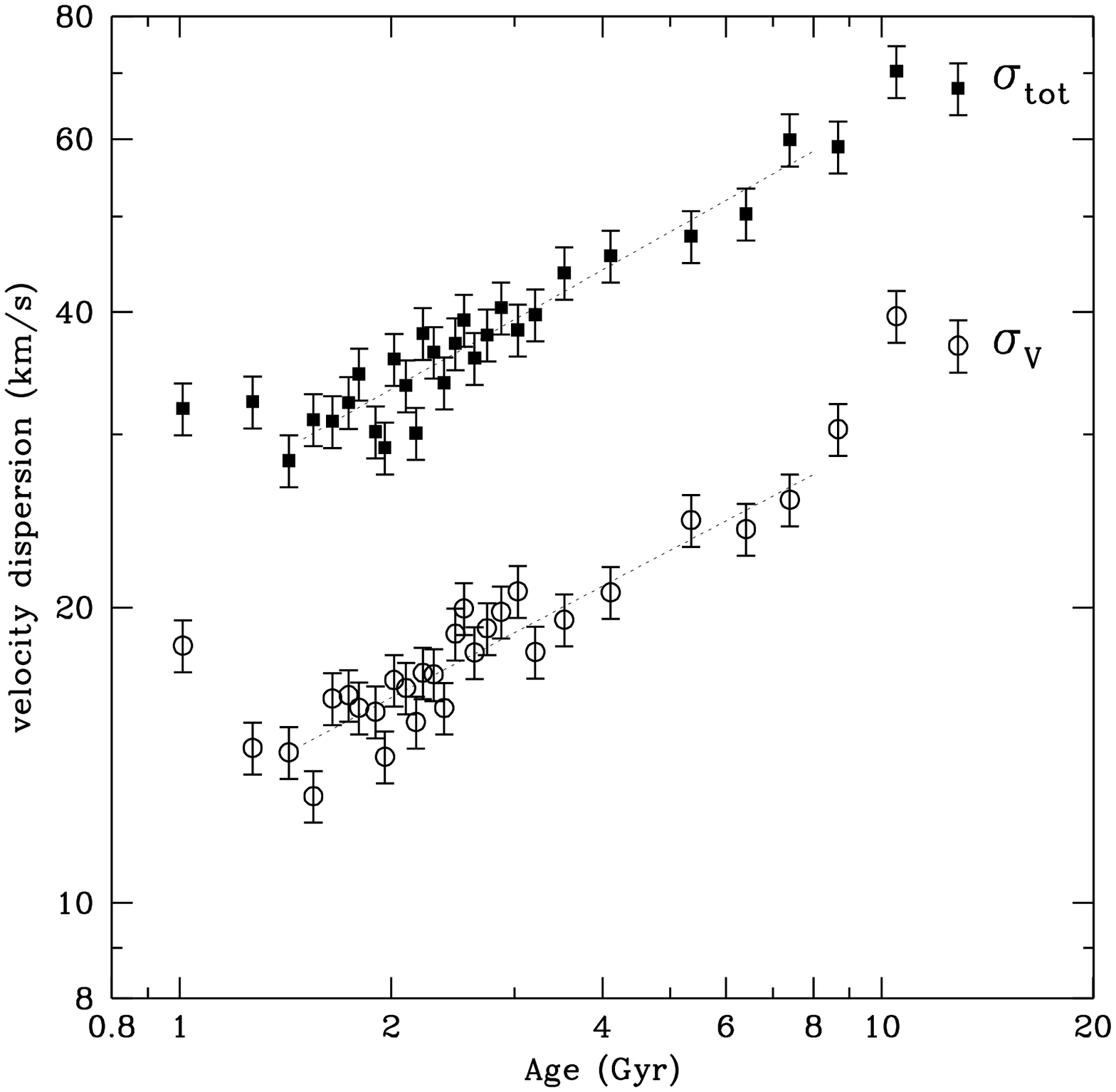}
                      \includegraphics[angle=0]{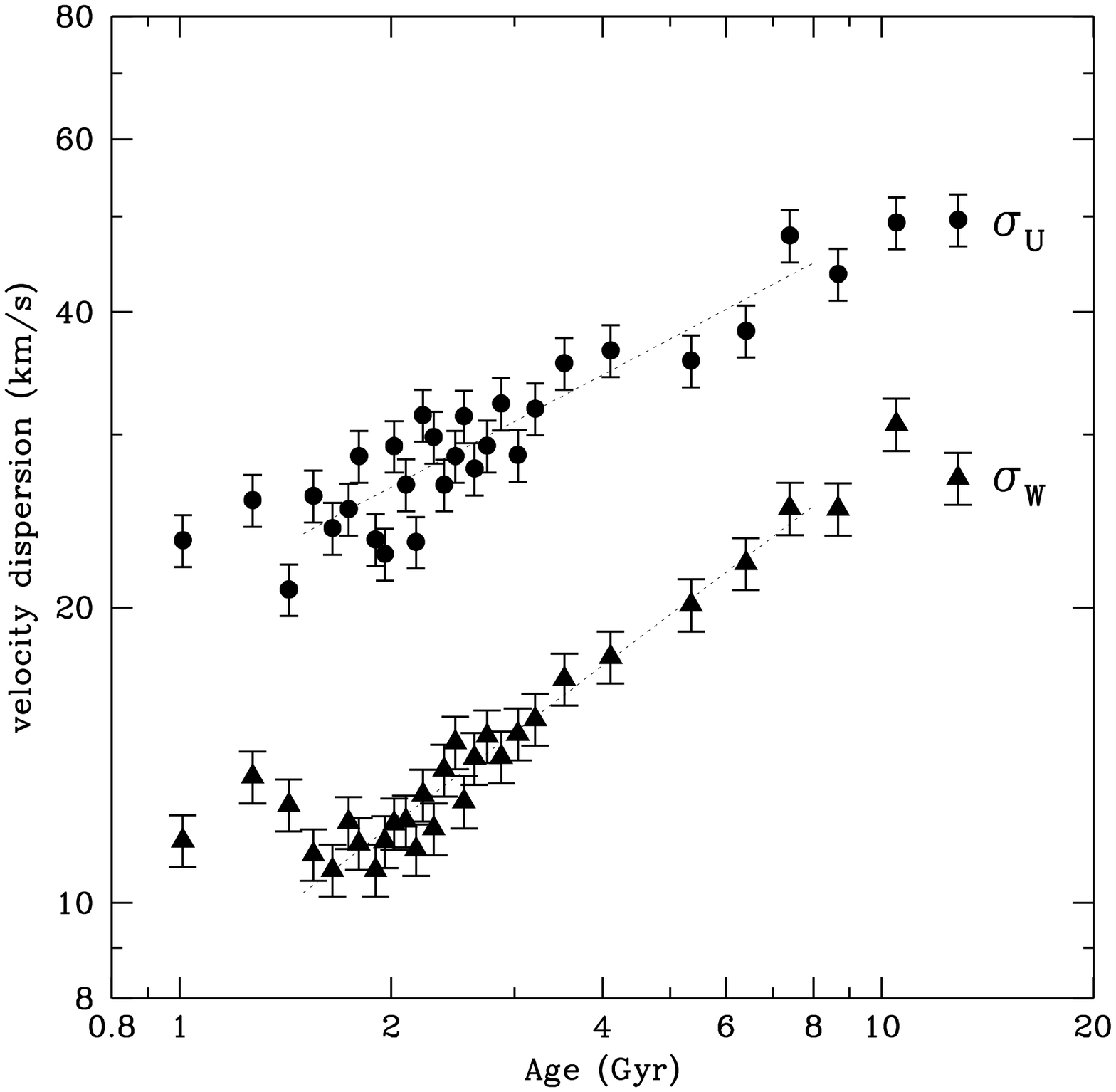}}
\caption{
Velocity dispersions vs. age for the GCS subsample of Fig. \ref{avrnew} 
(4065 single stars with $\sigma(Age)<25$\%; new calibrations). The 30 
bins have equal numbers of stars; the lines show fitted power laws. The 
3 youngest and oldest bins have been excluded from the fit.} 
\label{signew}  
\end{figure*}

Fig.~\ref{avrnew} shows the observed space velocity components as functions
of age, while Figs.~\ref{uv} and \ref{vw} show them in the {\it U-V} and 
{\it V-W} planes, separated into four groups by age. Like Figs. 20 and 30 
of the GCS, they illustrate the significant substructure in the $U$ and 
$V$ velocity distributions that persists over a wide range of ages (see 
also Famaey et al. \cite{famaey05} and the study of the Hercules stream 
by Bensby et al. \cite{bensby07}). 
In contrast, the $W$ velocities show no deviation from a random distribution 
in any age group (Fig. \ref{whist}), suggesting that different heating 
mechanisms are at work in the plane of the disk and perpendicular to it. 
However, it must also be noted that phase mixing is stronger for vertical 
motions due to the Galactic potential, and the orbital periods shorter, 
resulting in a more efficient smoothing of structure.

A detailed discussion of the mechanisms affecting disk star orbits in and 
perpendicular to the plane is beyond the scope of the present paper, but 
we can consider two basic questions, i.e. {\it (i)} whether the velocity 
dispersion of disk stars continues to increase during the lifetime of the
disk or whether it shows signs of a plateau or saturation in some age 
interval, and {\it (ii)} what functional form appears appropriate for 
the rising parts of the age-velocity relation (AVR). For this discussion, 
we divide the sample by age into 30 bins with equal numbers of stars (i.e. 
135 in each bin), against the only 10 bins used in the GCS; this allows us 
to follow the evolution in greater detail.

Fig.~\ref{signew} shows the resulting AVR, which shows a smooth, general 
increase of the velocity dispersion with time in both U, V, and W. Fitting 
power laws while excluding the three youngest and three oldest bins, we 
find exponents of 0.38, 0.38, 0.54 and 0.40 for the U, V, W and total 
velocity dispersions -- slightly larger than the values derived in the GCS, 
as expected from the change in age scale. 

This, however, is clearly not the full story of the phenomenon known 
as ``disk heating''. If that term is taken to refer to an increase in the 
{\it random} motions of disk stars, it is clear from Fig. \ref{uv} that 
the present distribution of the U,V velocities is not the result of pure 
``heating''; non-random processes are at work as well. Several possible 
mechanisms have been described in modern literature, based on detailed 
simulations, but all result in velocity dispersion increases with time that 
can be approximated by a power law. This is true, e.g. for the simulations 
that use transient spirals as the heating agent (De Simone, Wu \& Tremaine 
\cite{desimone}) or produce velocity structure and heating of the disk by 
interaction between two spiral systems (Minchev \& Quillen \cite{minchev}) 
or between a spiral and the bar (Chakrabarty \cite{chakrabarty}). 

To identify any underlying ``pure'' heating mechanism within the disk, one 
would have to remove the major non-random substructures, such as the 
Hercules stream or the Sirius–UMa, Coma, and Hyades-Pleiades branches
(GCS; Famaey et al. \cite{famaey05}; Bensby et al. 
\cite{bensby07}). This, however, introduces a certain arbitrariness as 
to the choice of stars to be removed; and it appears more interesting to 
us to identify the cause(s) of the substructures that dominate the U,V 
plane than to quantify any remaining minor effects.

On the other hand, the observed heating of the vertical (W) velocities, 
which show {\it no} signature of non-random mechanisms, is more efficient 
than found in simulations. E.g. H\"anninen \& Flynn (\cite{hanninen}) find 
an exponent of only 0.26 using molecular clouds and need massive black holes 
to reproduce the observed heating rate, in agreement with several earlier 
studies. The black-hole hypothesis is inconsistent with other observational 
data, but the extra vertical heating might be due e.g. to infalling 
satellites with dark matter substructure (Benson et al. \cite{benson}). 
In any case, further theoretical work on the heating mechanisms affecting 
the vertical velocities in the disk will have to satisfy the observational 
constraint illustrated in Fig. \ref{signew}. 

This conclusion contrasts with that by Quillen \& Garnett (\cite{quillen01}), 
who claimed that the heating of thin-disk stars saturates at a maximum 
velocity dispersion $\sim$3 Gyr after the birth of a star; the oldest stars 
were assumed to belong to the thick disk and heated by a different mechanism. 
This led Freeman \& Bland-Hawthorn (\cite{freeman02}) to suggest that 
remnants of early mergers might still be traced as dynamical groups in the 
disk today (see Helmi et al. \cite{helmi06} for an actual application). 

Fig. \ref{signew} shows no such saturation in our data, but the 
levelling-off of the velocity dispersion in the two oldest bins might be 
taken as weak evidence for a constant velocity dispersion in the thick disk. 
When judging the merits of the two results, it should be remembered that 
the sample discussed here is some 20 times larger than that discussed by 
Quillen \& Garnett (\cite{quillen01}) and the sampling errors correspondingly 
smaller.

Finally, we want to verify whether our age computation techniques might 
bias the slopes seen in the age-velocity relation (AVR). To do so, we have
imposed a fixed AVR on our synthetic catalogue, assuming purely Gaussian 
velocity distributions and adopting the slopes of $\sigma(U,V,W)$ vs. time 
as derived above. The thick disk contribution in the simulation is at the 
same level as in the observed sample, about 3\% after applying the 
apparent-magnitude cutoff. We then recomputed the ages and age errors for 
all stars in the synthetic sample, using our new calibrations and retaining 
only stars with (new) ages better than 25\%, sorted the stars into 30 
equal-size bins, and recomputed the velocity dispersion in each bin.

\begin{figure}[htbp] 
\resizebox{\hsize}{!}{\includegraphics[angle=0]{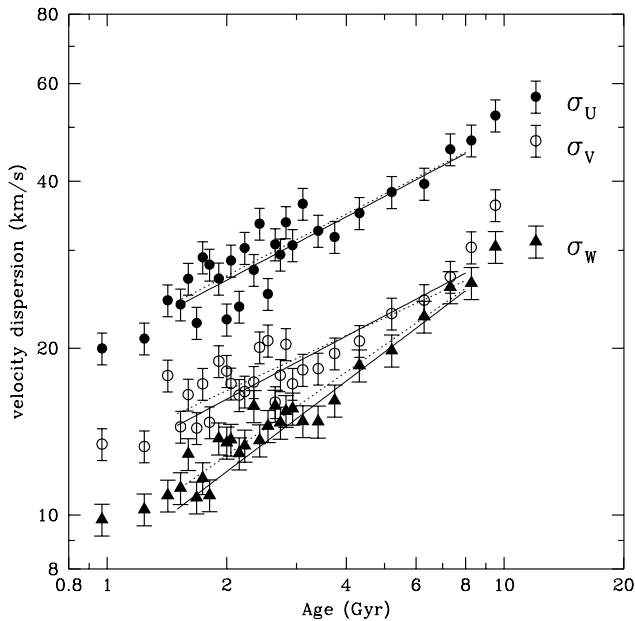}}
\caption{
Age-velocity relations for the synthetic sample. {\it Solid lines:} 
Adopted input relations for thin-disk stars. {\it Symbols and error 
bars:} Recomputed ages and velocity dispersions in 30 equal-size bins 
for stars with ages better than 25\%. {\it Dotted lines:} Fitted power laws.
} 
\label{avr2} 
\end{figure}

The input and reconstructed AVRs are compared in Fig.~\ref{avr2}. Symbols 
with error bars show the ``observed'' points with recomputed ages and age 
bins. As can be seen, all the reconstructed velocity dispersions are quite
consistent with the input AVRs over the range 1.5 -- 8 Gyr. There is thus 
no evidence for any error or bias being imprinted on the AVR from the age 
estimation process.

\section{Thick disk vs. thin disk}\label{TDtd}

In the literature, different criteria are used to distinguish thick-disk 
stars from the more numerous thin-disk stars in the Solar neighbourhood, 
based on kinematics, detailed chemical composition, or age. There 
is, however, no real consensus on what limiting values to adopt in any of 
these parameters, and the allocation of any individual star to either disk 
component is often ambiguous. Questions of current interest include the 
fraction of thick-disk stars in the Solar neighbourhood as well as the 
existence or otherwise of an AMR in the thick disk. 

The thick disk of the synthetic sample is well visible in the reconstructed
AMR and AVRs, especially among the oldest stars, where it dominates. 
In order to estimate the fraction of thick-disk stars in the GCS, 
we calculated the fraction of thick-disk stars in the simulated catalogue 
with V-velocities below -100 km/s and [Fe/H] $>$ -1. This sample contained
27\% of the total number of thick-disk stars in the catalogue, and no 
thin-disk stars were found in this velocity-metallicity domain. In the real, 
observed GCS we find 67 stars in this region. Thus, if we assume the same 
thick-disk velocity distribution as in the simulation, we find a total of 
248 thick-stars; as the total number of single stars with complete space 
velocities in the GCS is 8479, the thick-disk fraction is thus 2.9\%.

As regards a possible AMR in the thick disk we note that, interestingly, 
the ``observed'' AMR for the simulated thick-disk stars in Fig.~\ref{amrsim} 
shows a pronounced AMR of slope -0.022 dex/Gyr, although the input ``true'' 
AMR had a constant metallicity between the assumed thick-disk age limits 
of 11 and 12 Gyr. This effect would have been even more pronounced if our
strict quality criteria on accepted ages had not been applied (as in H06), or 
if the thick-disk sample had contained a small fraction of stars with younger 
ages and/or higher metallicity. 

Samples of field thick-disk stars can be constructed by combining different 
membership criteria. For nearby field stars, a kinematic criterion is most 
commonly used, based on a decomposition of the local velocity distribution 
into Gaussians corresponding to the thin and thick disk and assuming values 
for the asymmetric drift and {\it U,V,W} velocity dispersions for each. 
Membership probabilities for either disk are then assigned for each star, 
based on the observed space motion.  

First, as shown in Fig \ref{uv} and GCS Fig. 20, the actual local velocity 
distribution is nothing like a two-Gaussian model; in particular, the 
$\zeta$ Herculis stream occupies a position near the overlap between the 
thin and thick disks. Second, Fig. \ref{avr2} shows that the practice of 
assuming a single set of velocity dispersions for the thin disk, regardless 
of age, ignores the effects of the continuing dynamical heating of the disk. 
As general thin-disk samples are typically dominated by relatively young 
stars (see Figs. \ref{agecomp2} and \ref{avrnew}), the mean values of 
$\sigma(U,V,W)$ will tend to be underestimated, enhancing the probability 
that thin-disk outliers will be classified as thick-disk stars. 

If the thick disk is indeed very old, as is most commonly assumed, the 
kinematic parameters characterising the thin disk should be chosen to 
correspond to the oldest thin-disk stars (see Fig. \ref{avr2}). It 
is interesting to note that Bensby et al. (\cite{Bensby04}), who identify 
thick-disk stars with intermediate ages and chemical properties, assume 
significantly lower values of $\sigma(U,V)$ for the thin disk than Reddy 
et al. (\cite{reddy06}), who find a smaller fraction of such stars and 
consider them to be outliers from the thin disk. 
 
Given the inherent ambiguity of kinematic population indicators, the  
enhanced [$\alpha$/Fe] ratio observed in thick-disk stars may be 
the best membership criterion for individual field stars. In order to 
study the AMD of the thick disk from the cleanest possible sample of stars, 
we therefore selected all stars in common between the GCS and those 
kinematically classified thick-disk stars from Reddy et al. (\cite{reddy06}) 
that also have $\rm [Ti/Fe]>0.25$.

\begin{figure}[htbp] 
\resizebox{\hsize}{!}{\includegraphics[angle=-90]{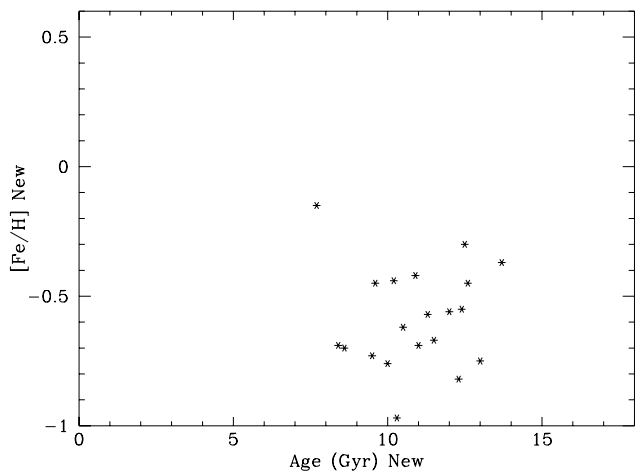}}
\caption{AMD for thick-disk stars with $[\rm Ti/Fe]>0.25$ from Reddy et 
al. (\cite{reddy06}). Only single stars with ages better than 25\% and with 
{\it E(b-y)} $<$ 0.02 or $d<$ 40 pc are shown.} 
\label{amrtd} 
\end{figure}

Fig.~\ref{amrtd} shows the AMD for this ``clean'' thick-disk sample, 
using our new ages. We see a tight grouping of stars with ages around 
11 Gyr and $\rm [Fe/H] \sim$ -0.55, and no trace of a slope. Taken together, 
Figs. \ref{amrsim} and \ref{amrtd} put into question the claims of a 
significant age-metallicity relation in the thick disk by, e.g. H06 or Bensby 
et al. (\cite{Bensby04}). 

Note that a strong, spurious correlation of age with metallicity would result 
if the Padova isochrones had not been corrected for the increasing discrepancy 
between the original isochrones and the observed unevolved stars at 
decreasing metallicity (see Sect. \ref{why}). This is also the case if the 
$\alpha$-enhancement of metal-poor stars is ignored when choosing the 
Z parameter of the models.

\section{Conclusions}\label{endchap}

We have redetermined the basic calibrations used to infer astrophysical 
parameters for the GCS stars from {\it uvby} photometry. The result 
is a substantial improvement in the determination of effective temperatures, 
especially for the hotter stars, and of absolute magnitudes (i.e. distances), 
as well as minor corrections to [Fe/H].
The GCS reddening calibration appears to be essentially correct. 
These new results are obtained by drawing on the large body 
of 2MASS K photometry, high-resolution spectroscopic abundance analyses, and 
accurate Hipparcos parallaxes that are now available. 

Using the improved astrophysical parameters, we have recomputed the ages and 
age error estimates for the GCS sample (Table \ref{table1}). In the 
process, we have compared results from a variety of stellar models, finding no 
large differences between models for typical GCS stars. We have also 
recomputed the temperature corrections needed for the models to agree with the 
observed unevolved main sequence at metallicities below solar. 

The resulting ages correlate well
with those published in the GCS, but are on average $\sim$10\% lower. The recent 
ages by VF05 for $\sim400$ stars in common are some 10\% lower still, while 
those by Takeda et al. (\cite{takeda}) are essentially in perfect agreement 
with ours, although they were also based on the spectroscopic temperatures and 
metallicities by VF05. This independent verification places the determination 
of isochrone ages on a very firm basis, provided adequate precautions are taken.

We note that, because of a small offset in the GCS {\it uvby} 
photometry for the Hyades stars and the non-standard He abundance of this 
cluster, the Hyades cannot be used to check metallicity or age scales for 
field stars such as those in the GCS, as assumed by H06. The revised results 
given here for Hyades and Coma stars are based on standard photometry 
whenever available.

The revised [Fe/H] values change the observed metallicity distribution only 
marginally; it is still in strong disagreement with the prediction of 
closed-box galactic evolution models. 

In preparation for the discussion of Galactic relations involving stellar 
ages, notably the age-metallicity and age-velocity diagrams, we have 
performed extensive simulations of the effects of our selection and 
computation procedures by applying them to synthetic catalogues with 
properties closely resembling those of the GCS, but with a variety of 
specified intrinsic properties, such as the AMR and AVR. We find that our 
methods faithfully recover the input relations within the observational 
errors, without introducing spurious trends or other systematic effects. 

The observed AMR retains the general features of that of the GCS, i.e. 
little or no variation in mean metallicity with age in the thin disk, 
plus an admixture of perhaps 3\% thick-disk stars, and with a large 
and real scatter in [Fe/H] at all ages. We find no evidence for a 
significant AMR in the thick disk. Note that these conclusions are only 
robust when backed by careful end-to-end simulations of the properties of 
the sample and associated selection criteria (notably the limits by apparent 
magnitude and the blue colour cutoff), and by meticulous attention to the 
details of the computation of the stellar ages and their uncertainties. 
Investigations using simpler approaches (e.g. H06 or Reid et al. \cite{reid07}) 
may well reach different conclusions.

A significant result of the GCS was the demonstration that the dynamical 
heating of the thin disk continues throughout its life. 
We confirm this from our revised data set, and with substantially higher time 
resolution than in the original GCS. Our simulations also confirm that no 
bias has been introduced in the slope of the AVR for the individual {\it 
U,V,W} velocity components by our age computation technique. 

We conclude that kinematic parameters taken to represent the thin disk in 
thick-disk/thin-disk separations based on observed velocities should be 
chosen to correspond to the oldest thin-disk stars. Using average values 
for thin-disk samples dominated by stars much younger than the thick disk 
may lead to contamination of the thick-disk samples by kinematic outliers 
from the thin disk, especially when such non-Gaussian kinematic features 
as the Hercules stream are present. 

\begin{acknowledgements}
We reiterate our grateful thanks to our many collaborators on the original 
GCS project from Observatoire de Gen{\`e}ve, Harvard-Smithsonian Center 
for Astrophysics, ESO, and Observatoire de Marseille. The GCS was made 
possible by large amounts of observing time and travel support from ESO, 
through the Danish Board for Astronomical Research, and by the Fonds National 
Suisse pour la Recherche Scientifique. Poul Erik Nissen and our referee, Gerry 
Gilmore, are thanked for valuable comments. We also gratefully acknowledge the 
substantial financial support from the Carlsberg Foundation, the Danish 
Natural Science Research Council, the Smithsonian Institution, the Swedish 
Research Council, the Nordic Academy for Advanced Study, and the Nordic 
Optical Telescope Scientific Association. This publication makes use of
data products from the Two Micron All Sky Survey, which is a joint project of
the University of Massachusetts and the Infrared Processing and Analysis
Center/California Institute of Technology, funded by the National Aeronautics
and Space Administration and the National Science Foundation. We have also 
made use of the SIMBAD and VizieR databases, operated at CDS, Strasbourg, 
France. 
\end{acknowledgements}

\end{document}